\documentclass[twocolumn]{aastex63}

\newcommand{\teff}{$T_{\mathrm{eff}}$}
\newcommand{\logg}{log $g$}
\newcommand{\loggf}{log $gf$}
\newcommand{\ebv}{$E(B-V)$}
\newcommand{\feh}{[Fe/H]}

\newcommand{\kms}{km-s$^{-1}$}

\newcommand{\vrad}{$V_{RAD}$}
\newcommand{\vrot}{$v\sin i$}
\newcommand{\vturb}{$\xi$}
\newcommand{\BVp}{$(B-V)'$}
\newcommand{\Vp}{$V'$}
\shorttitle{Photometry and Spectroscopy in NGC 7789}
\shortauthors{Anthony-Twarog, Brunker, Deliyannis, Rich, Steinhauer, Sun, \& Twarog}

\begin{document}


\title{WOCS XCIII: NGC 7789: the Evolution of Li, Stellar Rotation,\\ and Extended Main Sequence Turnoffs}


\author{Barbara J. Anthony-Twarog}
\affiliation{Department of Physics and Astronomy, University of Kansas, Lawrence, KS 66045-7582, USA}
\email{bjat@ku.edu}
\author{Samantha W. Brunker}
\affil{Department of Physics, University of Connecticut, Storrs, CT 06269, USA}
\email{samantha.brunker@uconn.edu}
\author{Constantine P. Deliyannis}
\affiliation{Department of Astronomy, Indiana University, Bloomington, IN 47405-7105, USA}
\email{cdeliyan@iu.edu}
\author{Evan Rich}
\affiliation{Department of Physics \& Astronomy, University of Nebraska-Lincoln, NE 68588-0299, USA}
\email{erich3@unl.edu}
\author{Aaron Steinhauer}
\affiliation{Department of Physics and Astronomy, State University of New York, Geneseo, NY 14454, USA}
\email{steinhau@geneseo.edu}
\author{Qinghui Sun}
\affil{Department of Astronomy, Tsinghua University, Beijing, 100084, China}
\email{qingsun@tsinghua.edu.cn}
\author{Bruce A. Twarog}
\affiliation{Department of Physics and Astronomy, University of Kansas, Lawrence, KS 66045-7582, USA}
\email{btwarog@ku.edu}

\begin{abstract}
Precision $UBVRI$ photometry of NGC 7789 is combined with Gaia data to map reddening variations across the cluster face.  Hydra spectra, Gaia astrometry, and isochrone fitting constrain the absolute reddening, apparent modulus, and age to \ebv\ $= 0.30 \pm 0.02$, $(m-M) = 12.51 \pm 0.06$, and $1.46 \pm 0.02$ Gyr for [Fe/H] between $-0.2$ and solar; the spectroscopic [Fe/H] $= -0.13 \pm 0.068$ (MAD) from 156 single-star members. Corrections for variable reddening reduce the scatter in the unevolved main sequence below the turnoff. A(Li) is derived for only single star members from the G-dwarf Li-Plateau to the tip of the red giant branch. Giants separate into two distinct groups, probable first-ascent giants with detectable Li that declines with evolution toward the red giant tip and stars within the clump and the asymptotic giant branch which only exhibit upper limits. A(Li) structure from the turnoff to the unevolved main sequence, including the Li-Dip, and the presence of an extended color spread among the upper main sequence stars are attributed to the \vrot\ distribution, indicating the wall of the Li-Dip as the true hot boundary of the Kraft break. Differences in the color-magnitude diagram topology of NGC 7789 and NGC 752 are explored and attributed to differences in the individual cluster \vrot\ distributions. Prior indications that main sequence stars more massive than the Li-Dip evolve redward across the Li-Wall, undergoing rotational spindown and Li depletion like stars within the Li-Dip, are confirmed.   

\end{abstract}

\section{Introduction}
In many respects, NGC 7789 is typical of the open cluster population within 1.5 kpc of the Sun: a near-solar composition, with significant, but not excessive, reddening ($E(B-V)$ $\sim$ 0.3) for its distance ($(m-M) \sim 12.4$) and galactic anticenter location (($l, b$) = (115.5\arcdeg, -05.4\arcdeg)).  Its age is comparable to the evaporation timescale of most open clusters ($\sim$1.5 Gyr) \citep[see, e.g.,][]{TW97,FR02,NE16, CA18}.
What sets this cluster apart is its unusual richness, matched by only a few clusters older than the Hyades, e.g. NGC 2158 and NGC 6791, clusters initially studied as potential low density globular clusters \citep{AC62, KI65}. 
Compared to two well-studied, nearby open clusters of comparable age, NGC 752 \citep{BO15, BO20, TW15, TW23, SA23} and NGC 3680 \citep{NO96, AT04, AT09}, each with fewer than a dozen giant branch members \citep{CA18}, NGC 7789's giant branch is more than an order of magnitude richer in numbers, providing an extensive and homogeneous sample for studies of late-stage stellar evolution for stars of intermediate to low mass.  
The age and richness of NGC 7789 have led to comparisons with the young, numerically rich clusters of the Magellanic Clouds, clusters that have been the focus of ongoing discussion regarding color-magnitude diagram (CMD) topology and plausible sources of unexpected scatter among stars at the main sequence  turnoff, commonly referred to as eMSTOs \citep[see, e.g.,][and many references therein]{MI09, GI13, WU16, YA18}. 

Stellar masses on the giant branch of NGC 7789 straddle the critical range for which He-ignition at the tip of the first-ascent red giant branch (RGB) switches from quiescent to non-quiescent under degenerate conditions \citep{GI16}, potentially leading to significant changes in the distribution of stars in the red giant region if the RGB stars encompass a range of mass \citep{GR98, GI99, GI00a}.  As stars of this mass leave the main sequence turnoff (MSTO) and evolve toward the subgiant branch (SGB), their essentially radiative atmospheres switch to an increasingly convective structure culminating with the first dredge-up near the base of the RGB. 
The rapidity of this evolutionary phase accounts for the paucity of stars between the MSTO and the luminosity level of the red clump for poorly populated clusters like NGC 752 and NGC 3680. 

Given the above, NGC 7789 was an obvious choice more than two decades ago for inclusion in the WIYN Open Cluster Survey (WOCS) \citep{MA00}.  Observations in the WOCS program have included pre-Gaia astrometry, broad- and intermediate-band precision photometry, high-resolution spectroscopy, and radial velocity analysis for membership and binarity, all aimed at further elucidating stellar evolution in the varying contexts of age and metallicity that characterize the Galactic disk.

The relevance of NGC 7789 was further enhanced by the increased focus within WOCS on probing stellar structure and evolution among intermediate-to-low-mass stars via atmospheric Li. Despite the greater distance and higher reddening, the richness of NGC 7789 offered an enticing counterpoint to the Li analyses in intermediate-age systems like NGC 752 and NGC 3680 \citep{AT09, BO22}, while being significantly older than the Hyades/Praesepe clusters \citep{CU17}. It was expected that the stars situated approximately 1.2 mag below the level of the turnoff would exhibit the sharp drop in atmospheric Li defining the high-mass edge of the so-called Li-Dip \citep{BT86}, a very striking feature in the Li-\teff\ plane not predicted by standard stellar evolution theory,  while better constraining the range and lower mass boundary of this exceptional feature as well as its dependence upon age and metallicity. 

In 2000, NGC 7789 was targeted by WOCS for reevaluation using a large set of CCD frames covering the full range of {\it UBVRI} filters.  As will be discussed in later sections and Appendix A, the resulting photometry achieved the exceptional precision desired, but exhibited radially-dependent zero-point variations across the field of the CCD frames. This issue was compounded by the discovery of comparable problems with the original NGC 7789 standard star photometry of \citet[][(hereafter ST00)]{ST00}. 

A decade later, HYDRA spectra were obtained for over 400 stars in the field of NGC 7789.
As regularly emphasized within the WOCS Li cluster studies \citep[see, e.g.,][and references therein]{AT21, SU22, AT24, SU26a}, proper spectroscopic analysis requires accurate temperatures and surface gravities, while interpretation of any resulting evolutionary patterns demands accurate and consistent stellar masses and ages. Both requirements emphasize the need for reliable determinations of the key cluster parameters of reddening, metallicity, age, and distance on, at minimum, an internally consistent scale, for which our analysis procedures require precise and accurate photometry.

Several research results published in the past five years have enabled progress on these fronts, beginning with 
a reanalysis of the original WOCS broad-band data made possible by an expanded and revised compilation of UBVI standards on the ST00 system.  In addition to the ongoing astrometric and photometric contributions of the Gaia satellite survey \citep{GA16, GA18, GA21, GA22}, we have benefited greatly from 
the multi-decade radial velocity survey of NGC 7789, identifying members, non-members, single stars, and binaries from hundreds of stars \citep{NI20} (NI20).
While the broad patterns developed from the NGC 7789 spectroscopic sample from the turnoff through the giant branch have been illustrated in a number of papers detailing other clusters within the WOCS survey \citep{DE19, TW20, AT21}, the individual results for the astrometrically restricted sample of single and binary systems in NGC 7789 have never been published until now.

The remainder of the paper is outlined as follows: Section 2 lays out the observational data, photometric and spectroscopic. Section 3 uses radial velocity and Gaia astrometric analyses to isolate cluster members and generate the cluster CMD. In Section 4 the cluster mean reddening, metallicity, and age are derived. Section 5 presents the detailed spectroscopic abundance analysis of 324 single members, focusing on metallicity estimation, while Section 6 emphasizes the Li abundances and the evolutionary interpretation of the emergent patterns, contrasted in Section 7 with a comparable analysis of the cluster twin, NGC 752. Section 8 probes the implications of the CMD topology for NGC 7789 and Section 9 summarizes our conclusions.  In Appendix A, the broad-band colors of ST00 and Section 2 are tested for spatially-dependent variation via comparison to Gaia photometry, while our precision $BV$ data, calibrated in Appendix B, and Gaia photometry are combined to map the variable reddening across the field of the cluster in Appendix C.

\section{Observations}
\subsection{CCD Data Collection and Merger}

Observations of NGC 7789 on the $UBVRI$ system were obtained with the WIYN 0.9m telescope at Kitt Peak National Observatory during a series of runs in 2003, 2004, and 2007,  for a total of nine nights of cluster data. The WIYN telescope at the time was equipped with an S2KB $2048\times2048$ CCD mounted at the $f$/7.5 focus of the telescope, providing a field of view $\sim$ 20\arcmin\ on a side. Nightly observations included both dome and sky flats for all filters, as well as bias frames. For NGC 7789, the telescope position was adjusted periodically during each night to ensure that stars located near bad pixels would not be excluded from the final photometric catalog. 

With the exception of one night, standard star observations were limited to occasional frames for a subset of the $UBVRI$ filters 
of both standard \citep{LA92} fields and open cluster fields.  The latter include NGC 6633, NGC 6791, NGC 6939, and NGC 188, as observed and compiled within ST00. This typically resulted in one to three cumulative frames in any given filter for any standard field. 
This somewhat nontraditional approach was possible because NGC 7789 was included in the original compilation of ST00, providing an internal calibration of the zero points for all filters except $U$ irrespective of the photometric conditions on any given night.
(Of the ST00 cluster standard fields at the time, only NGC 6791 included $U$ filter data.)

Preliminary analysis of the frames taken before 2007 \citep{GN04} exhibited the desired small internal scatter but indicated potential systematic variations with position across the face of the cluster.  To minimize any possible impact of the earlier processing procedures,  beginning in 2012, all frames were reprocessed through bias subtraction, trimming, and flat-fielding, using standard procedures within IRAF\footnote{IRAF is distributed by the National Optical Astronomy Observatory, which is operated by the Association of Universities for Research in Astronomy, Inc., under cooperative agreement with the National Science Foundation.}. A point of concern was the earlier reliance on dome flats necessitated by the slow readout of the S2KB chip in operation until 2013.  The limited sky frames were used to define an illumination correction to the highly precise but nonuniformly illuminated dome flats \citep[see, e.g.,][]{TW15}. Resulting PSF-based magnitudes derived for each frame exhibited a quadratic dependence on spatial variables, with a simple offset applicable to the magnitude difference between frames. 

Figure 1 shows the average standard-error-of-the-mean (sem) for $V$ and all color indices as a function of $V$. SEMs are only plotted for stars with at least three frames in each color for a given color index. Due to the large number of frames and the consistent reduction procedure, the derived precision of the final magnitudes and colors, as defined by the calculated dispersions of the magnitudes among all the frames of a given color, remains high to $V$ $>$ 17 in all indices. 

\begin{figure}
\figurenum{1}
\includegraphics[angle=0,width=\linewidth]{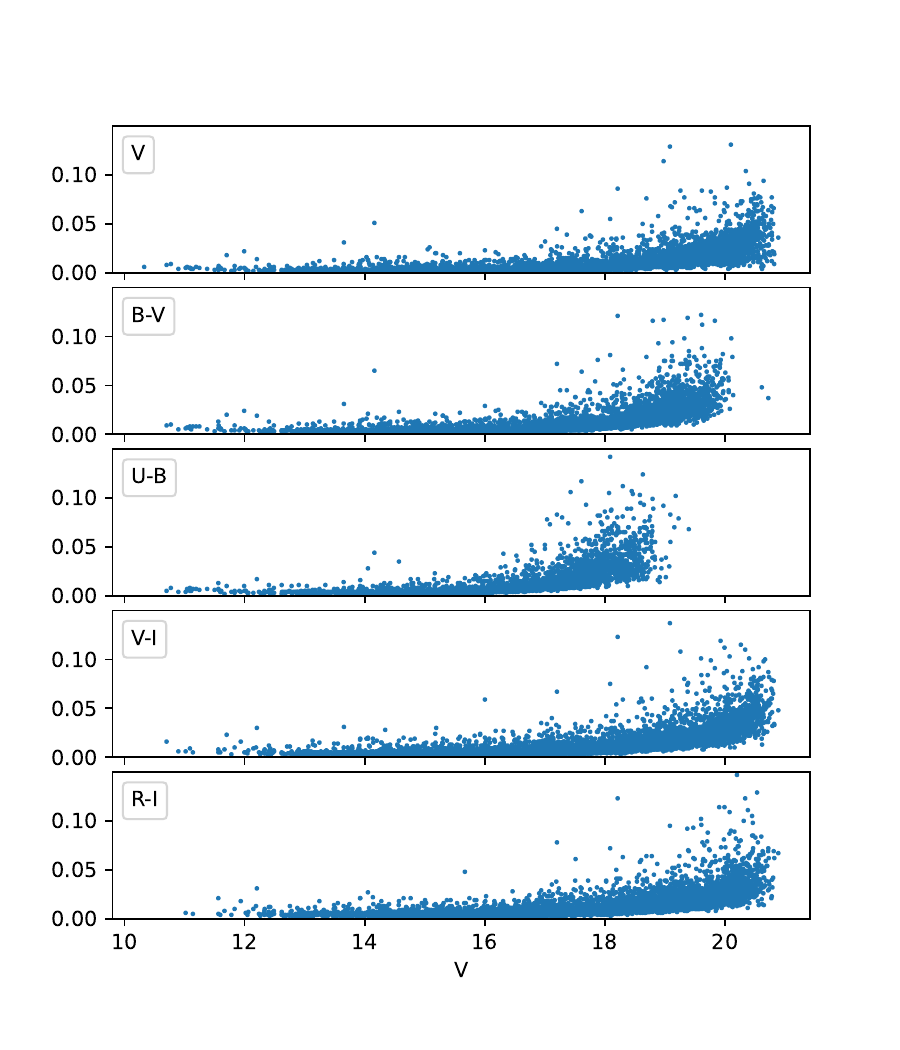}
\caption{Standard-errors-of-the-mean (sem) for each index and $V$ as a function of $V$ mag.}
\end{figure}

As expected, color indices that include a filter bandpass positioned farther toward the blue, i.e. $(B-V)$ and $(U-B)$, fall below high precision ($\sigma_{sem}$ $<$ 0.025) at brighter $V$ magnitudes.

\subsection{CCD $UBVRI$ Calibration Relations}

NGC 7789 exhibits an almost  two-magnitude range in $(B-V)$ from its blue stragglers to the coolest giants, making transformation from the instrumental to the standard system for any broad-band color index a challenge. The traditional approach of using a few dozen standards with linear color terms often proves inadequate when attempting to delineate the final colors for stars at the extreme but critical ends of the color distribution. 
Earlier attempts to use the ST00 standards across the 5 filters within NGC 7789 were severely limited  due to the variable range in the number of overlapping stars. For $U, B, V, R$, and $I$, the precision standards of interest numbered 0, 98, 599, 26, and 599, respectively. The more recent ST00 releases have raised the data base to a few hundred stars in all filters except $R$, a point discussed more fully in Appendix B. 

Standardization of the instrumental photometry followed general precepts as follows.  
Instrumental photometric indices for standard stars were retained only if there were three or more observations in each filter used to construct an index.   
For all indices and the $V$ magnitude, a general calibration equation of the form:

\begin{eqnarray}
INDEX_{stand} = a*INDEX_{instr} + b*(B-V)_{instr}^2 \nonumber \\
+ c*(B-V)_{instr} +  d \nonumber \\
\end{eqnarray}
\noindent
was adopted initially. However, the rich sample of standards coupled with the unusual level of precision for the instrumental data allowed a more nuanced and comprehensive calibration across the entire color range for all indices except $(V-R)$. 

To improve the calibration definition, cuts were also made based upon the calculated sem for the instrumental photometry.
These limits are explained within the more detailed discussion of the calibration of each index in Appendix B.

The final NGC 7789 photometry for 6372 stars is presented in Table 1, where the columns are self-explanatory. Coordinates are on the \citep{GA22} (DR3) system. With the exception of a few stars above $V$ = 13 where the internal errors for the frames are assumed to be small due to the brightness of the stars, photometric indices are only listed if both filters within an index have at least 2 observations. If not, the number of filter observations has been set to 0, while the index and index error have been eliminated.

\floattable
\begin{deluxetable}{rrrrrrrrrrrrrrrrr}
\tablenum{1}
\tablecaption{UBVRI Photometry in NGC 7789}
\tabletypesize\small
\tablewidth{0pt}
\tablehead{
\colhead{$\alpha(\rm{DR3})$} & \colhead{$\delta(\rm{DR3})$} &
\colhead{$V$} & \colhead{$B-V$} & \colhead{$V-I$} &  \colhead{$U-B$} & \colhead{$V-R$} &
\colhead{$sem_{V}$} &\colhead{$sem_{BV}$} &\colhead{$sem_{VI}$} &\colhead{$sem_{UB}$} &\colhead{$sem_{VR}$} &
\colhead{$N_V$} &\colhead{$N_B$} &\colhead{$N_I$} &\colhead{$N_U$} &\colhead{$N_R$} }
\startdata
 358.93106  & 56.79279  &  9.375 &  0.012 &  ..... &  ..... &  .....  & ..... & ..... & ..... & ..... & ..... &  1 &  1 &  0 &  0 &  0 \\
 359.18655  & 56.73523  & 10.302 &  0.191 &  0.243 &  0.111 &  0.121  & 0.006 & 0.007 & 0.008 & 0.006 & ..... &  3 &  2 &  2 & 11 &  1 \\
 359.38028  & 56.58853  & 10.457 &  0.749 &  0.699 &  0.294 &  .....  & 0.015 & 0.022 & ..... & 0.017 & ..... &  2 &  4 &  1 & 23 &  0 \\
 359.26364  & 56.76630  & 10.668 &  1.877 &  2.150 &  2.277 &  .....  & 0.008 & 0.009 & 0.016 & 0.005 & ..... &  4 & 18 &  4 & 26 &  0 \\
 359.29330  & 56.71404  & 10.739 &  1.827 &  2.080 &  2.242 &  1.054  & 0.009 & 0.010 & 0.015 & 0.008 & 0.010 &  9 & 19 &  2 & 26 &  2 \\
 359.37802  & 56.84773  & 10.871 &  1.848 &  2.033 &  2.196 &  1.055  & 0.004 & 0.005 & 0.006 & 0.004 & 0.013 &  8 & 19 &  4 & 26 &  2 \\
 359.65274  & 56.79558  & 10.886 &  0.581 &  ..... &  0.171 &  0.388  & ..... & ..... & ..... & ..... & ..... &  1 &  1 &  0 &  5 &  1 \\
 359.35596  & 56.66088  & 10.998 &  1.142 &  1.201 &  0.947 &  0.644  & 0.005 & 0.006 & 0.006 & 0.004 & 0.006 & 13 & 17 &  4 & 24 &  6 \\
 359.07410  & 56.66928  & 11.010 &  1.796 &  1.922 &  2.179 &  0.985  & 0.006 & 0.007 & ..... & 0.007 & 0.038 &  4 & 18 &  1 & 26 &  2 \\
 359.46490  & 56.64973  & 11.048 &  1.960 &  2.351 &  2.286 &  .....  & 0.005 & 0.008 & 0.024 & 0.008 & ..... &  6 & 10 &  2 & 23 &  0 \\
 359.46667  & 56.70765  & 11.058 &  1.760 &  1.880 &  2.117 &  0.994  & 0.005 & 0.006 & 0.009 & 0.005 & 0.010 &  8 & 17 &  5 & 23 &  2 \\
 359.21890  & 56.72523  & 11.082 &  1.718 &  1.849 &  1.929 &  1.027  & 0.004 & 0.005 & 0.011 & 0.005 & ..... &  6 & 21 &  2 & 26 &  1 \\
 359.35486  & 56.63406  & 11.108 &  0.165 &  0.263 & -0.079 &  0.114  & 0.004 & 0.008 & 0.005 & 0.007 & 0.006 & 11 & 10 & 15 & 20 & 13 \\
 359.28351  & 56.72298  & 11.166 &  1.616 &  1.671 &  1.937 &  0.930  & 0.006 & 0.008 & 0.013 & 0.007 & 0.022 &  7 & 15 &  2 & 21 &  2 \\
 359.29639  & 56.74175  & 11.220 &  1.709 &  1.773 &  2.068 &  0.955  & 0.005 & 0.008 & 0.006 & 0.006 & 0.009 & 10 & 16 &  2 & 24 &  2 \\
 359.28094  & 56.69526  & 11.348 &  1.650 &  1.731 &  1.961 &  0.957  & 0.004 & 0.005 & 0.017 & 0.007 & 0.009 & 15 & 21 &  2 & 23 &  8 \\
 359.35184  & 56.80865  & 11.479 &  1.612 &  1.657 &  1.820 &  0.908  & 0.003 & 0.003 & 0.023 & 0.006 & 0.007 & 18 & 24 &  2 & 26 &  7 \\
 359.60623  & 56.70774  & 11.520 &  0.116 &  0.309 & -0.304 &  0.179  & 0.005 & 0.013 & 0.008 & 0.013 & 0.020 &  5 &  7 &  8 & 14 &  5 \\
 359.43301  & 56.66245  & 11.535 &  0.236 &  0.248 &  0.144 &  0.121  & 0.003 & 0.006 & 0.005 & 0.007 & 0.005 & 12 & 13 & 15 & 16 & 10 \\
 359.61517  & 56.83525  & 11.580 &  1.351 &  1.340 &  1.303 &  0.775  & 0.007 & 0.009 & ..... & 0.006 & 0.017 &  8 & 11 &  1 & 16 &  3 \\
\enddata
\tablecomments{This table is available in its entirety in machine readable form in the online article.} 
\end{deluxetable}

\subsection{HYDRA Spectra Acquisition}
Stars were initially selected for spectroscopic study using a combination of astrometric membership from \citet{MS81}, radial velocity surveys, primarily \citet{GI98a} and preliminary data from the comprehensive study by NI20, as well as position within the CMD. For stars below $V$ $\sim$ 15, where the \vrad\ data were sparse and the astrometric uncertainties rapidly increased, reliance upon CMD position using earlier versions of the $BV$ photometry \citep{GN04} detailed in this paper became the primary methodology. To minimize contamination from binaries, stars located along the blue side of the main sequence were emphasized. 

Spectra for 433 stars in the field of NGC 7789 were obtained in six observing runs from September 2010 to December 2013 using the Hydra multi-object spectrograph on the WIYN 3.5-meter telescope.\footnote{The WIYN Observatory was a joint facility of the University of Wisconsin-Madison, Indiana University, Yale University, and the National Optical Astronomy Observatory.} To cover different magnitude and color ranges of candidate stars, seven separate fiber configurations were developed. The range in $V$ and the cumulative integration time for each configuration are as follows: 10.7 - 13.6, 40M, 12.2 - 14.8, 3H, 12.4 - 13.4, 3H 25M, 13.0 - 15.2, 6H 30M, 13.1 - 15.2, 7H 20M, 13.8 - 15.1, 7H 35M, and 15.1 - 16.5, 19H 12M. The fiber configurations also incorporated dozens of unassigned fibers from which simultaneous spectra could be used for sky subtraction. 

The adopted spectrograph setup is the same as that detailed in \citet{LB15} for NGC 6819 since, when possible, both clusters were observed in the same observing runs. The spectra were centered on 6650 \AA\ with a dispersion of 0.2 \AA\ per pixel and a range of $\sim400$ \AA. Examination of thorium-argon lamp spectra, used for wavelength calibration, indicates lines 2.5 pixels wide (FWHM), yielding an effective spectral resolution of 13,300. In addition to longer Th-Ar lamp spectra obtained during the day, comparison lamp spectra were obtained before and after object exposures in the course of the night.  Comparison lamps, dome flats and day-time sky spectra were obtained with the same fiber configurations used for program observations. These daytime solar spectra were used to calibrate the individual fiber sensitivities. The daily sky spectra obtained for each fiber configuration also provide solar reference spectra from which \loggf\ values may be adjusted to reproduce solar abundances for each line. Adopted \loggf\ values for these data are included in Table 4.

Our IRAF-based processing and reduction steps have been described in past papers \citep[see, e.g.,][]{LB15, AT18a, AT18b, AT21, AT24} and include the typical application of bias subtraction, flat-fielding using dome flats for each configuration, and wavelength calibration using comparison lamp exposures. Our strategy for cosmic ray cleaning uses ``L. A. Cosmic''\footnote{http://www.astro.yale.edu/dokkum/lacosmic/, an IRAF script developed by P. van Dokkum (van Dokkum 2001); spectroscopic version.} on the long exposure frames after the flat field division step. As will be demonstrated in Section 2.5 through comparison with NI20, absolute calibration of the wavelength scale for the spectra using comparison lamp exposures bracketing the program observations proved more than adequate for reliably defining the wavelength scale for individual stars. 

Final composite spectra were obtained by co-addition of multiple exposures to obtain the highest possible Signal-to-Noise per pixel ratio (S/N). We estimate the S/N for each star's spectrum from the signal above sky for the co-added spectra prior to continuum fitting.  In all cases, S/N for our spectra exceeds 125 with a median value of nearly 400.  

Spectra were examined for year-to-year radial velocity variations before co-addition.  No evidence was found for variations within the errors, which are large for some of the stars, especially those near the MSTO.  Our sample had been designed to include as few binaries as possible, relying on previous \vrad\ work for the evolved stars and photometric criteria for the fainter stars.

\subsection{Radial and Rotational Velocities: {\it fxcor}}
The Fourier-transform, cross-correlation utility {\it fxcor} in IRAF was used to assess kinematic information for each star from the summed composite spectra. In {\it fxcor}, program stars are compared to zero-velocity stellar templates of similar \teff.  The {\it fxcor} utility characterizes the cross-correlation-function (CCF), from which estimates of each star's radial velocity are inferred, easily enough for stars with sufficiently narrow lines (cooler and/or slow rotation). Templates are usable for spectral types F5 through K6, implying reduced efficacy for the coolest stars in NGC 7789. Larger \vrad\ errors for stars 8799 and 7029 are clearly due to their cooler spectra with M-star like molecular bands. (Unless otherwise noted, stellar identification numbers refer to the WEBDA\footnote{http:// webda.physics.muni.cz} number.) A few other cool stars exhibit unusually large \vrad\ errors; of the likely members, two are suspected or confirmed binaries (9728 and 5237) while another star, 13623, shows wide lines typical of rapid rotation.

Two stars were serendipitously observed in separate configurations in September and October of 2010.  Differences between separate configurations' \vrad\ values were 1.6 and -0.1 \kms\ for the two stars, identified in the comments for Table 2 in which average values are reported.

Projected rotational velocities, \vrot, can also be estimated from the cross-correlation-function full-width-half-maximum (CCF FWHM).  We employ a procedure developed by \citet{ST03} which exploits the relationship between the CCF FWHM, line widths and \vrot\ using a set of numerically ``spun up" standard spectra with comparable spectral types to constrain the relationship. Measured values for rotational velocity necessarily include the effect of stellar inclination through 
a $\sin i$ term.

A basic limitation here is enforced by spectral resolution, implying that derived rotational velocities below $\sim$10 \kms\ are not meaningful, in the sense that one is essentially measuring the instrumental profile. It is also apparent that both velocity estimates are severely compromised by projected rotational velocities larger than 35 to 40 \kms. Formal errors implied by the CCF analysis with $fxcor$ suggest that typical formal \vrad\ errors for our sample in NGC 7789 are $\sim 1$ \kms\ for stars with \vrot\ $\leq 40$ \kms.

We have shown in past cluster analyses that it is instructive to track the \vrot\ characteristics of evolved and unevolved stars in clusters of different ages. The rotational characteristics of MSTO stars in NGC 7789 are particularly critical since rapid rotation has a detrimental effect on so many of the spectra for stars in this cluster. Examining the MSTO sample of 217 stars judged to be single members, a histogram of \vrot\ values is bimodal.  Bearing in mind that \vrot\ values $\geq 40$ \kms\ should be viewed cautiously, one subgroup in the cluster is conspicuous for a median \vrot\ of 90 \kms, the MSTO stars brighter than $V = 15$.  By contrast, MSTO stars below this level, the level of the main sequence Li-gap wall, have a median projected rotational velocity under 25 \kms, consistent with the narrower main sequence appearance in the CMD (see, e.g., Figure 9).  The cooler subgiants and giants have a median \vrot\ of less than 10 \kms, formally consistent with even smaller \vrot\ values near 1-3 \kms\ and lines dominated by the instrumental profile.
We will return to the question of \vrot\ derivation for all cluster members using spectrum synthesis, a more reliable approach for the higher rotation speeds, in Section 6.1.  

Table 2 contains the basic information on the initial sample of 433 stars observed spectroscopically. The values listed for \vrot\ are those obtained, when possible, from $fxcor$ alone.
\floattable
\begin{deluxetable}{rrcrrhrrrrr}
\tablenum{2}
\tablecaption{Hydra Spectroscopic Sample}
\tablewidth{0pt}
\tablehead{
\colhead{$\alpha(\rm{DR3})$} & \colhead{$\delta(\rm{DR3})$} & \colhead{WEBDA} &
\colhead{$V$} & \colhead{$(B-V)$} & \nocolhead{phsrc} & 
\colhead{Status} & \colhead{$V_{RAD}$} & \colhead{$\sigma_{Vrad}$}  &
\colhead{\vrot} & \colhead{$\sigma_{v \sin i}$} \\
\colhead{} & \colhead{} & \colhead{} & \colhead{} & \colhead{} & 
\nocolhead{} & \colhead{} & 
\colhead{\kms} & \colhead{\kms} & \colhead{\kms} & \colhead{\kms} }
\startdata
359.26354 & 56.76608 & 6345 & 10.668 & 1.877 & 0 & M & -53.1 & 1.1 & 12.0 & 0.4 \\ 
359.56984 & 56.47105 & 12776 & 10.722 & 0.332 & 1 & NM & .... & ... & .... & ... \\ 
359.29335 & 56.71370 & 7029 & 10.739 & 1.827 & 0 & M & -52.6 & 6.5 & $\leq 10.0$ & 2.3 \\ 
359.63928 & 56.58469 & 13862 & 10.797 & 1.885 & 1 & M & -54.2 & 1.4 & $\leq 10.0$ & 0.3 \\ 
359.37826 & 56.84761 & 8799 & 10.871 & 1.848 & 0 & M & -56.8 & 7.0 & 99.1 & 36.6 \\ 
359.24366 & 56.91919 & 5873 & 10.929 & 0.160 & 1 & NM & 63.9 & 8.0 & 51.0 & 6.4 \\ 
359.07430 & 56.66880 & 2740 & 11.010 & 1.796 & 0 & M & -53.7 & 1.1 & 16.5 & 0.6 \\ 
359.46486 & 56.64906 & 10746 & 11.048 & 1.960 & 0 & M & -51.7 & 2.3 & $\leq 10.0$ & 0.5 \\ 
359.46687 & 56.70710 & 10765 & 11.058 & 1.760 & 0 & NM & -56.9 & 1.3 & $\leq 10.0$ & 0.3 \\ 
359.28354 & 56.72266 & 6810 & 11.166 & 1.616 & 0 & M & -55.5 & 0.7 & $\leq 10.0$ & 0.2 \\ 
359.35245 & 56.50005 & 8400 & 11.184 & 1.828 & 1 & NM & -63.8 & 1.1 & 10.9 & 0.4 \\ 
359.29639 & 56.74146 & 7091 & 11.220 & 1.709 & 0 & M & -56.1 & 1.0 & 12.4 & 0.4 \\ 
359.29190 & 56.68237 & 7013 & 11.327 & 1.543 & 1 & MB & -51.9 & 1.0 & 13.1 & 0.5 \\ 
359.28100 & 56.69489 & 6767 & 11.348 & 1.650 & 0 & M & -54.8 & 1.2 & $\leq 10.0$ & 0.2 \\ 
358.92806 & 56.92572 & 896 & 11.475 & 1.631 & 1 & M & -55.4 & 0.8 & 12.0 & 0.3 \\ 
359.35198 & 56.80846 & 8293 & 11.479 & 1.612 & 0 & M & -55.5 & 0.7 & 12.1 & 0.3 \\ 
359.60648 & 56.70687 & 13273 & 11.520 & 0.116 & 0 & NM & -12.6 & 10.3 & 55.5 & 7.9 \\ 
359.20955 & 56.98390 & 5179 & 11.555 & 0.235 & 1 & NMB & -27.3 & 4.9 & 20.2 & 3.0 \\ 
359.56131 & 56.86296 & 12478 & 11.576 & 1.594 & 0 & MB & -57.9 & 0.7 & 10.9 & 0.3 \\ 
359.38269 & 56.68945 & 8957 & 11.641 & 1.459 & 0 & M & -55.1 & 0.6 & $\leq 10.0$ & 0.2 \\ 
\enddata
\tablecomments{This table is available in its entirety in machine readable form in the online article.} 
\end{deluxetable}

\subsection{Comparison to Previous Work}
The definitive radial velocity sample for NGC 7789 is that of NI20. Cross-matching our sample with theirs and eliminating all binaries (NI20), whether cluster members or not, generates a data base of 276 stars in common. A straight average among the residuals in the sense (NI20 - Table 1) produces $-0.80 \pm 4.9$ \kms\ (sd). Reducing the sample to those stars with derived errors in radial velocity at or below 5.0 \kms\ generates a mean residual of $-0.11 \pm 1.35$ \kms\ (sd) from 135 stars. The dramatic reduction in the sample produced by the tighter cut on the error in \vrad\ is due to the significant fraction of stars near the top of the main sequence which exhibit significant \vrot, boosting the uncertainty in \vrad, an issue absent from the rich but cooler sample of evolved stars. From either sample, however, it is clear that the \vrad\ for both data sets are on the same system within the errors.

\section{Constructing the CMD}
\subsection{Membership From Gaia Astrometry}
The richness of the cluster field is readily apparent in Figure 2 where the CMD of all the stars in Table 1 is plotted. While the presence of the cluster is detectable against a plethora of field stars, key features of the cluster CMD, e.g. the subgiant and first-ascent giant branches, become lost in the noise. At the turnoff, the line-of-sight field star distribution becomes a significant factor redward of $(B-V)$ = 0.7 fainter than $V$ = 16. To isolate the cluster from the field, we followed the same approach consistently applied in our past cluster work within the Li series, as exemplified by the recent analyses of M67 \citep{TW23} and NGC 2204 \citep{AT24}, beginning with member selection using the astrometric information supplied by Gaia DR3. We emphasize that the selection limits imposed below are driven by our primary objective, to construct as pure a sample of cluster members as possible, sacrificing, when necessary, completeness with increasing magnitude for elimination of possible field star contamination. 

\begin{figure}
\figurenum{2}
\includegraphics[angle=0,width=\linewidth]{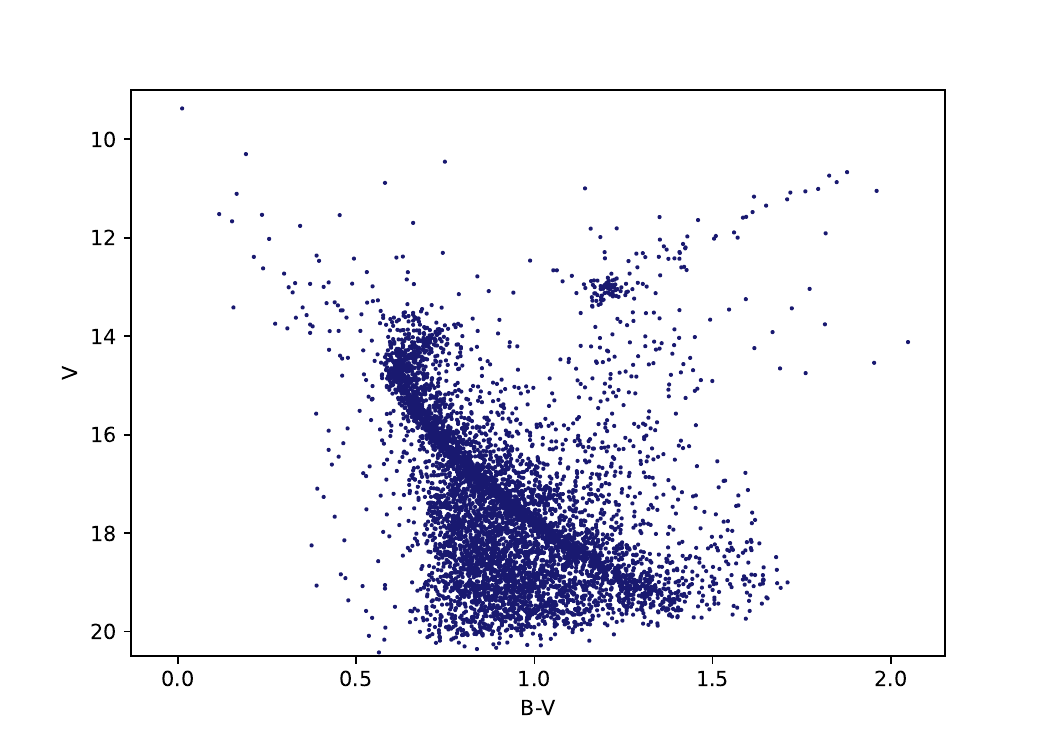}
\caption{CMD for all stars within the field of NGC 7789.}
\end{figure}

\citet{CA18} originally identified 3646 stars brighter than $G = 18$ with nonzero membership probabilities. Crossmatching this sample with DR3 data for the cluster area covered by our frames yields 2538 stars, of which 817 are initially eliminated since their membership probabilities are $\leq 0.9$. Of the remaining 1721, 400 stars with positional errors in DR3 $\geq 0.030$\arcsec\  in either RA or DEC and/or a ratio of parallax to parallax error  ($\pi$/$\sigma$$_{\pi}$) less than 10 were eliminated. The averages and standard deviations in $\pi$, proper motion in RA, $\mu$$_{RA}$, and proper motion in DEC, $\mu$$_{DEC}$, were then derived for the \citet{CA18} sample using DR3 data, producing (old vs. new) 
$0.452 \pm 0.038$ vs. $0.481 \pm 0.031$ mas, $-0.921 \pm 0.122$ vs. $-0.910 \pm 0.115$ mas-yr$^{-1}$, and $-1.932 \pm 0.118$ vs. $-1.959 \pm 0.117$ mas-yr$^{-1}$, respectively. It should be noted that an increased parallax of 0.029 mas based upon the more recent DR3 data is consistent with the pattern identified using a similar approach for NGC 752 (+0.040 mas) and M67 (+0.016 mas) \citep{TW23}.

With the revised cluster mean astrometric values in hand, the DR3 data for the stars of Table 1 were revisited. A star was eliminated from the analysis if it fell more than three $\sigma$ away from the mean cluster value in either component of the proper motion and/or the parallax. Any star with $\pi$/$\sigma$$_{\pi} < 10$ was eliminated, as were stars with potential variability noted by Gaia. Finally, use of the ($V$ - $G$) vs ($G$ - $R_{P}$) diagram allowed the identification and removal of 5 stars with excessively discrepant magnitudes for their colors, probably the result of image crowding and/or poor photometry. The initial membership sample based upon astrometry alone thus contained 894 stars.

\subsection{Radial Velocity Constraints}
A key goal of the WOCS program has been the collection of precise radial velocities over multi-year baselines.  In addition to identifying and characterizing binaries, membership can be established in conjunction with Gaia astrometry and in some cases, eclipsing binaries can lead to reliable stellar mass estimates.  
Examples of studies for which these elements have improved cluster CMD analysis may seen in studies of NGC 752 \citep{DA94, SA23, TW23} and M67 \citep{GE15, GE21, TW23}.

For NGC 7789 the radial velocity survey covers approximately all stars within 18\arcmin\  of the cluster center to $V$ $\sim$ 15 (NI20). As with M67, based upon the radial velocity means and scatter for each star, the number of observations, and the precision of the individual measures as impacted by the stellar rotation rate, stars were classed as members, nonmembers, likely members, binaries, SB1 binaries, SB2 binaries, and very rapid rotators. Our sample of astrometric members was crossmatched with the catalog of NI20, generating 411 stars. Stars were eliminated if, binary or single, their mean radial velocity classed them as nonmembers. Additionally, stars with probable membership based upon radial velocity, single or binary, were retained if the more recent Gaia data led to their classification as members according to our criteria, even though their astrometric membership was unavailable or questionable in the earlier analysis. Removal of the likely radial velocity nonmembers, as well as one star with insufficient data for classification, reduced our NI20 photometric sample to 389 stars.

Since the NI20 radial velocity survey only goes to $V\sim$ 15, just below the bluest point of the turnoff, it is crucial that probable binaries within the fainter CMD region defining the unevolved main sequence be removed, if possible, to minimize the scatter in the two-color diagrams used to constrain the reddening and metallicity. Since composite systems on the main sequence should scatter to the red of the single-star relation, the precision of the photometry should permit removal of the more extreme combinations producing scatter on the main sequence.  

Figure 3 shows the CMD for all probable members according to the criteria laid out in the previous discussion. Open black circles are stars with only astrometric membership, open red circles are single stars with both astrometric and radial velocity membership, and filled green triangles are members classed as binaries. 

\begin{figure}
\figurenum{3}
\includegraphics[angle=0,width=\linewidth]{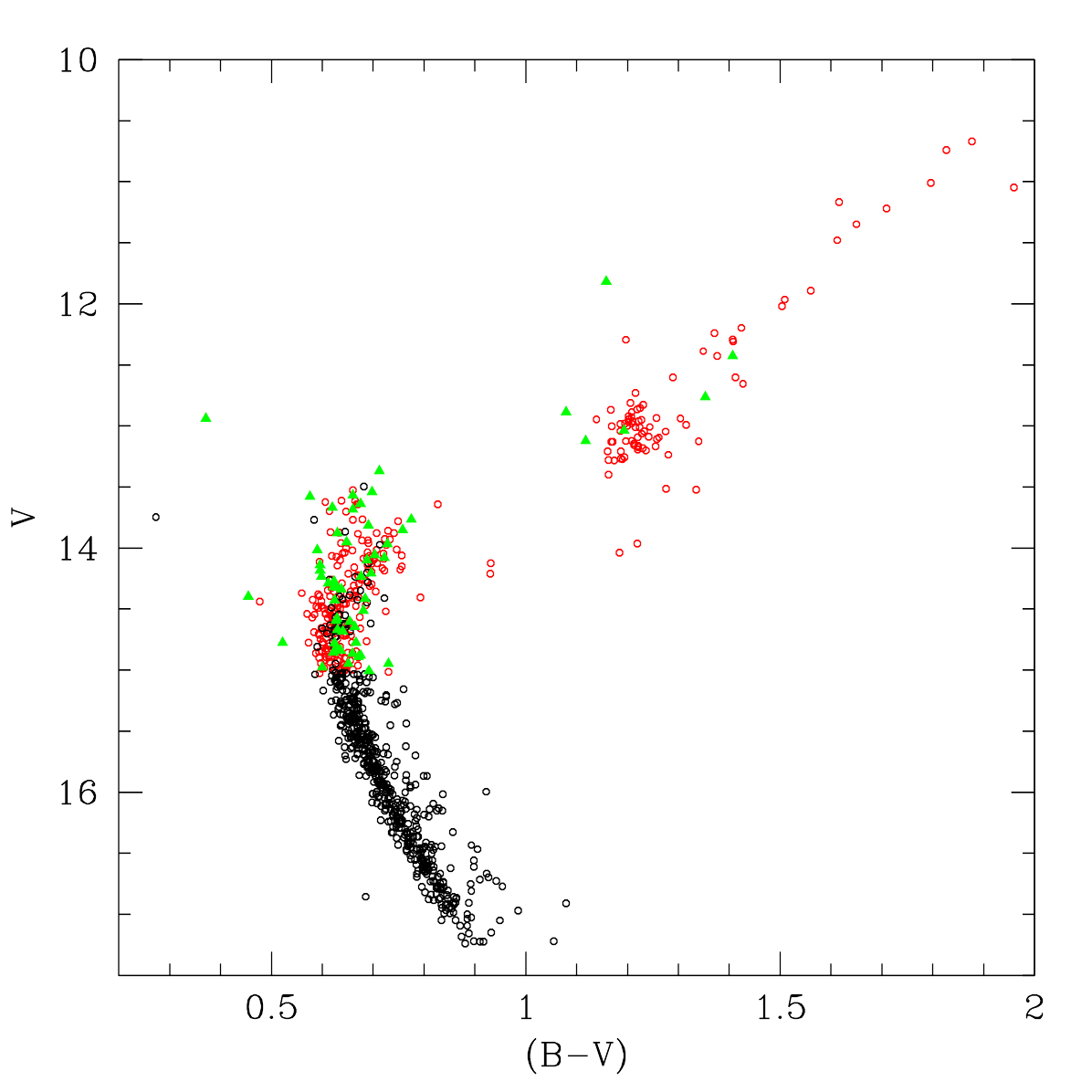}
\caption{CMD for stars classed as members of NGC 7789. Stars with only astrometric measures are open black circles, stars classed as radial velocity single and astrometric members are open red circles, and radial velocity binary members are filled green triangles.}
\end{figure}

A number of points are immediately obvious. First, despite the richness of the turnoff, red clump, and brighter giant branch, only six stars occupy the subgiant and RGB below the clump, indicative of an intermediate-age cluster, similar to poorly populated NGC 752 at an age of 1.45 Gyr \citep{TW23}, with non-degenerate He-core contraction still occurring among stars leaving the main sequence. Second, the dominant location of the binaries identified by radial velocity lies among the turnoff stars, specifically delineating the scatter caused by the crossover between the binary sequence and the redward hook of the stars evolving off the main sequence due to hydrogen exhaustion. Keeping in mind that no distinction is made between SB1 and SB2 systems, some of the identified binaries are positioned along the single-star sequence.

As already mentioned, the binary analysis cuts off almost exactly where the binary sequence composed of composite systems of identical mass begins to cross the vertical turnoff of the evolving main sequence stars. In Figure 4 we plot the CMD for the astrometric members of Figure 3 below $V$ = 15 without radial velocity information from NI20. Because the two-color analysis of a star is affected by surface gravity, stars which lie above $V$ = 15.20 will be excluded from the photometric analysis. This cut also has the benefit of removing possible binaries at the crossover region, stars which are likely responsible for the redward scatter of points between $V$ = 15.0 and 15.2. For the remainder of the CMD, stars that deviate by a significant amount in color from the blue edge of the unevolved main sequence should also be eliminated. On the blue side, only one star near $V$ = 16.85 plotted as a filled blue circle is removed. While the derived photometric errors for this star meet our restrictive criteria, fewer frames were used in determining the filter averages, often a signature of confusion with a nearby optical companion. Perusal of the field confirms this option. On the red side, stars identified by position in the CMD as likely photometric binaries and/or nonmembers are plotted as filled red circles. It should be noted that none of these stars were included in the spectroscopic sample of Table 2, eliminating a radial velocity check from our data. These stars will be eliminated from the multicolor analysis. 

To check if the red points in Figure 4 were defined by a real shift in CMD position rather than simple photometric error, we initially tested the CMD distribution using the Gaia ($G, B_p - R_p$) CMD. Not only did the red points as identified in Figure 4 also lie systematically to the red in the Gaia CMD, but it became apparent that the CMD morphology for stars (open circles in Figure 4) positioned around the non-binary sequence was very similar for both data sets, i.e. stars found systematically to the blue or red to a smaller degree than the red points of Figure 4 were in the same relative position in both CMDs. While this could be the product of an intrinsic variation in the \teff\ of the stars populating the main sequence, a more plausible explanation is a variation in reddening across the field of the cluster. This option is confirmed in Appendix C, allowing production of a positional map of the reddening change across the cluster field.

\begin{figure}
\figurenum{4}
\includegraphics[angle=0,width=\linewidth]{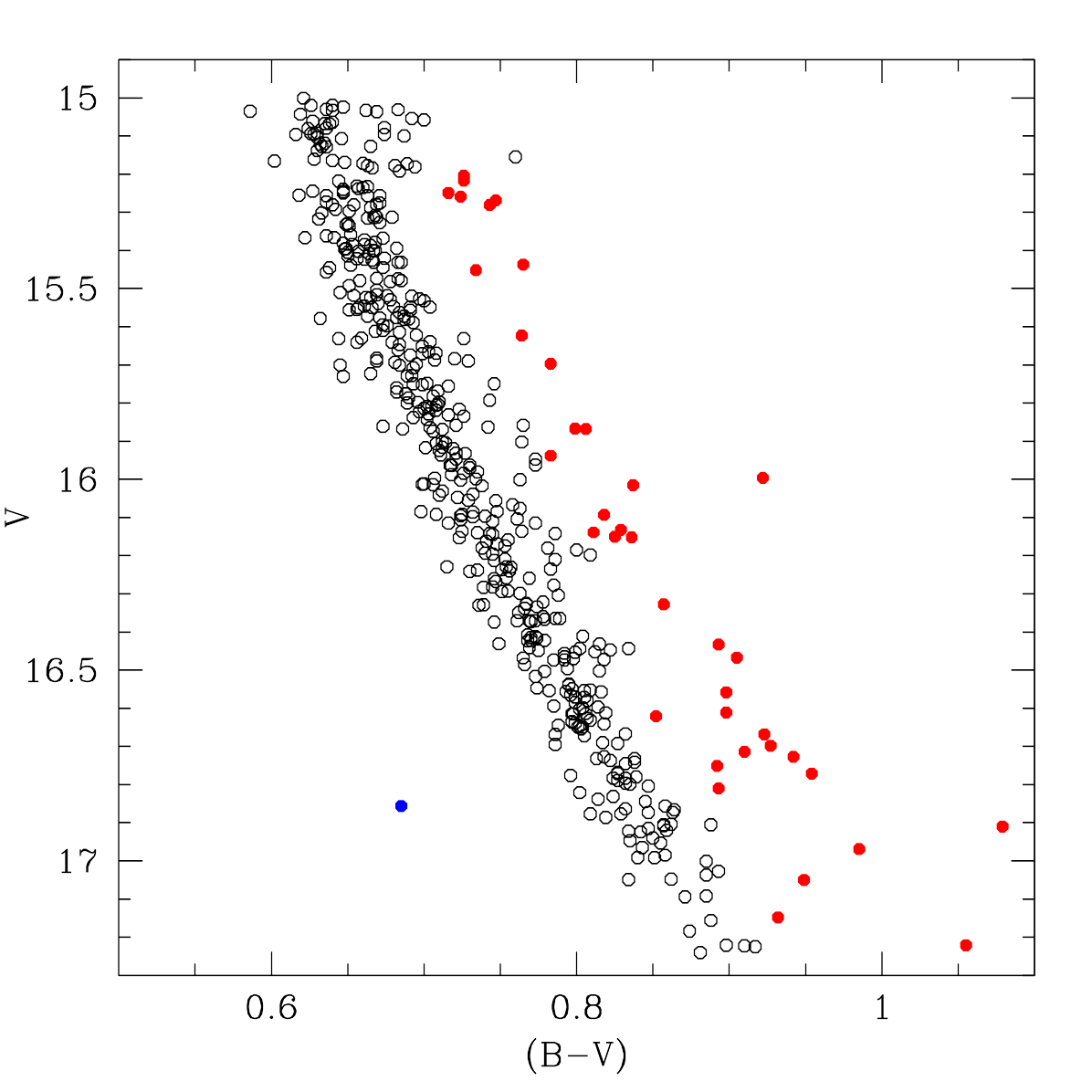}
\caption{CMD for stars fainter than $V$ = 15 with only astrometric membership credentials. Stars tagged as likely binaries or nonmembers from CMD position to the red of the main sequence are plotted as filled red circles while the one star positioned to the blue of the main sequence is plotted as a filled blue circle.}
\end{figure}

\section{Mean Reddening, Metallicity, and Age}
While precision broad-band photometry permits reliable derivation of the reddening gradient across the cluster, the more challenging issue remains:  determination of the mean \ebv\ value defining the zero-point of the scale. 
This issue plagues the existing all-sky \ebv\ maps discussed in Appendix C as well;  while all agree on a qualitative gradient across the field of the cluster, depending upon which all-sky \ebv\ map is adopted and/or the distance at which the cluster reddening is estimated, NGC 7789 could have a mean reddening value anywhere between \ebv\ = 0.21 and 0.37.

\subsection{Multicolor Broad-Band Derivation of \ebv\ and [Fe/H]}
With precision photometry in five colors, simultaneous derivation of \ebv\ and [Fe/H] should be possible since both parameters shift the positions of stars in two-color diagrams where at least one color index includes the highly sensitive $U$ filter. For an excellent illustration of the technique, the reader is referred to the evaluation of similar data for NGC 7142 \citep{SU20}, a cluster with high reddening comparable to that of NGC 7789. 
For our analysis, photometry was restricted to 388 unevolved stars ($V \geq 15.2$) assumed to be single (open circles of Figure 4). All indices were corrected {\it only} for the variable component of reddening, with corrections based on the map of Figure 13.  In the following discussions, photometric indices designated by primes denote indices corrected only for the amount of reddening above or below the cluster mean value.  
Two-color relations were then plotted using four combinations, $(U-B)'$, $(U-V)'$, $(U-R)'$, and $(U-I)'$ versus \BVp, shifted using the procedure outlined by Deliyannis et al. (2025, in preparation) relative to the mean relations for the Hyades cluster, assumed to have [Fe/H] = +0.15 \citep{MA13, CU17}.

The ([Fe/H], \ebv) results are reasonably consistent across the four color pairs, $(U-B)'$, $(U-V)'$, $(U-R)'$ and $(U-I)'$ versus \BVp\, generating ($-0.27$, 0.25), ($-0.30$, 0.26), ($-0.20$, 0.245), and (-0.30, 0.23), respectively, for a mean cluster value of [Fe/H] $= -0.27 \pm 0.03$ (sem) and \ebv = 0.246 $\pm$ 0.007 (sem). For reasons which will be discussed in the next section, various tests were applied to determine the sensitivity of the results to systematic alterations of the photometry. For example, if the original photometry had been used without the variable reddening correction, ([Fe/H], \ebv) becomes ($-0.26 \pm 0.02$, 0.258 $\pm$ 0.009), not a statistically significant change. A more plausible adjustment would be an error in the $(U-B)'$ zero point. It was found that while changes in $(U-B)'$ strongly impact the [Fe/H] estimation through $\delta$$(U-B)'$, \ebv\ remains relatively insensitive to moderate shifts in the $(U-B)'$ zero point. For example, if the mean $(U-B)'$ indices were increased by +0.02 mag, the cluster parameters become ($-0.16 \pm 0.02$, $0.243 \pm 0.006$). A more elaborate test of the two-color sensitivity to change is supplied by \citet{SU20}, who go through multiple iterations of their data for NGC 7142 by adopting a range of definitive [Fe/H] values for the cluster two-color relations, then determining the optimal \ebv\ for each metallicity. As the fixed [Fe/H] rises from $-0.4$ to $-0.2$ to 0.0, the required reddening also rises from 0.243 to 0.283 to 0.338. Thus, if the photometry is correct and the cluster metallicity is required to be closer to solar, the reddening will rise above 0.246 by an amount that increases with increments from [Fe/H] $= -0.27$. 

\subsection{\ebv\ and {\rm [Fe/H]} - An Alternate Approach}
Because the reddening correction affects the \teff\ scale for both main sequence stars and giants, abundance estimation from spectroscopy or multicolor photometry will be affected to varying degrees by the adopted \ebv.  Finding additional ways to jointly constrain the cluster reddening and metallicity is valuable.  
Thanks to the astrometric revolution triggered by Gaia, an alternative method is possible using a reliable set of theoretical isochrones covering a range of metallicity. The cluster parallax fixes the true distance modulus; then, for a given reddening, the apparent distance modulus will only match the unevolved main sequence of the cluster to an isochrone with a unique metallicity.  

As an initial constraint, the NGC 7789 cluster CMD has been cleaned of all known binaries (filled green triangles in Figure 3) and photometric binaries (filled red circles of Figure 4). The relative positions in the CMD for the remaining stars were adjusted for variable reddening using the map of Figure 13. Isochrones covering a range in metallicity from [Fe/H] $= -0.2$ to 0.0 and ages between 1.35 and 1.55 Gyr were constructed using the Victoria-Regina \citep{VA06, VA14} (VR) isochrones, allowing direct comparison with the results from comparable analyses of other clusters in the WOCS series. The VR isochrones of appropriate metallicity and adjusted Gaia parallax have consistently supplied excellent matches to the reddening-corrected cluster CMDs whether defined using $BV$ photometry as in NGC 6819 \citep{DE19}(DE19), NGC 2243 \citep{AT21}, and NGC 2204 \citep{AT24} or $by$ data as in NGC 752, M67 \citep{TW23}, and NGC 3680 (Deliyannis et al. 2025, in preparation). 

Adopting $(m-M)_{0}$ = 11.58, the isochrones at each metallicity were shifted to the red by \ebv\ and in $V$ by $(m-M)$ = 11.58 + (3.1*\ebv)) until the isochrone main sequence optimally matched the cluster CMD for single stars below $V$ = 15.2.  With the chosen \ebv\ value for that metallicity, the age of the isochrone demonstrating the best fit to the CMD in region of the red hook between $V$ = 14 and 15 was determined by visual inspection.
This procedure was repeated for three metallicities between [Fe/H] $= -0.2$ and 0.0. As expected, the higher the metallicity, the lower the coupled \ebv. By contrast, as [Fe/H] increases, the cluster age rises by a very modest amount, from 1.45 Gyr at [Fe/H] $= -0.2$ to 1.47 Gyr at [Fe/H] = 0.0, due to the redder intrinsic turnoff assigned to a higher [Fe/H], rather than a noticeably larger age. 

The isochrones found to provide the best fit at each metallicity are plotted against the restricted \BVp\ data in Figure 5. Open blue circles are single-star members from NI20, open green triangles are astrometric members not classed as probable photometric binaries, and open red circles are astrometric members without radial velocity data from NI20 but a CMD position inconclusive for binarity determination. All stars' colors and magnitudes, designated \Vp\ and \BVp, have been adjusted to correct only for the amount of differential reddening relative to the average reddening. 
The VR isochrones have ages of 1.45 (blue), 1.46 (red), and 1.47 (black) Gyr, corresponding to the (\ebv, [Fe/H]) combinations of (0.315, $-0.19$), (0.295, $-0.10$), and (0.26, 0.00), respectively. Note that while the lack of stars populating the giant branch below the RGC would make differentiation among the isochrone options difficult under normal circumstances, all three evolved tracks occupy very similar CMD locations. In summary, on the VR scale, the most probable age for NGC 7789 is 1.46 $\pm$ 0.02 Gyr, where the predominant uncertainty remains the absolute value for the foreground reddening \ebv. This makes NGC 7789 identical in age, within the uncertainties, to the approximately solar-metallicity cluster, NGC 752 \citep{TW15, TW23}, a point we will return to in Section 7.

\begin{figure}
\figurenum{5}
\includegraphics[angle=0,width=\linewidth]{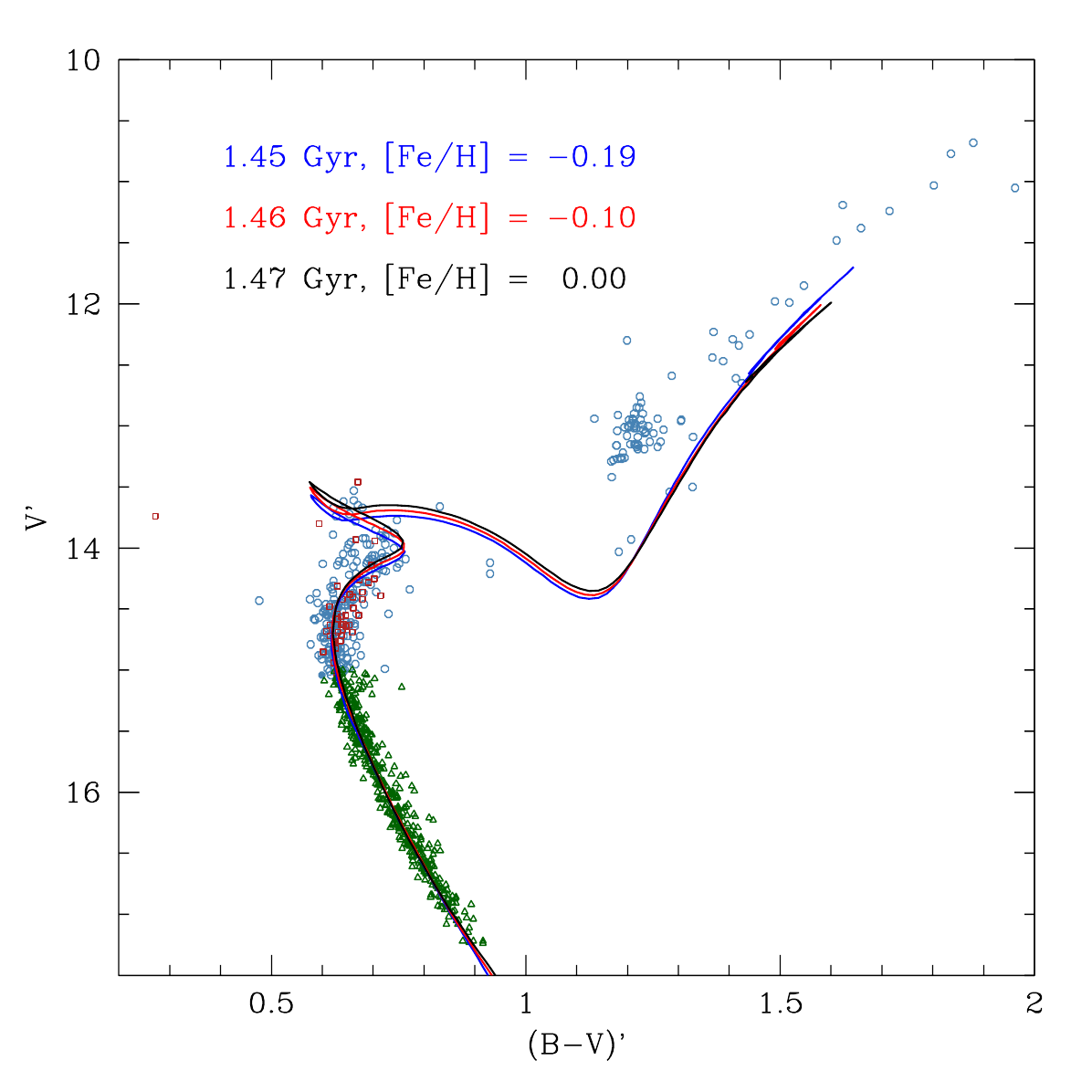} 
\caption{Isochrones from VR.  Blue circles are probable single members based on proper motion and \vrad\ information, while red squares have astrometry indicative of membership but lack information about binarity.  Fainter stars indicated by green triangles are likely single members based on astrometry and photometry.}
\end{figure}

As a second constraint, one can obtain the cluster metallicity by deriving spectroscopic abundances under different assumptions for \ebv. 
Higher \ebv\ values lead to bluer dereddened colors and higher \teff\ values,
typically leading to higher derived abundances.
As a sensitivity indicator, 10 single star members were selected from the full sample. The dwarfs were chosen from the faintest and coolest stars to avoid problems with rapid rotators above $V$ = 15.2 and to analyze stars as close in \teff\ to the Sun as possible. Giants were selected from a modest \teff\ range near the richly populated luminosity level of the RGC to minimize the impact of the rapidly growing complexity of the spectra among cooler stars along the more luminous giant branch. Spectroscopic metal abundances as detailed in the next section were derived for the two samples under the assumption that \ebv\ = 0.33, 0.31, and 0.29. The resulting mean [Fe/H] values are $-0.07$, $-0.105$, and $-0.135$, respectively. 

With two independent and non-parallel relations between acceptable \ebv\ and \feh\ values, only one parameter pair satisfies both relations, 
(\ebv\, [Fe/H]) = (0.30, $-0.12$). 
For the comprehensive derivation of the spectroscopic abundances in the following section, we have adopted \ebv\ = 0.30 and [Fe/H] $= -0.12$ as initial inputs for the model atmosphere construction. 

\subsection{Comparison With Previous Estimates: Reddening and Age}
Summaries of past reddening and age estimates for NGC 7789 are presented by \citet{WU07} for studies before 2007. More recent spectroscopic studies are summarized by \citet{OV15} and \citet{NA23} (NA23). All spectroscopic analyses to date have assumed uniform reddening across the face of the cluster, typically adopting values of 0.27 or 0.28.  The historic range for \ebv\ presented by \citet{WU07} was somewhat larger, 0.22 to 0.35, a range that has not been exceeded by more recent studies.    

For age estimation, \citet{GI98b} used VR isochrones, based on models with extensive convective core overshooting parameterized to match the MSTO red hook feature and the extended giant branch and concluded that NGC 7789, if its metallicity is between [Fe/H] $= -0.2$ and 0.0, couldn't be younger than 1.6 Gyr.  
This comparison, however, also implied an apparent distance modulus $\leq 12.2$ for \ebv\ in the range of 0.27 to 0.29, clearly incompatible with a distance modulus consistent with the cluster's parallax from Gaia, 12.42 to 12.48 for these reddening values.  

Likewise, \citet{VA00} used near-infrared photometry and the isochrones of \citet{BE94} to derive \ebv\ = 0.30, [Fe/H] $= -0.25 \pm 0.11$, and an age of 1.4 Gyr. Following a common pattern with older isochrone fits, $(m-M)_0$ = 11.25, too small by 0.3 mag.  

\section{Spectroscopic Analysis of Hydra Data}
\subsection{Equivalent Width Measurement with ROBOSPECT}  

As described in \citet{AT24} and earlier work, we employ the automated line-measurement program, ROBOSPECT \citep{WH13}, for large spectroscopic samples where interactive measurement of equivalent widths (EW) is impractical. ROBOSPECT uses a gaussian line profile to obtain EW measurements.  Each star's spectrum was individually corrected for its radial velocity, then passed through 25 iterations of continuum fitting and line estimation, with most input parameters set to default values. Our limited wavelength range (6400 - 6800 \AA) implies a rather small potential line-list of fewer than 20 isolated iron lines, with a few lines of Si, Ni and Ca.  We typically exclude from consideration measured EW larger than 150 m\AA\ out of concern that they lie beyond the linear portion of the curve of growth.  Very weak lines ($\leq 15$ m\AA) are not retained since, for a typical line width and S/N $\sim 100$, a 3-$\sigma$ EW error is approximately 15 m\AA\ \citep{CA88, DP93}. Our candidate line-list includes 20 lines (15 Fe, one each for Si and Ca, three Ni lines).  Two Fe lines near 6609 \AA\ consistently yield unreliable results, probably due to blending in less than optimal resolution conditions, and are no longer used.  Two additional Fe lines appear to be problematic in the present sample. A previously used line at 6837 \AA\ lies very near the red edge of our spectra where the continuum fitting is less than ideal and will not be included in the analysis. An additional line at 6750.15 \AA\ appears to yield strikingly discrepant results for giants and dwarfs; it will be dropped from consideration as well. 

Our set of 324 single-star members, whittled down from the original spectroscopic sample of 433 stars, quickly separated itself into two sets demarcated by the degree of success for EW measurements by ROBOSPECT. For convenience, we will use RO$\alpha$ to designate stars for which ROBOSPECT results were straightforward, with RO$\beta$ describing stars for which ROBOSPECT struggled.  

For the 167 stars in the RO$\beta$ category, EW for fewer than five lines from our line-list were successfully measured.
This is most likely due to a combination of significant rotation and/or binarity which renders lines more shallow and broad, particularly near the turnoff region.  
Of the stars near the MSTO region of the CMD, over 90\%\ fall in the RO$\beta$ class.  These include the seven blue stragglers, whose high \teff\ values also place them out of range for model atmosphere construction.  

Of the remaining 157 RO$\alpha$ stars, a third are bluer than \BVp $ = 0.8$, including a few stars near the turnoff region but more typically on the lower main sequence.  For these stars, ROBOSPECT produces EW measurements for four to five iron lines.  By contrast, most Fe lines are consistently measurable for the over 100 cooler stars in the RO$\alpha$ category with the exception of the typically strong line at 6677\AA\ for which EW are often above 150 m\AA\ for giants.

Figure 6 illustrates examples of stars in the RO$\beta$ and RO$\alpha$ categories. The two MSTO stars have nearly identical CMD positions and Li EW. The noticeably higher rotation for 14999 renders all but strong lines shallow as well as broad, a problem for automated or manual EW measurements.  
Star 13623 is the only RG in the RO$\beta$ group of stars.  Its Li EW is nearly identical to that for the more typical RG, 11777.

\begin{figure}
\figurenum{6}
\includegraphics[angle=0,width=\linewidth]{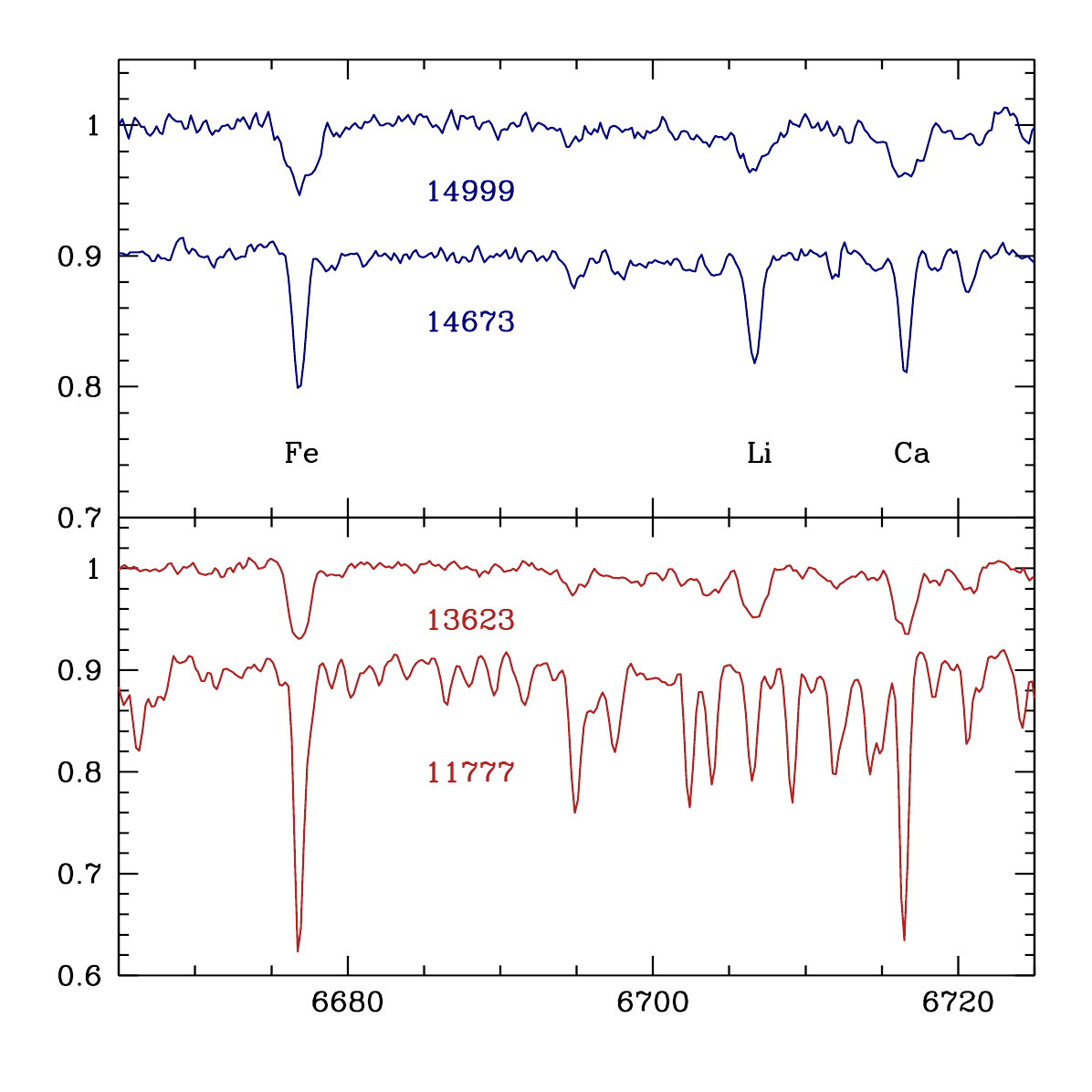} 
\caption{Representative spectra for MSTO and RG stars classed as RO$\alpha$ and RO$\beta$. In each panel, the RO$\alpha$ star's spectrum is offset by 0.1 downwards.  }
\end{figure}

\subsection{Atmospheric Parameters: \teff, \logg\ and \vturb}
Each star's measured EW values were evaluated using MOOG \citep{SN73}\footnote{Available at  http://www.as.utexas.edu/~chris/moog.html} in the context of an appropriately constructed model atmosphere, for which reliable estimates of \teff, \logg\ and \vturb\ are required.  Surface gravity estimates (\logg) were obtained by comparing \Vp\ magnitudes and \BVp\ colors to VR isochrones (see the red isochrone of Figure 5). For estimates of \teff, we utilize the modified color-temperature relation of \citet{HU15} developed and described in \citet{AT24}.  Table 3 of the latter paper summarizes the coefficients for the modified \citet{HU15} color-temperature calibration, but included a typographical error in the assignment of coefficients to terms. A corrected table is included here as Table 3. For a precision of $0.01$ in \BVp, estimated \teff\ errors are $\sim 20$ K for giants and $\sim 30$ K for turnoff stars.  

\floattable
\begin{deluxetable}{rrrrrrr}
\tablenum{3}
\tablecaption{Revised H15 Color-Temperature Relations}
\tablewidth{0pt}
\tabletypesize\small
\tablehead{
\colhead{Class} & \colhead{Num} &\colhead{$(B-V)_{min}$} &\colhead{$(B-V)_{max}$} 
&\colhead{Max.$T_{\mathrm{eff}}$} &\colhead{Min.$T_{\mathrm{eff}}$} &\colhead{std.dev} \\
\colhead{ } & \colhead{$a0$} &\colhead{$a1$} &\colhead{$a2$} &\colhead{$a3$} & \colhead{$a4$} &\colhead{$a5$}  } 
\startdata
MS/MSTO &    99  & 0.21 & 1.51 & 7734 & 3618 &  139 \\
\nodata & 8099.9436 & -4343.1089 &  959.1094 &  256.6719  &  -39.6315 &     52.6089 \\
SGB/RGB & 72 & 0.43 & 1.62 & 6478 & 3602 &  100 \\
\nodata & 7702.3968 & -3601.1812 &  695.6553 &  431.7820 &  -12.5042 &  -194.4370 \\
\enddata
\tablecomments{\teff $ =  a0 + a1\cdot X + a2\cdot X^2 
+ a3\cdot [Fe/H] + a4\cdot [Fe/H]^2 + a5\cdot X\cdot[Fe/H] $ }
\end{deluxetable}

Using the color-temperature relation with an assumed initial input value for \feh\ of $-0.12$, each star's \teff\ was derived from its individually dereddened color by subtracting the cluster mean reddening, \ebv\ $= 0.30$, from its \BVp\ color.
Together with the estimated \logg, the \teff\ permits the estimation of the microturbulent velocity parameter, \vturb, using the formulation by \citet{BR12} for stars within the valid ranges of \logg\ and \teff.  For a few very luminous giants with \logg\ $\leq 2$, \vturb\ $= 2.0-0.2$ \logg\ was used.

\subsection{Metal Abundance Determinations}
For abundance determinations, we employ the $abfind$ driver in the 2014 LTE version of the MOOG software suite, for which the solar iron abundance is set at 7.50. MOOG requires the specification of each spectral line's \loggf\ value.
Excluding the coolest giant star, for which no model atmosphere could be constructed, appropriate model atmospheres were constructed for 156 RO$\alpha$ stars using the grid of ATLAS9 models crafted by \citet{KU95}\footnote{http://kurucz.harvard.edu/grids/ }. Grid models are available with \feh\ values from $-5.0$ to $+1.0$; our software performs a linear interpolation between models with bracketing \teff\ and \logg\ to produce each model.  

For each eligible star, MOOG utilizes the measured EW values and model atmosphere to produce individual elemental abundance estimates, [A/H], for each measured line of element A, producing hundreds of individual estimators of the cluster metal abundance. We employ median statistics to summarize cumulative information for each star and for each utilized metal line, as well as global estimates of the cluster metallicity. The median absolute deviation (MAD) provides a robust estimate of the dispersion among measured values.  For typical distributions of values without extreme outliers, a traditional standard deviation is $\sim 1.48 \times$MAD.  

Considering all 156 single members in the RO$\alpha$ category, 1490 separate estimates of the Fe abundance were obtained, yielding an estimated [Fe/H] $= -0.118$ with a MAD value of 0.086.  The median of separate stellar abundance estimates is similar, $-0.130$ with a MAD value of 0.068.  Direct comparisons of derived abundances for two serendipitously re-observed stars are gratifyingly similar, with values different by 0.086 and 0.004 dex.

The set of 1490 separate Fe abundance measurements is heavily biased towards the more numerous cooler stars for which the number of measured Fe lines is at least 10, with a median value of 12. In sharp contrast, for the stars with \logg\ $\geq 3$, the median number of successfully measured iron lines is four with a maximum value of only eight. Potential relationships between [Fe/H], MAD, \teff\ and the number of measured lines were explored; no trend between [Fe/H] and \teff\ was detected, although five stars with \teff\ below 3900 K exhibit both anomalously low [Fe/H] values and exceptionally high MAD values.  

Table 4 reiterates the wavelengths, species and \loggf\ values for our line-list along with the median abundance, MAD, and the number of stars upon which the estimates are based. Results for elements other than iron are included. Ca and Si, both based on a single line, are consistent with scaled-solar abundance ratios. Ni, defined by three lines, exhibits weak evidence for a subsolar ratio with [Ni/Fe] $\sim -0.1$.   

\floattable
\begin{deluxetable}{rcrcrr}
\tablenum{4}
\tablecaption{Lines Used for EW Analysis}
\tabletypesize\small
\tablewidth{0pt}
\tablehead{
\colhead{$\lambda$(\AA)} & \colhead{Element} & \colhead{log $gf$} &
\colhead{$N_{Star}$} &\colhead{[X/H]} & \colhead{MAD} } 
\startdata
6597.56  & Fe & -1.04 & 140 &  0.00 & 0.07 \\
6627.54  & Fe & -1.57 & 114 & -0.09 & 0.05 \\
6646.93  & Fe & -3.87 & 108 & -0.20 & 0.06 \\
6653.91  & Fe & -2.44 &  93 & -0.12 & 0.08 \\
6677.99  & Fe & -1.50 &  51 & -0.28 & 0.08 \\
6703.57  & Fe & -3.02 & 116 & -0.12 & 0.07 \\
6710.32  & Fe & -4.77 & 108 & -0.07 & 0.05 \\
6725.36  & Fe & -2.35 & 109 & -0.18 & 0.08 \\
6726.67  & Fe & -1.16 & 140 & -0.22 & 0.07 \\
6733.15  & Fe & -1.62 & 108 & -0.10 & 0.10 \\
6806.86  & Fe & -3.14 & 110 & -0.16 & 0.07 \\
6810.27  & Fe & -1.08 & 154 & -0.11 & 0.06 \\
6820.37  & Fe & -1.20 & 139 & -0.03 & 0.06 \\
6717.68  & Ca & -0.22 &  44 & -0.16 & 0.07 \\
6721.85  & Si & -1.09 & 132 & -0.07 & 0.07 \\ 
6643.63  & Ni & -1.83 &  89 & -0.40 & 0.11 \\
6767.77  & Ni & -2.23 & 140 & -0.24 & 0.08 \\ 
6772.31  & Ni & -1.00 & 141 & -0.19 & 0.07 \\
\enddata
\end{deluxetable}

\floattable
\begin{deluxetable}{cccrcccccccch}
\tablenum{5}
\tablecaption{Li and Fe Abundances for stars in NGC 7789}
\tabletypesize\small
\tablewidth{0pt}
\tablehead{
\colhead{ID(Webda)} & \colhead{$V'$} & \colhead{$(B-V)'$} &
\colhead{\vrot} & \colhead{\teff} & \colhead{\logg} & \colhead{\vturb} &
\colhead{A(Li)} & \colhead{D/L} & \colhead{[Fe/H]} & \colhead{MAD} & 
\colhead{No.Fe} & \nocolhead{} \\
\colhead{} & \colhead{} & \colhead{} &
\colhead{\kms} & \colhead{K} & \colhead{} & \colhead{\kms} & 
\colhead{} & \colhead{} & \colhead{} & \colhead{} &
\colhead{Lines} & \nocolhead{} }
\startdata
6345 & 10.674 & 1.879 & $\leq 10$ & 3735 & 1.18 & 1.76 & -3.50 & L & -0.26 & 0.41 & 11 & RLl \\
7029 & 10.795 & 1.845 & 11 & 3783 & 1.22 & 1.76 & 0.20 & D & -0.53 & 0.49 & 11 & RDd \\
13862 & 10.825 & 1.894 & $\leq 10$ & 3715 & 1.25 & 1.75 & 0.10 & D & -0.44 & 0.43 & 11 & RDd \\
8799 & 10.874 & 1.849 & 39 & 3778 & 1.30 & 1.74 & 0.20 & D & -0.37 & 0.46 & 12 & RDd \\
2740 & 11.032 & 1.803 & 15 & 3844 & 1.38 & 1.72 & 0.30 & D & -0.16 & 0.39 & 12 & RDd \\
7091 & 11.254 & 1.720 & 11 & 3973 & 1.50 & 1.70 & 0.40 & D & -0.18 & 0.19 & 13 & RDd \\
6810 & 11.259 & 1.646 & $\leq 10$ & 4095 & 1.47 & 1.71 & -1.90 & L & -0.23 & 0.12 & 12 & RLl \\
6767 & 11.410 & 1.670 & $\leq 10$ & 4054 & 1.58 & 1.68 & 0.48 & D & -0.13 & 0.11 & 12 & RDd \\
896 & 11.411 & 1.610 & $\leq 10$ & 4157 & 1.65 & 1.67 & 0.13 & D & -0.20 & 0.09 & 12 & RDd \\
8293 & 11.491 & 1.616 & 11 & 4147 & 1.65 & 1.67 & 0.58 & D & -0.19 & 0.14 & 12 & RDd \\
8957 & 11.666 & 1.467 & $\leq 10$ & 4423 & 1.75 & 1.65 & -1.65 & L & -0.14 & 0.13 & 12 & RLl \\
10652 & 11.884 & 1.557 & $\leq 10$ & 4252 & 1.90 & 1.62 & 0.73 & D & -0.10 & 0.09 & 12 & RDd \\
13089 & 11.914 & 1.543 & $\leq 10$ & 4278 & 1.96 & 1.61 & 0.70 & D & -0.14 & 0.16 & 12 & RDd \\
11413 & 11.963 & 1.486 & $\leq 10$ & 4386 & 1.97 & 1.61 & 0.88 & D & -0.10 & 0.08 & 12 & RDd \\
1269 & 11.981 & 1.514 & 10 & 4332 & 1.94 & 1.61 & -1.20 & L & -0.17 & 0.11 & 12 & RLl \\
3798 & 12.012 & 1.441 & 10 & 4474 & 1.94 & 1.61 & 0.90 & D & -0.13 & 0.08 & 12 & RDd \\
15337 & 12.134 & 1.419 & $\leq 10$ & 4518 & 2.07 & 1.59 & -0.60 & L & -0.03 & 0.14 & 12 & RLl \\
7617 & 12.139 & 1.421 & 11 & 4514 & 2.03 & 1.59 & -1.15 & L & -0.18 & 0.09 & 12 & RLl \\
1135 & 12.225 & 1.433 & $\leq 10$ & 4490 & 2.07 & 1.59 & 0.70 & D & -0.19 & 0.22 & 11 & RDd \\
5440 & 12.227 & 1.380 & $\leq 10$ & 4598 & 2.06 & 1.59 & 2.60 & D & -0.29 & 0.07 & 12 & RDd \\
\enddata
\tablecomments{This table is available in its entirety in machine readable form in the online article.} 
\end{deluxetable}

Individual [Fe/H] values are reported only for stars with at least ten measured iron lines, effectively the giant stars in the sample, in Table 5 along with \Vp, \BVp, and the adopted atmospheric parameter estimates. 

\subsection{Comparison With Previous Abundance Estimates}
Past spectroscopic determinations of [Fe/H] for NGC 7789 have focused almost exclusively on cool stars, {\it i.e.} giants, a concern in this significantly and variably reddened cluster. Summaries of past work are compiled in \citet{WU07} for pre-2007 determinations, including photometric and spectroscopic approaches, and in \citet{OV15} and NA23 for more recent spectroscopic data. NA23 compare their spectroscopic abundance estimate of [Fe/H] $= -0.02 \pm 0.05$ (sd) from 14 red giants using roughly 49 Fe I and 10 Fe II lines per star to the earlier work of \citet{OV15}, \citet{TA05}, \citet{PC10}, and \citet{JA11} who respectively derive cluster mean values of [Fe/H] of $0.03 \pm 0.07$ (sd) from 32 giants, $-0.04 \pm 0.05$ (sd) from nine giants, $0.04 \pm 0.07$ (sd) from three giants, and $0.02 \pm 0.04$ (sd) from 28 giants. 

\citet{LB15} showed how [Fe/H] derived by our spectroscopic pipeline is dominated by \teff\ differences over variations in \logg\ and \vturb, implying that [Fe/H] differences with our result are likely due to different \teff\ scales. Of the four spectroscopic surveys with significant overlap with our sample, \citet{JA11} derived \teff\ from reddening-corrected color indices ($(B-V), (V-K), (J-K)$), adopting \ebv\ = 0.28. \vturb\ was set to a constant 1.5 \kms, while \logg\ was calculated from the derived \teff, coupled with an adopted distance modulus and turnoff mass from the cluster age (1.8 Gyr adopted).  \citet{OV15} and \citet{TA05} incorporate spectroscopic refinement of \teff\ and \logg\ values for which starting values were based on photometry and color-temperature relations, coupled with isochrone comparisons. Finally, NA23 derived \teff\ through the use of atomic line depth ratios, then obtained the remaining parameters through constraints imposed by the  spectroscopic line analysis. 

\begin{figure}
\figurenum{7}
\includegraphics[angle=270,width=\linewidth]{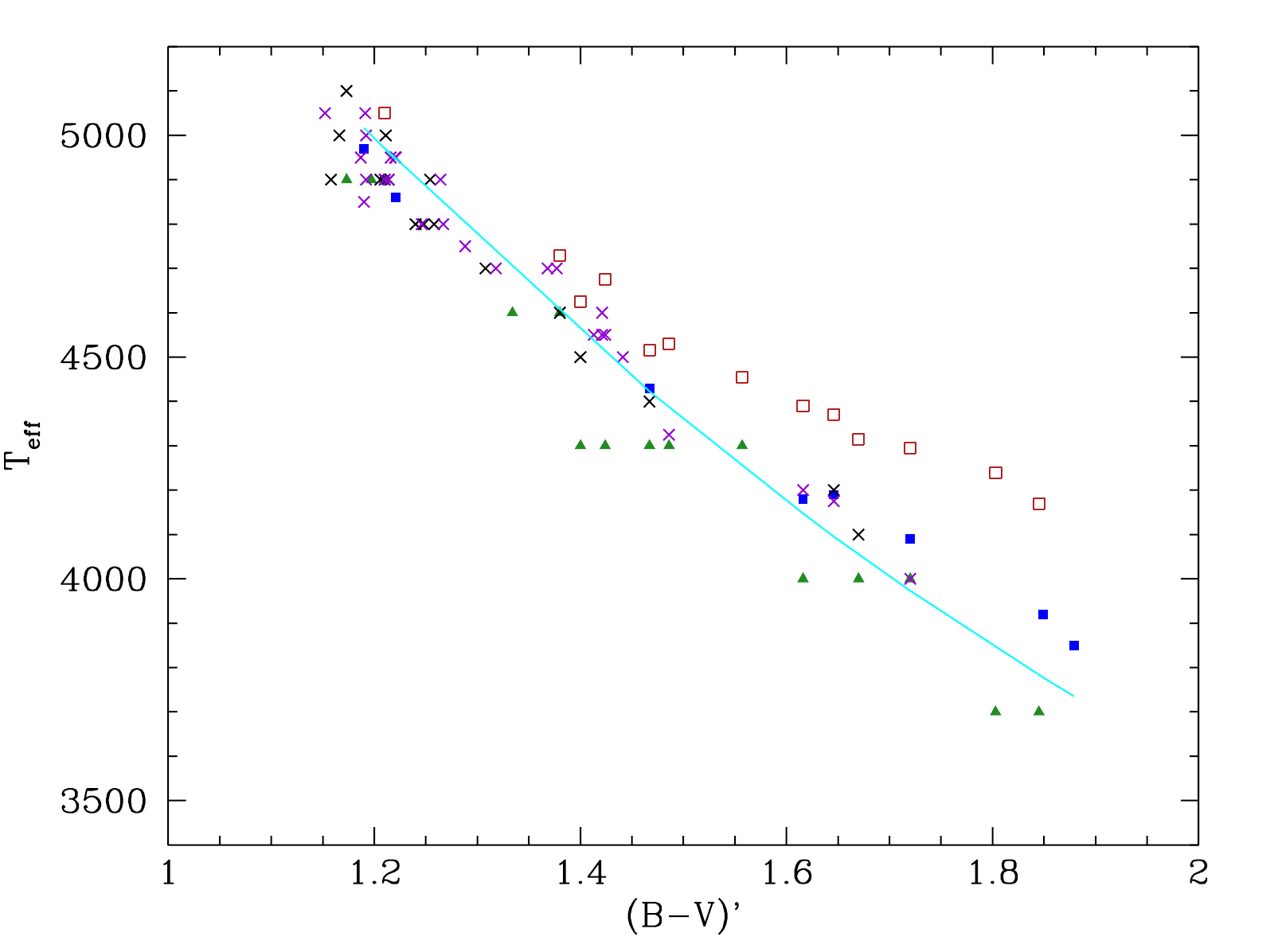} 
\caption{For stars in common with the present survey, adopted \teff\ values from \citet{JA11} are indicated by black crosses, from \citet{PI86} by green triangles, from \citet{TA05} by blue filled squares, by \citet{OV15} by violet crosses, and by NA23 using red open squares.  The \teff\ values employed in the present survey are indicated by the cyan continuous line. }
\end{figure}

Figure 7 illustrates the \teff, \BVp\ trends for our adopted relation (solid line) and the four spectroscopic surveys discussed above. (We will discuss the \citet{PI86} data in Section 6.) With the striking exception of NA23, the studies show a consistent pattern across the color range from \BVp\ = 1.15 to 1.85. At the hot end, our \teff\ scale is about 25 K hotter than the typical estimate derived in the other analyses. Near \BVp\ $\sim$ 1.5, our systems are essentially the same. Finally, at the cool end, the sample is small, but it appears that our scale is $\sim$ 125 K cooler. Given the distribution on the giant branch dominated by the bluer stars, particularly the clump, the average differential in \teff\ should be negligible for a large representative sample. For example, from 27 stars in common with \citet{OV15}, the average \teff\ difference is $ 3 \pm 90$ K (sd), using the Arcturus-linked \teff\ data due to its reduced scatter. No significant dependence of \logg\ or \vturb\ as a function of \BVp\ can be discerned, although our \vturb\ values are higher, on average, by 0.15 \kms, which should result in lower [Fe/H] values by $\sim 0.04$ dex. None of these parameter comparisons can account for [Fe/H] values that are 0.17 dex lower, on average, than values derived for these 27 stars. A similar conclusion emerges for the nine stars common with \citet{TA05}. The lack of a discernible trend with \BVp\ for the other differences in parameter choices (\logg\ or \vturb) leaves the  0.10 dex residual in [Fe/H] for the nine stars in common largely unexplained. We should note that our comparisons were constructed with respect to the abundances derived using the Gaia ESO survey line-list \citep{GI12} in \citet{OV15}.

The most striking feature of Figure 7 is the growing deviation of the \teff\ data of NA23 with increasing color. At the hot end, their scale is about 50 K hotter, reaches $\sim$ 100 K near \BVp\ = 1.45, and grows to $\sim$ 400 K among the reddest giants. Note that every study, including NA23, exhibits no clear trend of [Fe/H] with \teff. We will return to a discussion of the crucial impact of the \teff\ pattern on A(Li) in Section 6.

We close this discussion of spectroscopic parameter choices with a comparison to the APOGEE Stellar Parameters and Chemical Abundances Pipeline (ASPCAP) results \citep{GP16} in the field of NGC 7789. ASPCAP simultaneously determines abundance and parameter information by comparing APOGEE spectra to grids of model spectra. From DR17, we found 236 stars within $30\arcmin$ of the cluster center, of which 53 are likely single members with [Fe/H] based on ROBOSPECT analysis of our spectra.  We observe no clear patterns of dependence on \BVp\ for residuals in our choices for atmospheric parameters or derived [Fe/H] and a modest average \teff\ difference of only $-12 \pm 8$ K (sem). Our [Fe/H] values are $0.10 \pm 0.006$ (sem) lower, on average, than the ASPCAP values. Our [Fe/H] results could be pushed to values closer to solar by adoption of a {\it higher} mean \ebv, but at the expense of an inconsistent fit between the observed CMD and the parallax-adjusted isochrones.  

In summary, precision photoelectric photometry and moderate resolution spectroscopy have been analyzed for NGC 7789 for decades \citep{WU07}. An unweighted average of the 8 studies since 1975 in this category produces [Fe/H] $= -0.15 \pm 0.11$ (sd) and \ebv\ = $0.26 \pm 0.03$ (sd). Reanalysis \citep{TT85} of 16 and 19 giants on the $UBV$ and DDO systems, respectively, constrained \ebv\ to $0.31 \pm 0.01$ (sem) and [Fe/H] to $-0.10 \pm 0.03$ (sem) (assuming $[Fe/H]_{Hyades}$  = +0.15 \citep{CU17}). With a revised DDO calibration \citep{TW96, TW97}, [Fe/H] $= -0.089 \pm 0.035$ (sem).

\section{Li: Abundance Estimation and Evolutionary Implications}
\subsection{Li From Spectrum Synthesis}
A primary goal of this program has been the delineation of Li abundances as a function of mass and evolutionary phase based upon the trends identified within star clusters of varying age and metallicity. For consistency, we will follow the same pattern laid out in previous papers, most recently the discussion of NGC 2204 \citep{AT24}.

As described in the previous section, our sample of 324 single member stars is almost equally divided into categories of stars for which ROBOSPECT analysis yields plentiful abundance information (RO$\alpha$) and those (RO$\beta$) for which rapid rotation or undiagnosed binarity impedes automated EW measurement for all but the stronger lines.
As discussed extensively in \citet{AT24}, the direct measurement of Li abundance from the 6707.8 \AA\ line is complicated by the presence of a usually very weak Fe I line that is nearby at our spectral resolution. Measurement of the Li line may be further complicated in hotter stars where the Li line is weaker for a given abundance, or in stars with significant rotational broadening. Brighter MSTO stars in NGC 7789 are more negatively affected by rapid rotation than any stars in NGC 2204. In an effort to measure Li abundances for as many of the 324 single members stars as possible, we turned to the procedure adopted in our previous cluster discussions, spectrum synthesis.  

Use of spectrum synthesis to define the individual stellar Li abundances is especially critical when the Li line is weak to nonexistent and upper limits to the potential abundance are the best that one can achieve. Our standard procedure is as follows: each spectrum is compared to the relevant model atmosphere using the $synth$ driver in MOOG, where the relevant model atmospheres are constructed using the stellar parameters derived as detailed in the previous section ([Fe/H], \logg, \teff, \vturb). For the seven blue stragglers and the coolest red giant star, model atmospheres were not constructed because the \teff\ estimates fall outside of model limits, leaving 316 stars for which Li abundances were derived by two independent synthesis measurers. 
  
When the model \teff\ has been appropriately chosen, lines other than Li show consistent levels of agreement between the spectrum and model, where ``consistency'' means that the line profile characterizing the spectrum is also appropriately characterized by MOOG. The line profile incorporates the known instrumental line width (0.55 m\AA\ for Hydra spectra) as well as corrections for limb-darkening (coefficient taken uniformly to be 0.5) and broadening due to rotation. For the last parameter, projected rotational velocities may be interactively estimated. For all completed syntheses, the rotational velocity which appeared to best fit the entire observed spectrum was recorded, as well as the resulting A(Li) value, noted as a detection or an upper limit, by each of the independent measurers. Subsequent comparison of the two interactively determined \vrot\ values with the estimates from $fxcor$ analysis shows remarkable consistency up to 40 \kms, above which $fxcor$ values are understood to be unreliable \citep{AT21}. 

Where the Li line is distinct and clearly present, designated by a D in Table 5, an abundance for Li is discerned by comparing the fit for different assumed A(Li) values. In other cases, the Li line is not obviously distinguishable from noise or nearby spectral features, designated by an L in Table 5, a problem prevalent among the warmer and often rotationally broadened turnoff stars. In such cases, an upper limit is determined by noting the value of A(Li) below which changes to the spectrum-model fit are no longer discernible.  
A(Li) determinations, including upper limits, for all measurable spectra are listed in Table 5.  Table 5 also includes the average \vrot\ value for each star, where the averages for values above 40 \kms\ solely incorporate the interactively determined values. Comparison between the two separate A(Li) measures generated residuals uniformly close to 0.0 dex at all \teff, but with the dispersion increasing predictably as a function of spectroscopic difficulty, {\it i.e.} $\pm$0.10 dex for slowly rotating giants and fainter main sequence stars, increasing to more than $\pm$0.2 dex for increasingly broadened stars in the brighter MSTO region.

In the accompanying figures, different symbol types are used to highlight the 
RO$\alpha$ (open) and RO$\beta$ (filled) sub-samples, with the latter group dominated by rapidly rotating stars, potential binaries or both.  Point colors and shapes further distinguish between Li detections (red circles) and upper limits (blue triangles).

\subsection{Li Structure and the CMD: Evolved Stars}

\begin{figure}
\figurenum{8}
\includegraphics[angle=270,width=\linewidth]{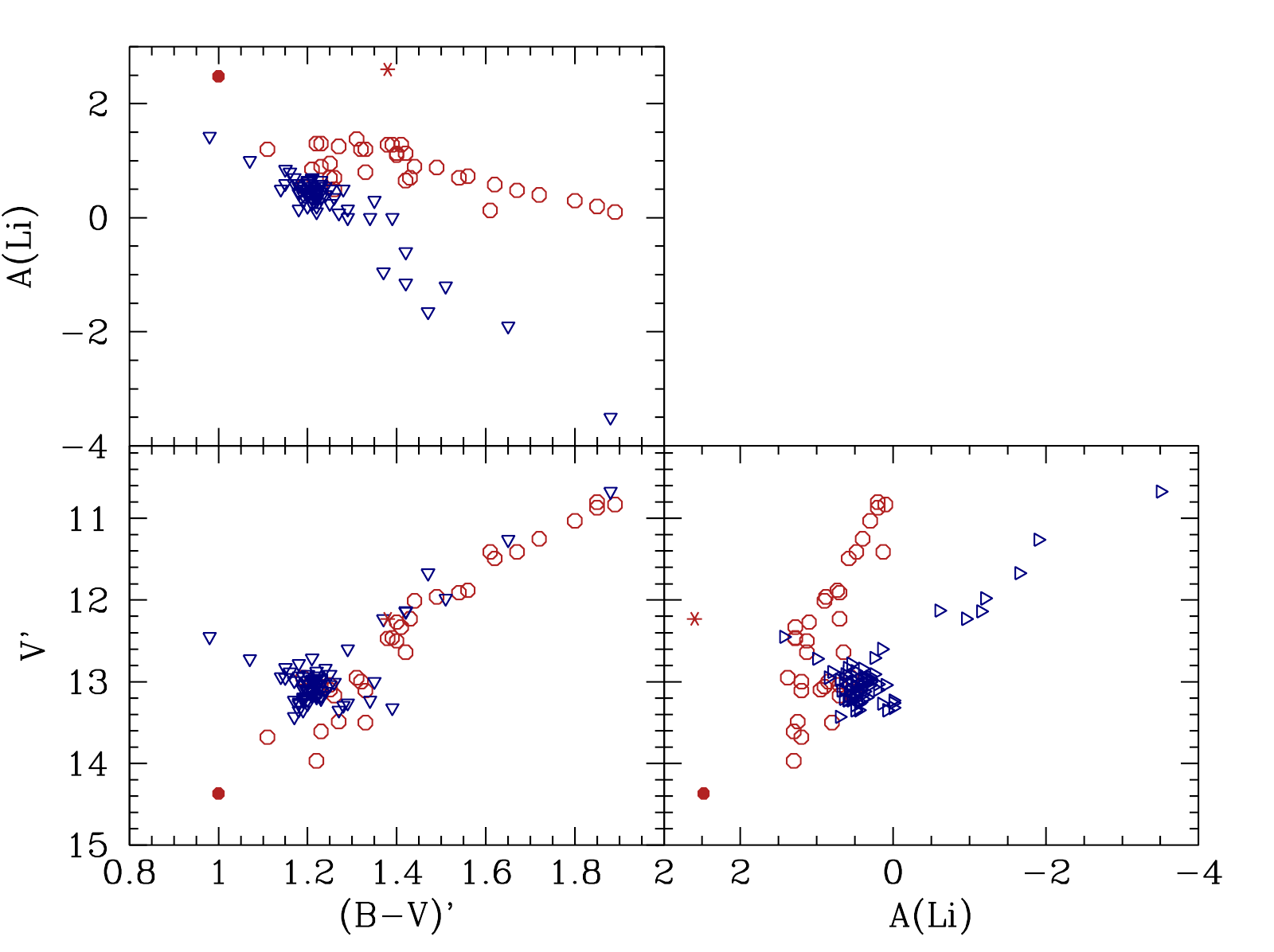} 
\caption{Li abundance patterns among the giants. Red circles indicate stars with Li detections, blue triangles those with upper limit determinations. Nearly all of these stars are in the RO$\alpha$ category with narrow lines denoted by open symbols.
The known Li-rich giant 5440/K301 is denoted by an asterisk; the single RG in the RO$\beta$ category (filled red circle) is the significantly rotating star 13623.}
\end{figure}

Figure 8 presents the CMD and Li abundances for stars from the middle of the SGB to the tip of giant branch, excluding the reddest giant which fell outside the model boundaries. 
This figure includes a number of stars not part of Figure 5 due to positions outside the CCD field.  Photometry for these stars has been obtained by conversion of the Gaia ($G, (B_{p}-R_{p})$) data to ($V, (B-V)$) of ST00 using transformation relations defined by stars within the CCD field. While a model atmosphere could not be constructed for the complex spectrum of the coolest giant, this star clearly exhibited a measurable Li line, though slightly weaker than the lines detected among the giants between \BVp\ = 1.8 and 1.9. 

Leaving aside for the present the two conspicuously anomalous stars in Figure 8, the first obvious feature is the separation of stars above the clump luminosity into two distinct groups: stars with Li detections, starting at A(Li) $\sim$ +1.1 near \BVp\ = 1.4 and declining to A(Li) $\sim$ 0 near \BVp\ = 1.9, and the 7 stars with upper limits only, all falling below A(Li) $= -0.5$. 
We caution that the seeming color trend of limiting values for A(Li) is generally an artifact of improving abundance sensitivity with decreasing stellar \teff\ rather than evidence for an evolutionary trend, in contrast with the sample of stars with Li detections.

While less striking, it is also apparent that with one exception, the stars with Li upper limits fall to the brighter (blueward) side of the giant branch defined by the stars with measurable Li. It should be noted that this trend occurs within a sample corrected for variable reddening as detailed in Appendix C.  This trend is also confirmed by NA23, even in the absence of variable reddening corrections. 

The similarity to NGC 2204, an older but more metal-deficient open cluster with \ebv = 0.07, is striking (see Figure 7 of \citet{AT24}). The NGC 2204 CMD isn't as well-populated, but the separation into declining detections with increasing luminosity versus non-detections is obvious, as well as the color separation in the CMD between detections and non-detections.  Among giants with Li detections above the RGC, A(Li) in NGC 2204 declines from +0.6 at $(B-V)$ = 1.2 to $-0.4$ at $(B-V)$ = 1.7. The straightforward interpretation for both clusters is that the stars with Li detections are RGB members while the stars with upper limits are post-RGC stars on the AGB, though the evolutionary status of the evolved stars is best determined through asteroseismology, unavailable for NGC 7789.

Returning to the CMD, the RGC is clearly dominated by stars with upper limits to A(Li) near 0.6 though, as predicted, the bluest stars within the clump have higher limits due to \teff\  sensitivity. Of the 61 stars between \Vp\ = 12.7 and 13.5 with upper limits to A(Li), there are six stars whose CMD positions place them on or redder than the probable location of the RGB at the level of the RGC. If these stars are truly RGC stars, their peculiar location remains unexplained, though similar stars have been identified in other less populated clusters, e.g. two stars in NGC 2506 (Figure 5, \citet{AT18b}) and one star in NGC 2204 (see Figure 7 of \citet{AT24}). The one star in NGC 2204 is evidently a rapid rotator for its luminosity (\vrot\ = 35 \kms), but only star 6693 in NGC 7789 at \vrot\ = 13 \kms\ has a higher speed than typical for the giant branch stars at its luminosity (\vrot\ = 9 - 10 \kms).

On the other side of the Li ledger, of the 12 stars between \Vp\ = 12.9 and 13.5 with Li detections, 7 overlap with the RGC location. These 7 giants have mean A(Li) = 0.74 $\pm$ 0.14 (sd); the 5 giants in the giant branch region redward of the clump have mean A(Li) = 1.17 $\pm$ 0.22 (sd). Since there is no statistically significant difference in the average \vrot\ of the two groups, one possibility is that a bimodal distribution of \vrot\ among stars at the turnoff generates a bimodal A(Li) distribution for the RGB up to the level of the RG bump. The greater reduction in \vrot\ for the stars occupying the higher velocity peak triggers a greater diminution of A(Li). Alternatively, the lower A(Li) group could be the end result of binary evolution/interaction, particularly in a cluster rich in blue stragglers. For reasons discussed in the next section, determination of the mixing-dominated $^{12}$C/$^{13}$C ratio for the stars at the luminosity level of the clump could prove invaluable. Without more information, the question remains open. 

Of the two anomalous stars seen in Figure 8, the red filled circle is star 13623, one of only a handful of stars between the MSTO and the giant branch at the level of the RGC. 
Its spectrum in the region of the Li line is one of several illustrated in Figure 6.
Its Li abundance near 2.5 and \vrot\ $\sim\ 60$ \kms\ are consistent with a star approaching the red edge of the subgiant branch, having partially spun down from a significantly higher rotation speed as it expands, while reducing its presumed initial A(Li) near 3.3 to its current value via partial subgiant dilution, as the surface convection zone (SCZ) deepens past the Li preservation region.  The further evolved stars (red circles) with \BVp\ = 1.2 to 1.45 show a Li-Plateau near A(Li) = 1.2, illustrating the full extent of subgiant dilution following the maximum penetration of the SCZ. This Li-Plateau probably also owes its existence to the composition discontinuity left behind from the maximum penetration of the SCZ, which likely inhibits non-convective mixing with layers below.  After the outward-moving hydrogen burning shell erases this composition discontinuity, mixing can take place, as evidenced by the declining Li abundances for \BVp\ $>$ 1.45.

The second star of note (red asterisk) is 5440/K301, which was originally flagged as a Li-rich giant star by \citet{PI86} and confirmed by NA23. As seen in Figure 5, its luminosity places it well above the RGC in a location normally associated with the red giant bump \citep[see, e.g.,][]{AT18b}. NA23 provide an excellent summary of the possible origins of its anomalous A(Li), taking into account the fact that it is tied for the lowest $^{12}$C/$^{13}$C ratio of any star in their sample, indicative of high levels of mixing, while maintaining an excessively large He I EW, a feature normally associated with rapid rotation and Li enhancement \citep{SN22}. Neither our data nor that of NA23 exhibit evidence for significant rotation or the presence of a binary companion.

\subsection{Comparison to Previous Li Results}
Before turning to Li among the stars in the CMD turnoff, a comparison of the current Li results with past work makes sense for two reasons. First, the two large Li studies in NGC 7789, \citet{PI86} and NA23, are almost exclusively focused on evolved stars.  Second, the more recent and more precise study (NA23) appears to contradict both \citet{PI86} and the current analysis in a way which, if correct, would have serious implications for our understanding of giant branch Li evolution. 

The critical early Li compilation by \citet{PI86} included 23 mostly evolved stars.  Four stars do not overlap with our single member Hydra sample; an additional star does but is too cool for our analysis pipeline.  Of the remaining 18 stars, four have upper limits for Li in both studies and three are warmer stars near the MSTO.  For the 11 cool stars with Li detections in both studies, our values for A(Li) for the giant stars in common are insignificantly larger by $0.08 \pm 0.08$ (sem) dex. From Figure 7, despite the lower \teff\ resolution of \citet{PI86}, it is apparent that the \teff\ scales of our two systems follow the same trend with color, with the older data typically cooler by about 100 K, quantitatively consistent with the slightly lower mean A(Li) estimates. Consideration of other parameter differences, (OURS - PI86), as a function of \BVp\ color, shows no significant dependences except perhaps for \vturb.  On average, we employ microturbulent velocity values about 1 \kms\ smaller, with even greater discrepancies for the warmer giant stars. 
 
From NA23 we focus on the results for 13 giants in common to our study. In contrast with the comparison to \citet{PI86}, our overall values for A(Li) are, on average, lower by $0.44 \pm 0.22$ (sd) dex compared with those of NA23. Equally important, the residuals in A(Li) are strongly correlated with \BVp, indicating an effect likely tied to \teff. As noted earlier, unique among the NGC 7789 spectroscopic studies to date, NA23 obtain \teff\ via atomic line depth ratios. The differences in \teff, with the discrepancy increasing for progressively cooler stars, are readily apparent in Figure 7.  As expected, the differences in A(Li) with \BVp\ mirror the trend in \teff\ with color. 

Tests of synthesis measurements were conducted by constructing model atmospheres 100 K higher/lower than our adopted \teff\ values to judge the impact on measured A(Li).  Such a \teff\ difference implies an increment/decrement in A(Li) of +0.15 dex/$-0.20$ dex for stars near the base of the giant branch and +0.10/-0.25 for the coolest giants in our sample. To confirm that the differences in A(Li) are attributable to differences in adopted \teff\ and not due to differences in synthesis measurement, we recomputed model atmospheres and remeasured A(Li) in our spectra by synthesis, using the same atmospheric parameters employed by NA23.  The Li residuals essentially disappear, as does any trend in A(Li) residuals as a function of \BVp. 

NA23 supply A(Li) for 14 cluster members. One star is the Li-rich anomaly, 5440/K301. Two others exhibit upper limits. One of these sits within the RGC and the second is tagged as a probable AGB star, i.e. both are post He-ignition and should exhibit diminished A(Li) due to mixing. The remaining 11 stars, ranging in luminosity from \Vp\ = 12.6 to 10.7, show no trend in A(Li) with decreasing \Vp. In fact, within the uncertainties, all 11 stars have the same A(Li) = 1.08 $\pm$ 0.15 (sd). If one star, 6810/K468 is dropped from the discussion due to a significant discrepancy between the derived values, A(Li) = 0.8 for NA23 vs an upper limit of $-1.9$ from our data, their mean value becomes 1.11 $\pm$ 0.12 (sd). 

While the A(Li) differences may be tied to the differences in adopted parameters, especially \teff, we are still faced with the task of deciding which, if either, scale is likely to be more realistic. We offer two indications that the NA23 scale may be in error. The first is the apparent collapse of the \teff\ - \BVp\ scale at the cool end. Between \BVp\ $\sim$ 1.45 and 1.85, \teff\  for NA23 changes by about 400 K; all the remaining scales, independent of the zero point, change by roughly 700 K. Second, if the NA23 scale is correct, stars evolving on the RGB beween the red giant bump and the tip of the RGB undergo no change in A(Li), indicative of little to no atmospheric mixing for stars evolving toward the red giant tip. This lack of a trend is contradicted by the $^{12}$C/$^{13}$C ratio for 10 of the 11 giants under discussion; one of the 11 stars has no $^{12}$C/$^{13}$C measure. Assuming that $^{12}$C/$^{13}$C is relatively immune to \teff\ changes (NA23), the 10 stars exhibit a significant mean trend of $^{12}$C/$^{13}$C declining with $V$, dropping from $\sim$35 at \Vp\ = 12.7 to 14 at \Vp\ = 11.0. As already noted, the two post-He-ignition stars have $^{12}$C/$^{13}$C = 15.

\subsection{Li Structure and the CMD: MSTO Stars}

Before examining the A(Li) trends among the MSTO stars, a question of parameterization should be addressed. For main sequence stars the most common approach for quantifying the pattern of A(Li) is to correlate the abundance with \teff, i.e. the Li-Dip is often discussed as occupying a specific range in \teff. This makes sense since it is well known that the boundaries of the Li-Dip change mass and color with changing [Fe/H]. While this approach works well for clusters younger than the Hyades since the stars occupying the Li-Dip have low enough mass that their position within the CMD remains relatively unchanged, by the age of NGC 7789 and NGC 752 and beyond, this is no longer the case. We will return to this point in Section 7.3. 

Equally important, errors in color indices are naturally larger than either of the errors for the individual magnitudes that make up the color index. Since, for example, the change in $V$ along the unevolved main sequence is approximately six times larger than the change in $(B-V)$, $V$ is more sensitive to small changes in mass along the main sequence than the the typical color index and less sensitive to errors of comparable size. As we have done in all discussions of A(Li) within clusters older than a Gyr, we will retain $V$ as our primary variable for sorting stars along the main sequence and the turnoff region. 

In Figure 9, filled symbols highlight the prevalence of stars in the RO$\beta$ class due primarily to rapid rotation. 
Fortunately, the rapid rotators are primarily found above \Vp\ $\sim$ 15.2. (It should be noted that the decline in stars between \Vp\ = 14.9 and 15.2 is an artifact of the original selection process designed to minimize inclusion of probable binaries in the spectroscopic sample below where the NI20 sample cuts off.)  The sharp ``Wall'' feature on the warm side of the Li-Dip, defined as the magnitude where Li detections start to decline steeply, appears at \Vp\ $\sim$ 15.2 to 15.4, with the cool side of the Li-Dip appearing near \Vp\ $\sim$  16.0. For the stellar sample above the Li-Wall, the apparent limiting A(Li) = 3.3, the initial value adopted in the iterative fitting procedure for the spectra. Combined with the rapid \vrot\ implied by the filled symbols, the scatter between A(Li) = 3.3 and 3.0 becomes understandable as a likely byproduct of the reduced \vrot\ accuracy with rotationally broadened lines. Note, however, that there are stars well above the Li-Wall with limits to A(Li) significantly below the 3.0 to 3.3 detections. A wide range in A(Li) above the Li-Wall, including non-detections, is a common property in clusters the age of NGC 7789 and older. For a composite example of data from three clusters showing the same effect relative to the Li-Dip structure as defined by the Hyades, the reader is referred to Figure 10 of \citet{AT09}.  

Due to its rich MSTO population, NGC 7789 represents a much better link between the well-studied younger clusters, Pleiades \citep{MA07, BO18}, M35 \citep{ST04, AT18a, JE21}, M48 \citep{SU23}, and Praesepe/Hyades \citep{CU17} and the older disk clusters (e.g. NGC 6819 (DE19), NGC 2204 \citep{AT24}, M67 \citep{TW23}, NGC 7142 \citep{SU20}, NGC 188 \citep{SU25, SU26a}) than the closer but sparsely populated NGC 752 \citep{BO22}. Throughout the series of cluster papers dealing with predominantly older but less populated systems than NGC 7789, an oft-repeated theme of the analyses has been the role of stellar rotation, specifically the spin-down of main sequence stars as a potential driving force behind the reduction in atmospheric Li in stars of high enough mass where normal atmospheric convection should prove nonexistent or inadequate to the task. The most recent evidence in support of this concept as applied to the cool side of the Li-Dip can be found in the discussion of the Li abundance among the subgiants in the older open cluster, NGC 188 \citep{SU25}. Given the rich data set of Figure 9, can we ascertain any indication of the potential origin of the Li-Dip?

\begin{figure}
\figurenum{9}
\includegraphics[angle=270,width=\linewidth]{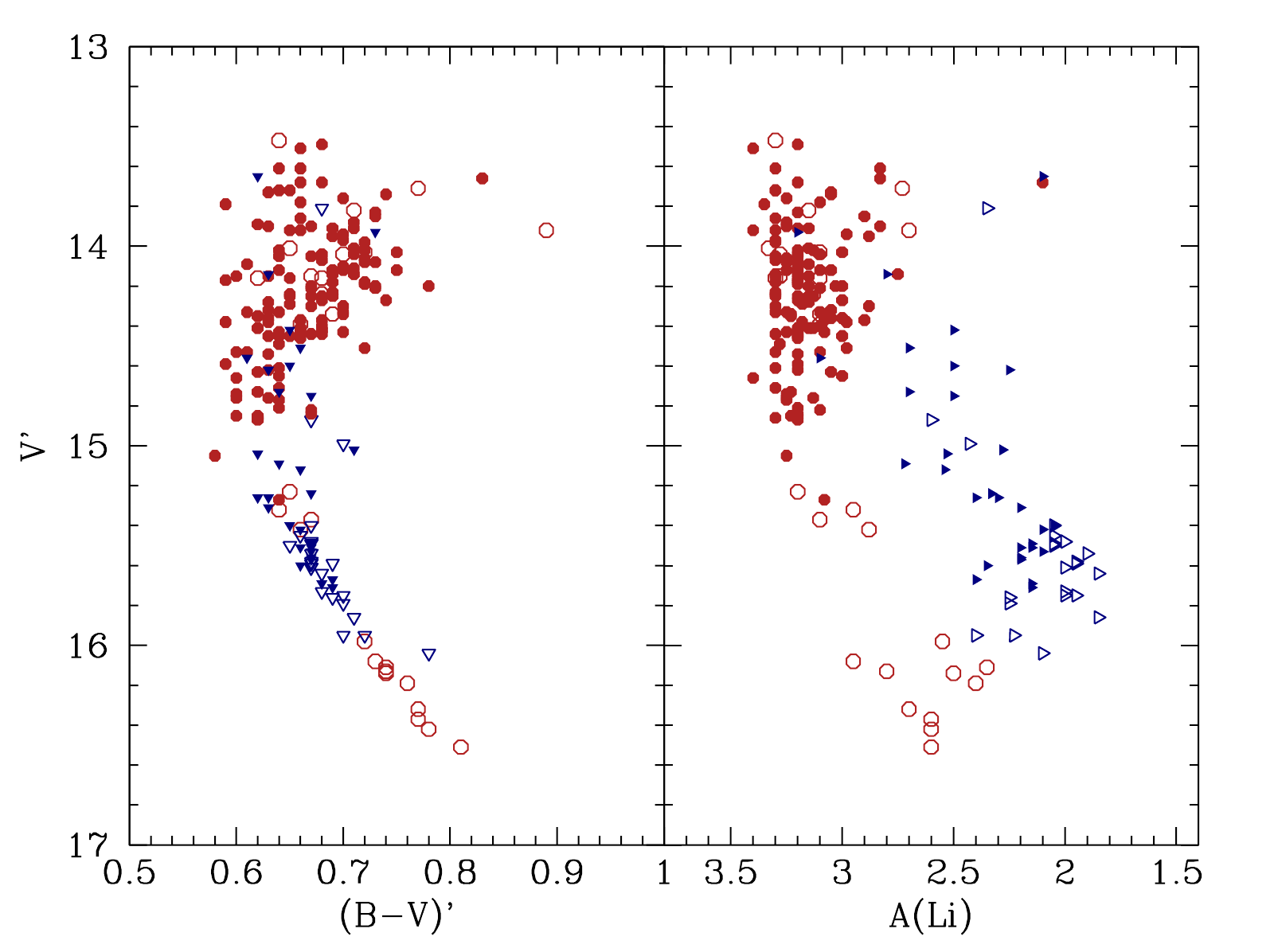}
\caption{Li abundance patterns among MSTO stars. Symbols have the same meaning as in Figure 8. Red symbols indicate Li detections, blue triangles upper limits to Li. Open symbols are used for stars in the RO$\alpha$ category while filled symbols denote stars in the RO$\beta$ category due to broadened lines. This latter class of stars dominates the turnoff region sample.}
\end{figure}

To illustrate the strong connection between the main sequence A(Li) structure and stellar rotation, Figure 10 shows \vrot\ as a function of \Vp\ for all stars with Li detections (circles) and upper limits (triangles). The \vrot\ distribution of Figure 10 confirms in more detail the preliminary results discussed for the cluster in DE19 and strengthens the pattern that first emerged in comparison with NGC 6819, NGC 2506 \citep{AT18b}, and NGC 3680 \citep{AT09}, recently expanded to include NGC 2243 \citep{AT21}, NGC 2204 \citep{AT24}, and NGC 188 \citep{SU25}. While there are hints of a bimodal distribution in \vrot\ for stars in Figure 10 brighter than 15.25, with a possible minimum near 70 \kms, the increasing uncertainty in the measured \vrot\ above 50 \kms\ weakens this claim.
Along the \Vp\ axis, the plot can be separated into three distinct groups. First, for \Vp\ fainter than 15.75, all \vrot\ values are at 21 \kms\ or lower, declining gradually with increasing \Vp\ to a limiting value near 12 \kms at \Vp\ = 16.5, keeping in mind that our spectroscopic lower limit for \vrot\ is approximately 10 \kms. 

The second group contains the stars between \Vp\ = 15.25 and 15.75, the range that includes the Li-Wall and the brighter half of the stars in the Li-Dip. For this middle group the \vrot\ range suddenly expands to just under 60 \kms\ from the 20-25 \kms\ limit among the fainter stars. The cooler \vrot\ transition matches the approximate location of the Li-Dip center. (Given that the Li-Dip is populated by stars with predominantly upper limits in A(Li), exact definition of the Li-Dip center is a challenge. The choice here is made relative to the brighter and fainter boundaries where Li detections begin to predominate over upper limits.) The Li-Dip center has been tied to the break in the Kraft curve for \vrot\ \citep{KR67} (hereafter referred to as the KB), the sharp division between the hotter F stars that spin fast and take a very long time to spin down (DE19) and the cooler F stars that spin down much more quickly. This striking feature was first pointed out for the Hyades by \citet{BO87} and since has been identified in multiple clusters younger than the Hyades, e.g. NGC 2516, M34, and NGC 6633 \citep{JE02, TE02}. 

Finally, between \Vp\ = 15.25 and 13.5, \vrot\ covers the full range from 20 \kms\ to above 140 \kms, with no apparent change in the distribution with \Vp. To clarify the possible significance of these three \vrot\ zones, as well as the structure among the evolved giants, the added insight supplied by complementary data within NGC 752 will be presented next.

\begin{figure}
\figurenum{10}
\includegraphics[angle=0,width=\linewidth]{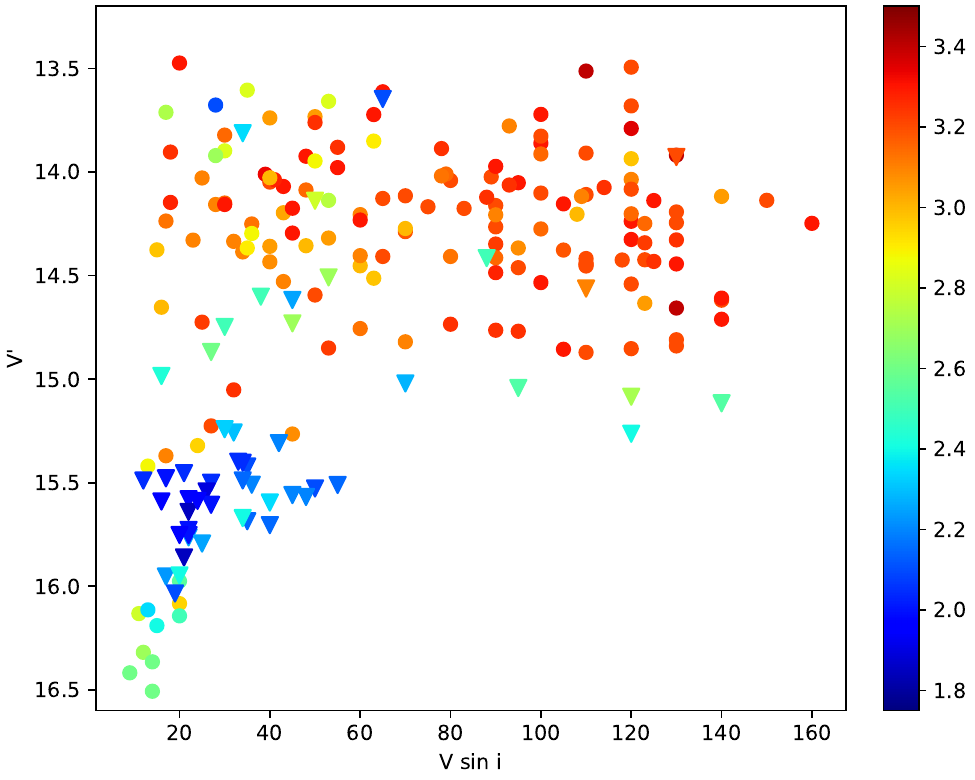}
\caption{\vrot\ distribution with increasing \Vp\ for stars at the MSTO. A(Li) for each star is indicated by the color coding, with detections drawn as circles and upper limits as triangles.}
\end{figure}

\section{NGC 752: NGC 7789 Twin?}
The open cluster NGC 752 is, in most ways, a much better studied system than NGC 7789, primarily due to its proximity ($(m-M)_{0}$ = 8.22) and much lower reddening ($E(B-V)$ = 0.035) \citep{TW23}. The one surprising weakness in its data base is the lack of a uniform broadband $UBVRI$ survey covering the wide area defined by the cluster, though precision $uvbyCa$H$\beta$ photometry \citep{TW15} is available for the core regions of the cluster and Gaia photometry is available for almost all cluster members. When membership is restricted in a fashion identical to that of NGC 7789 and the cluster CMD is compared to the same isochrones used in the current investigation, though $(b-y)$-based rather than $(B-V)$, NGC 752 is found to have an age of 1.45 $\pm$ 0.05 Gyr for [Fe/H] $= -0.032$. While other analyses have generated ages ranging from approximately 1.3 to 1.6 Gyr, the spread can be attributed to differences in the adopted set of isochrones coupled with varying assumptions for the reddening/metallicity. (For the most recent comprehensive discussions of the derived cluster parameters, the reader is referred to \citet{SA23} and \citet{TW23}.)  

Unfortunately, the one shortcoming of the NGC 752 data set is the paucity of members; the original Gaia astrometric survey by \citet{CA18} identified only 253 members to the astrometric limit of the data base, in contrast with a few thousand for the much more distant NGC 7789, despite selection to an intrinsically much brighter limit on the unevolved main sequence. The apparent dissolution of the cluster, postulated for decades, has been evaluated by \citet{BH21}, who mapped the cluster's tidal tails and demonstrated the significant effects of mass segregation on the mass function between the core and extended cluster environs. More recently, \citet{BF22} and \citet{LI23} have taken the tidal tail analysis a step further, leading to the discovery of 467 possible members spread over 37\arcdeg\ of the sky!  This expanded sample offers potential candidates for followup study to fill in many of the gaps discussed below that currently exist within the spectroscopic surveys of the cluster. (Stellar identifications will be given using the systems of \citet{HE26} and tagged with an H or \citet{PL91} and tagged with PL.)

\subsection{NGC 752: The Red Giants}
Despite the limitations, can NGC 752 provide any insight into the Li status of NGC 7789, or vice versa? We begin with a discussion of the red giants. Following the same precepts applied to NGC 7789, the data are restricted to only single-star members isolated via astrometry from Gaia and from radial velocity studies as detailed in \citet{DA94, ME98, ME08, ME09}. 
Gaia astrometry tags 15 possible red giant members in the cluster core, with an additional four identified by \citet{LI23}. Of the 15 core RG, two are binaries. Two others have radial velocities that make cluster membership questionable; one of them also has an emission line spectrum.  Of the remaining 11 giants from the cluster core, \citet{JA24} have concluded from UV observations that star H295 has an extremely low mass white dwarf companion, implying an evolutionary history that includes probable stellar interaction coupled with mass loss and/or transfer. Without further information, we opt to retain it in our Li discussion. 

There have been multiple studies of Li abundances for main sequence stars, but only three compilations for the giants. The first was part of the same program which revealed the Li pattern in NGC 7789 and included 11 red giants in the field of NGC 752 \citep{PI88}, with upper limits only for all but two stars. 
\citet{GI89} studied six stars in the cluster field of which four are single members, all of which had detectable Li.  
The more recent work is that of \citet{BO15} analyzing 10 cluster giants. Within these studies, all 11 giants of interest have A(Li) measures.  In all cases where multiple measures are available, we have defaulted to the \citet{BO15} value given the higher resolution and more uniform S/N of their spectroscopy. Only one star with an upper limit has been taken from \citet{PI88}.

To optimize precision, photometry from Gaia ($G,(B_{P}-R_{P}$)) and ($V, (b-y)$) \citep{TW15} were transferred to the ($V, (B-V)$) system of \citet{DA94} and the values averaged. Note that all the giants have excellent photoelectric $V$ magnitudes and ($B-V$) indices. The data were then adjusted to the reddening (\ebv\ = 0.30) and apparent modulus ($(m - M)$ = 12.51) of NGC 7789 and plotted on the same figures as the data of NGC 7789. In Figure 11, we repeat two of the plots of Figure 8, with the symbols having the same meaning (open circles for detections, triangle for upper limits), but the NGC 7789 data are light red while the NGC 752 points are drawn as dark blue circles and triangles. Note the change in scale relative to Figure 8 since all the stars fall at the hot end of the red giant distribution defined by NGC 7789.

The first obvious difference between the clusters is the total lack of stars above the luminosity level of the clump in NGC 752. (Note: of the four additional giants identified by \citet{LI23}, only one is brighter than the RGC. A second star has the same luminosity as the clump but sits $\sim$0.25 mag redder.) This asymmetry in the distribution of red giant stars compared to clusters of similar or slightly older age, e.g. NGC 3680 and IC 4651,  has been noted before \citep{AT04, AT09}. From NGC 7789, one quarter of the red giants should sit brighter than the clump, indicating that four stars should occupy the region for a sample list of 15 red giants, rather than one, keeping in mind that we have no binary information about the four stars from the cluster tidal tails. 

The more striking result is the distribution of stars in the CMD and the Li abundance. As in NGC 7789, the giants separate into two distinct groups, those with A(Li) detections between 0.8 and 1.4, and those with A(Li) at or below 0.2. All the stars in the first group fall to the red along a path expected for the RGB. The Li-deficient stars form a tight clump of four almost identical stars in the CMD, with the remaining 2 positioned fainter than the clump and overlapping slightly with the RGB. One of these is H295, the red giant with a potential low mass white dwarf companion. While the sample is small, the clump stars occupy a very limited range in color compared to the extensive spread among the RGC stars in NGC 7789. We will return to this point after dealing with the MSTO stars.

\begin{figure}
\figurenum{11}
\includegraphics[angle=0,width=\linewidth]{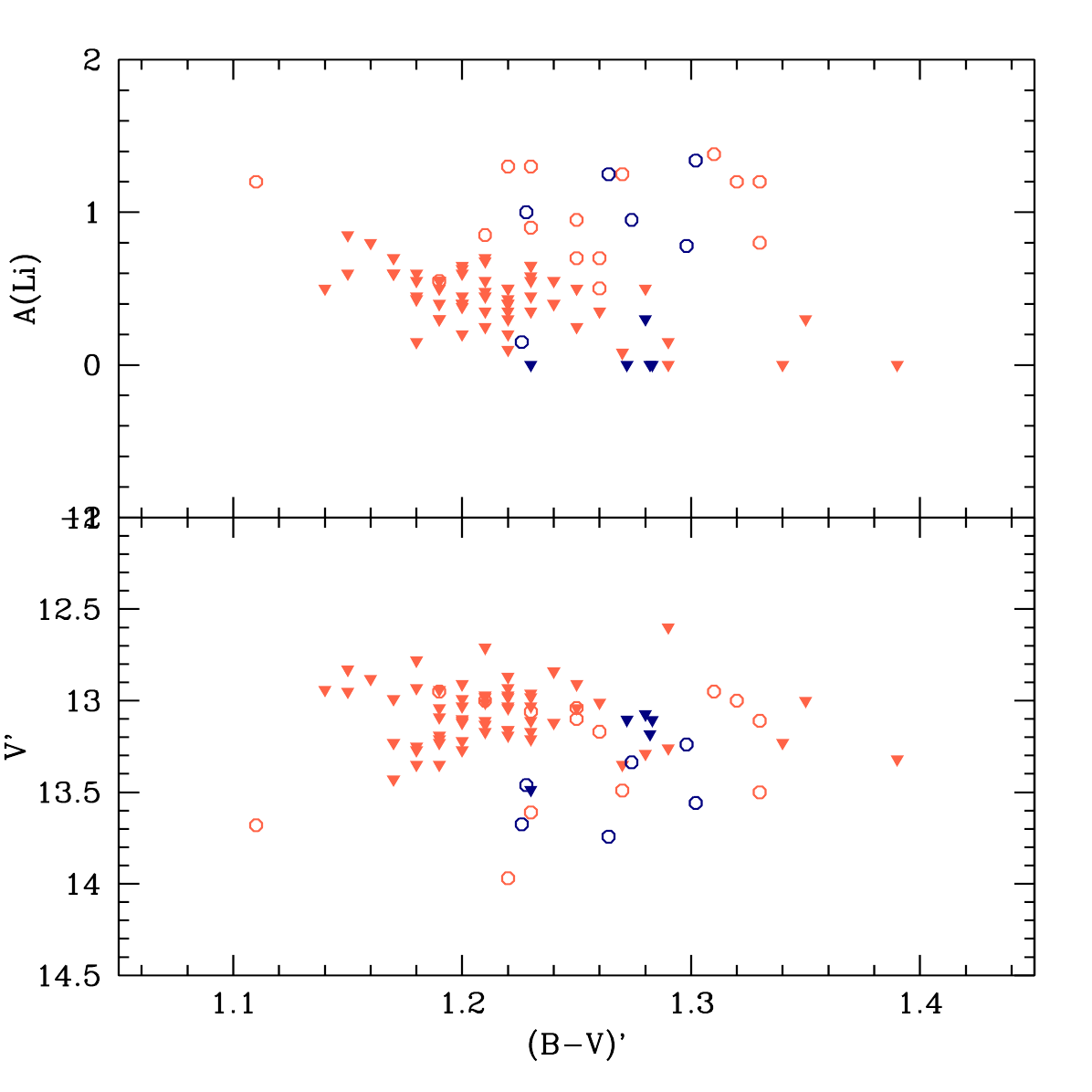}
\caption{Red giant single stars within NGC 752, shifted to the reddening and distance of NGC 7789. Light red symbols are the data of Figure 8, with circles denoting Li detections and triangles tagging upper limits to A(Li). Blue symbols define the same information for NGC 752.}
\end{figure}

\subsection{NGC 752: The MSTO Stars}
Since the first look at Li among the turnoff stars of NGC 752 \citep{HP86, PI88}, the cluster has undergone periodic reevaluations and additions, usually as part of a comparative analysis with other clusters, often including M67 and/or the Hyades \citep{DA94, BA95, SE04, AT04, AT09, CA16}. The most recent expansion of the sample comes from \citet{BO22} with the addition of Be abundances and a uniform merger of the past Li data for stars at the turnoff, a sample which forms the core of our discussion. Before the comparisons are made, a few corrections are required for consistency with the NGC 7789 analysis. \citet{BO22} compiled a uniform catalog of 39 MSTO stars with A(Li) measures in NGC 752. Of these, 8 are astrometric nonmembers based upon Gaia analysis. A ninth, PL1118, is just beyond the 3-$\sigma$ limit for astrometric membership but will be retained given the fact that the cluster is disintegrating and some probable members are likely to be found in a kinematic transition phase. Of the remaining 31, six are spectroscopic binaries. H183, located well below the Li-Dip, sits more than 0.5 mag above the unevolved main sequence in a position indicative of a composite binary system, though no radial velocity evaluation of this possibility exists; it will be deleted. Finally, three stars, H120, H146, and H207, exhibit photometric peculiarities. As discussed earlier, comparison of precision $V$ and $G$ magnitudes normally generates a well-defined relation between $(V-G)$ and $(B_{P} - R_{P})$. The same holds true for the stars in NGC 752, with four exceptions among the A(Li) catalog. H10 is a known binary and is already eliminated, leaving H120, H146, and H207. H120 is actually H120A and H120B, separated by $\sim$0.5\arcsec. The other two stars have not been resolved as optical pairs, but their positional precision within Gaia is excessively poor, often an indication of image distortion caused by unresolved companions. H120 will be retained but the others, located well below the Li-Dip, will be dropped from the analysis.

The results for NGC 752 (filled blue symbols) are shown in Figure 12a, superposed upon the data for NGC 7789 (red symbols). Triangles distinguish upper limits to A(Li). The clear benefit of the addition of NGC 752 comes in the definition of the Li pattern fainter than the A(Li) minimum, i.e. below the center of the Li-Dip, since all but two of the  NGC 752 stars in the composite sample above this point have been eliminated as either binaries or nonmembers. A modest number of likely single stars do populate the CMD brighter than the Li-Wall in NGC 752, including one blue straggler (see Figure 1 of \citet{LI24}) but, unfortunately, these have yet to be included in Li surveys. Overall, it is apparent within the uncertainties that NGC 752 follows the same evolutionary trend as NGC 7789, with a likely Li-Wall between \Vp\ = 15.2 and 15.4, a Li-Dip center, based upon the position of the detection boundaries, between \Vp\ = 15.6 and 15.8, and a return to a Li-Plateau centered near \Vp\ = 16.5.

\begin{figure}
\figurenum{12}
\includegraphics[angle=0,width=\linewidth]{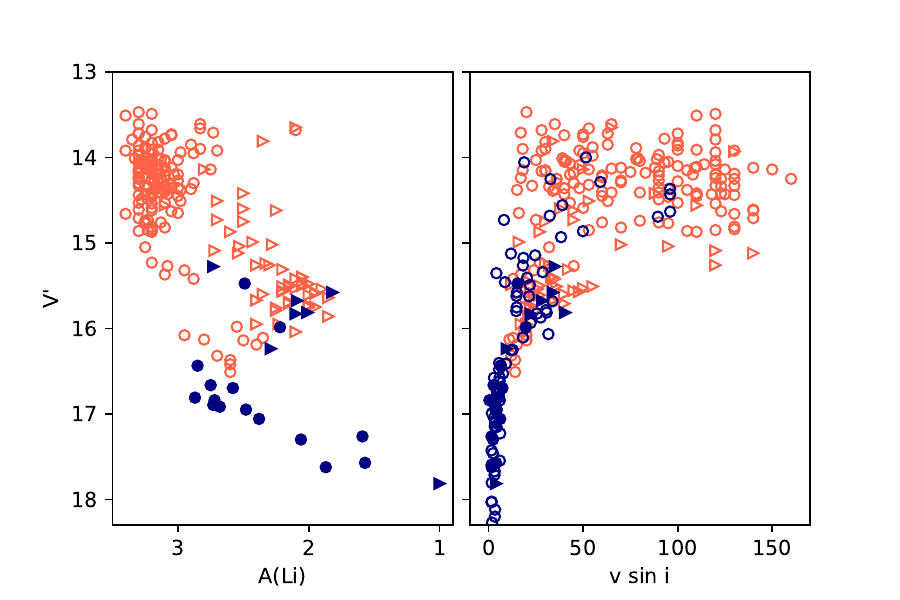}
\caption{(a). A(Li) for MSTO single stars within NGC 752 (blue symbols), shifted to the reddening and distance of NGC 7789 and superposed on the NGC 7789 data (red symbols). Triangles mark upper limits in A(Li). (b). Comparison of the \vrot\ distribution for NGC 752 (blue) and NGC 7789 (red). Added are stars within NGC 752 with \vrot\ but no A(Li) data (open blue circles).}
\end{figure}

How does the \vrot\ distribution for this sample compare with that for NGC 7789? With only 22 MSTO single stars evaluated for Li, the sample is inadequate for our needs. However, the lack of Li data doesn't imply a lack of \vrot. To get a better sense of the stellar distribution in rotational velocity, all astrometric cluster members \citep{CA18} within 3\arcdeg\ of the the cluster center were mapped to the results from the DR17 catalog of stellar parameters from APOGEE \citep{DR17}, generating \vrot\ for a sample of 116 stars. This catalog was then mapped to the radial velocity catalog of \citet{ME09}, allowing elimination of any known radial velocity binaries and/or nonmembers. From 30 single stars in common below \vrot\ = 40 \kms, an excellent correlation is found between the two catalogs, following the relation \vrot (ME09) = 0.94*\vrot (APO) - 0.431. Three discrepant stars were not included in the calibration; two have \vrot\ well above 50 \kms\ in the APOGEE catalog while the third lies near the upper bound of the relation. These three single stars have the largest radial velocity uncertainties within the \citet{ME09} catalog for NGC 752, implying that they likely have unusually broader lines due to \vrot\ and/or binarity, even if a binary signal hasn't been detected statistically. 

All \vrot\ values from APOGEE were converted to the system of \citet{ME09} and, for those in common, averaged. The final sample, plotted in Figure 12b, consisted of 139 stars from all sources ranging from \Vp\ = 14.0 to 23. Symbols have the same meaning as in Figure 12a. For NGC 752, only the filled symbols represent stars with A(Li) measures; open blue circles are NGC 752 stars with only \vrot\ values. As noted, the NGC 752 sample with \vrot\ extends another 5 magnitudes below \Vp\ = 18; excluding 2 stars with anomalously large \vrot, the remaining 47 stars have an average \vrot\ of 3.7 \kms, with a dispersion of only 2.2 \kms.

As with NGC 7789, the NGC 752 distribution falls into three distinct zones. First, for \Vp\ brighter than $\sim$ 15.1, the full range of \vrot\ is displayed, though the range is clearly smaller for NGC 752 than NGC 7789. The upper bound for NGC 752 may be even smaller, i.e. closer to 60 \kms. There are four stars with \vrot\ above 60 \kms, three of which are assigned 96 \kms. These values are questionable in that the APOGEE survey assigns this numerical value to stars with the apparently broadest detectable lines, for whatever reason, not necessarily due to pure rotational speed.  A similar example is H64, one of the three stars excluded from the \vrot\ transformation relation between APOGEE and \citet{ME09}, where the respective \vrot\ measures were 88.4 and 8.1 \kms. With that caveat in mind, the four highest \vrot\ candidates will be retained in the discussions that follow. 

The second zone between \Vp\ $\sim$ 15.1 and 15.85 contains 24 stars with \vrot\ between 4 and 41 \kms, though only 2 stars sit below 10 \kms, not unexpected if all the stars actually have rotational speeds well above 10 \kms\ but appear smaller because we are equating \vrot\ with $v_{rot}$ without correcting for the unknown inclination effect. Excluding two rapid rotators near the bright boundary, the data for NGC 7789 rises above 50 \kms, but its total sample (42) is larger than that for NGC 752, so the significance of this difference is marginal. Within the third zone, \Vp\ below 15.85, the \vrot\ limit shows a gradual decline from $\sim$ 25 \kms\ with increasing \Vp, approaching the limiting value of 3.7 $\pm$ 2.2 \kms\ by \Vp\ = 17.0. Note that all the stars below \Vp\ = 16.5 come from NGC 752 and that the observational limiting \vrot\ for these stars is well below 10 \kms.

\subsection{The Cluster MSTO Li Pattern and Implications for Stellar/Li Evolution}

While there are important differences between the MSTO data of NGC 752 and NGC 7789, Figure 12 emphasizes their complementary nature, with NGC 752 delineating the trend among stars fainter than the Li-Dip minimum to a location beyond the Li-Plateau. Using \Vp\ = 15.7 for the approximate location of the center of the Li-Dip, the matched isochrone indicates mass of the stars at this point in the MSTO $\sim 1.36$ M$_{\sun}$. 
The composite data of Figure 12 demonstrate that the key factor defining the distribution of stars as a function of \Vp, in particular the transitions among the three defining zones, is \vrot. This conclusion is based not only on the analysis of NGC 7789 but multiple clusters over a range in age (DE19). Figure 12 displays a combined sample of extraordinary richness
at a pivotal transition age to confirm and expand the proposition that {\it stellar rotation, in particular the spindown of stars with age, is a critical signature of Li evolution both within and outside of the Li-Dip}.  Even more recent corroboration may be found in NGC 188 \citep{SU25} and field stars \citep{SU26b}. 

What can be learned from the critical boundaries for NGC 7789/752? On the hot edge of the Li-Dip, the position of the Li-Wall between \Vp\ = 15.20 to 15.4 translates to stars with reddening-corrected $(B-V)_0$ between 0.32 and 0.35, or \teff\ = 6776 to 6664 K using our color-temperature calibration.
Thanks to the rich population of NGC 7789, the \vrot\ range is well populated through at least 130 \kms, greatly enhancing the precipitous drop in \vrot\ as one crosses from the hot to the cool side of the Li-Wall (Figure 12b). 

As noted earlier, discussions of the \vrot\ pattern among intermediate and low mass stars often focus on the Kraft curve \citep{KR67}, with special emphasis on the Kraft break (KB).  While the \vrot\ transition defined by the KB, supposedly starting near the center of the Li-Dip, has been flagged as evidence for the role of rotation, the Li-Wall has generally been ignored. This is partly because many clusters used to map the Li-Dip are modestly populated above the Li-Wall and/or inadequately sampled in Li studies. 
With a wide range in \vrot, possible contamination by binaries, and confusion caused by nonmembers, a boundary as sharp as that identified within the NGC 7789/752 sample is difficult to discern in a small sample. 

The compound analysis of the Hyades and Praesepe by \citet{CU17} provides an example. The two clusters are shown to be similar enough in age and metallicity to be classed as twins. Removing binaries from the sample, the Li-Dip is reasonably well mapped, except near the likely position of the Li-Wall. For the Hyades, only one star is positioned blueward of this point; its \vrot\ is 62 \kms. Two stars at the color of the Li-Wall have \vrot\ = 62 and 68 \kms. 
There are stars bluer than the Li-Wall in the Hyades that have \vrot\ between 70 and 150 \kms\ \citep{BO87}; unfortunately, they don't have Li measures. 

With the addition of Praesepe to the data set, the Li-Dip structure becomes much more apparent. The Li-Wall is significantly strengthened by the addition of three stars at $(B-V)_0$ = 0.40 $\pm$ 0.01. At the Li-Wall and blueward, 17 stars now define the pattern, with 16 having \vrot\ = 70 \kms\ to 210 \kms. This places the Hyades/Praesepe Li-Wall, on the \teff\ scale of \citet{CU17}, at 6724 $\pm$ 40 K. 

A recent attempt to define the location of the KB by \citet{BE24} uses a sample of 278 nearby single field stars, 217 of which lie cooler than the KB. The KB itself is defined as the region where the \vrot\ distribution plateaus with a range approaching a limit between 50 and 60 \kms. Beyond this plateau toward lower \teff, \vrot\ declines precipitously to less than 20 \kms. On the \teff\ scale of \citet{BE24}, the plateau extends from 6650 to 6450 K, with an uncertainty in each boundary of $\pm$50 K. Based upon the combined NGC 7789/752 data, the KB plateau as defined by \citet{BE24} is equivalent to our intermediate \Vp\ zone, extending from 15.3 $\pm$ 0.1 to 15.8 $\pm$ 0.1. Contrary to past associations of the KB with the center of the Li-Dip, {\it the NGC 7789/752 data imply that the Li-Wall defines the hot boundary of the \vrot\ zone that is the KB.} 

The implications of this empirical constraint are twofold.
First, the Li-Dip boundaries among unevolved main sequence stars, including the Li-Wall, have consistently exhibited correlated dependences on mass and [Fe/H], producing similar boundaries in \teff\ (see \citet{AT21} and references therein). Despite the differences in both age and metallicity among the four clusters discussed above and the inhomogeneous field star sample of \citet{BE24}, within the modest uncertainties attached to all the key parameters, {\it the Li-Wall appears at approximately the same \teff, 6725 K}. 

This has significant implications for the evolution of the Li-Dip with time. The Hyades/Praesepe stars currently populating the Li-Wall will be $\sim$125 K cooler by the age of NGC 7789/752 and the stars of slightly higher mass currently hotter than the Li-Wall will have evolved redward of the \teff\ boundary. If the boundary remains unchanged, these stars will spend the remainder of their main sequence lives, with the possible exception of a short dash to the blue within the hydrogen-exhaustion phase, spinning down and, in theory, depleting Li until they enter the subgiant phase. 

The Li-Dip itself is almost nonexistent within the Pleiades (120 Myr) but has become a very distinct and clear feature no later than the age of M48 (420 Myr \citep{SU23}). It becomes deeper and wider by the age of the Hyades and Praesepe (650 Myr \citep{CU17}). This implies a timescale approaching 0.4 to 0.5 Gyr for significant changes in \vrot\ and A(Li) for stars near the Li-Wall. Thus, for stars near and above the Li-Wall boundary at the age of NGC 7789, the range of the \vrot\ distribution should begin to collapse and A(Li) should begin depletion within the next 0.4 to 0.5 Gyr. As noted in past discussions, direct correlations between \vrot\ and A(Li) for stars above the Li-Wall over time remain elusive since \vrot\ measures the instantaneous rotation speed. Without knowing the original \vrot, it is impossible to know whether the current speed indicates significant spindown or a total lack of spindown. What one can expect is a statistical reduction in the range of \vrot\ and a correlated decrease in both the average A(Li) and the lower bound on the A(Li) distribution with increasing age. This is exactly what is seen in DE19 and confirmed in \citet{AT21} and \citet{AT24}. 

A second implication of the empirical constraint highlighted above is to winnow the traditional set of candidates proposed to explain the dramatic decline in Li (as well as Be and B)  over the \teff\ range of the Li-Dip.  These have included diffusion, mass loss, or mixing of the atmosphere through convection to a hotter layer where these three sensitive elements are destroyed and depleted gas is returned to the surface, with the convection driven by rotation and/or gravity waves. 
It is clear from the expansive studies of MSTO and subgiant stars in clusters \citep[see, e.g.,][among many]{DE98, BO04, BO05, RA07, BO16, BO22, SU25} that 
{\it 
only rotational mixing--mixing induced as instabilities are triggered over time when stars spin down--is consistent with the observations. } 
The dominant process driving the spindown must originate within the interior rather than the atmosphere to explain the lack of sensitivity of the Li-Wall to metallicity effects. 

Put simply, the Li-Wall represents the boundary where the structural changes in the stellar core lead to slower rotation and the declining internal spin is translated to the surface, driving mixing and reduced surface rotation rates.  With time, however, the hot boundary of the Li-Dip  becomes more apparent as the cumulative impact of mixing grows. 
Finally, by the age of the Hyades, the stars populating the Li-Wall on the cool (lower mass) side drive the surface Li to a level that makes only upper limits observable. Stars just hotter than the Li-Wall fail to develop surface  rotational spindown while on the zero-age-main-sequence (ZAMS) and therefore retain their primordial A(Li) with the surface \vrot\ distribution maintaining a range to 140 \kms \ or higher. This general pattern pertains until evolution off the ZAMS carries these stars to cooler \teff\ than the Li-Wall, altering the core composition and mass distribution during the hydrogen exhaustion phase, inducing rotational spindown and mixing to the surface. The long-term effect depends, among many things, on the competing timescales defined by the rate of spindown and the time spent evolving from the \teff\ zone of the Li-Wall to the subgiant branch, all of which are impacted by the internal alteration of the star's structure due to H-exhaustion on the main sequence.  As first stated in DE19 and confirmed by the clusters added since then \citep{AT21, AT24}, the stars hotter than the Li-Dip on the main sequence appear to recreate the processes of the Li-Dip as they age toward the subgiant branch, spinning down while destroying atmospheric Li. Moreover, spindown of these stars takes place on a longer timescale than the rapid spindown observed in G dwarfs.

It should be emphasized again that other processes may be at play beyond this simple outline. In the 420 Myr-old cluster M48, each of two pairs of stars near 7150 K and 6900 K exhibit differences in A(Li) of 0.4 dex, suggesting that some mechanism can create differences at this level prior to substantial MS evolution to the red  \citep{SU23}. Within the current investigation, while the weakness of the data set between \Vp\ = 14.9 and 15.25 dates back to the original sample selection, the few stars that populate this regime have a range in \vrot\ that mimics the pattern seen among the richer data set at brighter magnitudes. This would normally indicate that, despite the modest sample here, this narrow luminosity range deserves inclusion outside the Li-Wall on the hot, more luminous side. 
However, five of the stars in this zone with \vrot\ greater than 70 \kms\ have upper limits on A(Li) compared to only three stars among the richer MSTO sample brighter than \Vp\ = 14.9. If the argument above is correct, the rapid rotators with upper limits to A(Li) represent the stars which have most recently evolved into the Li-Dip regime. If the timescale for mixing processes is shorter than that for surface rotational spindown, these stars have been caught with reduced A(Li) but, at most, minor reductions in \vrot. Over the next 0.5 Gyr, these stars will become more Li-depleted but their \vrot\ speeds correspondingly will drop, on average, below 25 \kms, placing them within the fainter, slow rotator zone currently seen in the NGC 7789/752 CMD, and equivalent to the pattern seen within the 2.25 Gyr-old cluster, NGC 6819 (DE19). 

\section{CMD Structure and the Red Giant Distribution}
\subsection{Does NGC 7789 Have an eMSTO?}
As discussed in the Introduction, the CMD topology of the Magellanic Cloud star clusters has been a source of debate for almost two decades. For clusters near the age of NGC 7789, the primary feature of interest has been the eMSTO, defined as an uncharacteristically large spread in color among stars at or within the hydrogen-exhaustion phase prior to the subgiant branch. The possible sources of this dispersion include the usual suspects: age spread, metallicity variations, rotation, binarity or some combination thereof. 
The rich cluster, NGC 1846, among the first identified with an eMSTO \citep{MB07}, provides a relevant example of the many avenues typically explored to explain the CMD observations \citep[see, e.g.,][]{GO09, RU13, NI16, KA20, LI21, LI22, OH23}. 
However, the relevance of these systems for Galactic open clusters initially seemed questionable given the lack of star clusters of appropriate age and mass within the Galactic disk. 

That changed with the analysis by \citet{MA18} of four Galactic open clusters, two younger (0.3 and 0.5 Gyr) than the Hyades/Praesepe clusters and two older (0.8 and 1.1 Gyr). Of the four, only M11, the youngest of the four clusters, had precision Str\"omgren photometry and \vrot\ for an extensive sample of stars at the MSTO. The correlation between \vrot\ and color at the MSTO definitively demonstrated that the evolutionary position of a star at the turnoff in a young open cluster was controlled in part by \vrot\ and that, in the case of M11, a large range in \vrot\ created the eMSTO for this cluster, a result later confirmed by \citet{LM19}. For the remaining three clusters, Gaia data provided astrometric membership and precision photometry, leading to a broad range in color among the stars at the turnoff, inconsistent with limits on photometric scatter and variable reddening imposed by the fainter main sequence color range. While the investigation showed that some clusters of comparable age in the Galaxy shared the same CMD morphology as the more massive clusters in the Magellanic Clouds, with rotation serving as a key element in defining the CMD, questions remained regarding the origin of the large \vrot\ spread and the main sequence split among stars more massive than 1.6 M$_{\sun}$  for clusters younger than 0.7 Gyr. 

Since \citet{MA18}, numerous studies have identified potential examples of eMSTOs in open clusters, predominantly by using Gaia astrometry to identify members and Gaia photometry to estimate the impact of binaries, reddening, and photometric scatter \citep[see, e.g.,][]{CO18, PI19, DE24, CH24}. Some of these studies have used spectroscopically derived \vrot\ information to definitively measure any correlation with the rotation distribution \citep[see, e.g.,][]{BA18, SU19, MA24}.

It should be noted that the potentially distorting effect of a spread in \vrot\ on CMDs near the MSTO, especially for age estimation, was highlighted earlier by \citet{BR15a, BR15b, BR15c}, particularly in regard to the age of the Hyades and the existence of eMSTOs. While claims to a greater age for the Hyades and Praesepe based upon isochrones modified to mimic the impact of rotation have been refuted, in part by using models developed to include a rotational spread as an initial condition \citep{GO18}, there has been broad acceptance of the potentially impactful role of rotation in explaining eMSTOs in the Magellanic Clouds and our Galaxy.
What continues to be debated is the degree of impact, 
 i.e. what fraction of the extended sequences can be attributed to rotation, not whether it occurs at all \citep[see, e.g.,][]{GO14, GO17, LI22, CO24}. 

Of particular interest for the current investigation is the CMD synthesis by \citet{LI22} for the LMC massive cluster, NGC 1846. NGC 1846 has an age comparable to that of NGC 7789 but is more massive and more metal-poor.  
With binaries removed from the MSTO of Figure 5 for \Vp\ $\leq 15$, a qualitative comparison of the CMD turnoff distribution for NGC 7789 can be made with Figure 5 of \citet{LI22}, specifically the unaries sample with slow rotation, meaning that the \vrot\ distribution is assumed to be a wide half-Gaussian peaked at 0 \kms. The apparent \vrot\ distribution with luminosity is visible in Figure 6 of \citet{LI22}, with a range near 200 \kms\ at the top of the MSTO and narrowing to less than 100 \kms\ one magnitude below the top of the turnoff. The critical conclusion for NGC 1846 from the synthetic CMD analysis was that the color spread seen in the turnoff of NGC 1846 could be explained by binaries, an age spread, and rotation, but that the inclusion of rotation cut the age spread in half to 70 - 80 Myr relative to the non-rotating models. By contrast, the more poorly populated young open clusters of the Galaxy exhibit formation timescales more than an order of magnitude smaller, with the uncertainties in the final ranges set by the limitations of overly simplistic stellar models and the observational difficulties imposed by the complicated environments of star formation complexes \citep{CA22}.

Despite the detailed differences in the color indices defining the CMDs used in the NGC 1846 analysis and ours in NGC 7789 and the extremely rich MSTO population of NGC 1846, we conclude that the derived \vrot\ distribution of Figure 10 is adequate to explain the modest scatter in \BVp\ above \Vp\ = 15 (see Figure 9). The narrowing \vrot\ range at fainter magnitudes is more than consistent with the minor color spread at fainter magnitudes in NGC 7789, without evidence for (or need of) a spread in age (see Figure 5). By contrast, after correcting for the effects of variable reddening \citep{AT14} and removing probable binaries, the older (2.25 Gyr) cluster, NGC 6819, with a population of stars comparable in richness to NGC 7789, shows a well defined MSTO through the red hook and subgiant branch with minimal scatter compared to what is seen in NGC 7789. As noted earlier, NGC 6819 has a narrow \vrot\ distribution (all stars with \vrot\ less than 25 \kms) from the top of the turnoff through the Li-Dip (DE19), a preview of NGC 7789's appearance in another 0.5 Gyr.

\subsection{The eMSTO Impact on the Giant Branch}
Turning to the giant distribution (Figure 11), the contrast between NGC 7789 and NGC 752 is striking. Since the turnoff/subgiant region feeds the red giants, any distinction between the clusters in the latter category should be a reflection of an underlying dichotomy within the former. The turnoff of NGC 7789 differs from that of NGC 752 in two undoubtedly related ways. First, NGC 7789 has an eMSTO, i.e. a rich population of stars in and above the level of the red hook in the MSTO (\Vp\ $<$  14.7), in comparison with the tight pattern found among the single star NGC 752 CMD (see, e.g., Figure 3 of \citet{TW23} above $V$ = 10.2). Again, the lack of scatter is not a simple result of the smaller sample of stars populating the cluster CMD. The richer data set generated by \citet{LI24} shows a similar limited spread despite the fact that no attempt has been made to eliminate binaries from that CMD. The second striking difference in the two data sets is seen in Figure 12a, reemphasizing that the data set for NGC 752 is not a complete sample. Between \Vp\ = 15.15 and 13.45, NGC 7789 shows an almost uniform distribution of stars from \vrot\ = 20 to 135 \kms. Of the 158 stars in this magnitude range, 63 have \vrot\ at 60 \kms\ or less and 57 are at or above 100 \kms. For NGC 752 the comparable numbers are 10 of 14 below 60 \kms, with no star above 100 \kms. As discussed earlier, of the four stars above 60 \kms, three are assigned 96 \kms\ by APOGEE and are of questionable reliability.

How does this relate to the red giant distribution of NGC 752 relative to NGC 7789? \citet{ME98} first noted the distinctive structure of the giant branch in NGC 752, pointing out the apparent dominant RGC, with an extended population of fainter and bluer giants not predicted by standard stellar evolution models at the time. \citet{ME98} assigned only two red giants with detectable Li to the RGB, grouping all those with upper limits to the RGC. \citet{GI00a} expanded this discussion by adding NGC 7789 to the comparison and concluding that the cluster's giant branch exhibited features similar to those in NGC 752. Their explanation for the extended/double clump was a range in stellar mass populating the giant branch; the higher mass distribution ignited He under non-degenerate conditions and populated a lower luminosity clump while the lower mass end of the distribution experienced He-flash at a higher mass core and burned He within a higher luminosity clump. The suggested sources of this spread were variable mass loss rates on the giant branch or a range in rotation and/or convective core overshooting producing a range of core masses at He-ignition for stars of the same total mass. (For an excellent overview of the structural change of the RGC in the CMD as key parameters are varied, i.e. age, mass, metallicity, etc., the reader is referred to \citet{GI16}.)

Given what we now know about the differences in the \vrot\ distributions at the MSTO between NGC 7789 and NGC 752 and the lack of evidence for an extended star formation history, i.e. an age spread, within NGC 7789, the most plausible source of the differences among the red giants remains a range in mass among the stars leaving the main sequence due to the small differences in evolution caused by the wide range in \vrot. The mass spread produces a color spread in the RGC, with the MSTO rapid rotators having higher mass at the time of He-ignition under non-degenerate conditions while the slower rotators have lower mass which leads to He-flash and an RGC color/luminosity distribution more typical of older clusters and NGC 752. 

To close, while the trend in observation over the last few years has been the discovery of unexpected structure and dispersion within the CMDs of Milky Way open clusters, providing a potential link to the much richer, more massive clusters of comparable age within the Local Group, not all systems 
demonstrate the same patterns. Obviously, NGC 752 would never be cited as a cluster with apparent indications of an eMSTO, despite its similarity in age and metallicity to NGC 7789. As discussed above, our conclusion is that the CMD differences are directly linked to the differences in each cluster's \vrot\ distribution, but also potentially tied to the difference in total cluster mass, evidenced by the active dissolution of NGC 752. 

With this in mind, we highlight the insightful analysis of the enigmatic cluster, NGC 2509 \citep{DJ20}. What makes this cluster special is that, when viewed in the context of six other clusters of comparable age, including NGC 7789, it has an unusually tight main sequence from the turnoff to the unevolved main sequence, making it similar to NGC 752. It is about 0.6 Gyr younger and numerically richer, but still less populous than NGC 7789.  The structure of the CMD is so sharply defined that the binary sequence and its crossing point at the MSTO are easily identifiable. (A more realistic comparison with NGC 752 should be made using the binary-inclusive CMD from \citet{LI23}.) Directly relevant to our discussion, \citet{DJ20} conclude that the tight CMD can only be made compatible with their model isochrones if the range in rotation rates for the turnoff stars is narrow, covering approximately 0.4-0.6 of the critical rotation speed for a zero-age-main-sequence star. We emphasize again that the \vrot\ distributions plotted for NGC 752 and NGC 7789 actually underestimate the true rotation rates. 

Equally important, one can compare the CMD distribution of stars at the RGC. The distribution of stars at the RGC level and fainter bears a strong resemblance to that for NGC 752, with a likely RGB lying fainter and bluer that the red limit of the clump. While red giants do populate the CMD brighter than the clump, (a) the cluster sample in NGC 2509 is numerically richer, (b) binaries in NGC 2509 have not been removed, and (c) NGC 2509 is about 0.6 Gyr younger than NGC 752. As discussed in \citet{AT24}, there is a rapid transition in red giant CMD morphology as clusters age beyond the point of He-flash. In particular, the ratio of giants above and below the clump reverses, with the younger system dominated by the clump and an extended giant branch to the tip, as in NGC 7789, while older clusters have a severely reduced giant branch extension with more red giants populating the giant branch below the clump (see NGC 2506 \citep{AT16, AT18b} for an excellent example of this). From the fact that stars are beginning to populate the subgiant branch and potentially the base of the RGB, it is probable that within 0.6 Gyr the population of red giants above the clump will be severely reduced, enhancing the similarity to NGC 752.

\section{Summary and Conclusions}
High precision $UBVRI$ photometry, combined with comprehensive astrometric and equally precise Gaia photometric data, has been used to isolate highly probable members of the rich open cluster NGC 7789. After the elimination of probable binaries from the sample brighter than $V$ $\sim$ 15 via the radial velocity survey of NI20 and probable photometric binaries among the fainter sample, the unevolved stars are shown to exhibit a correlated photometric scatter in the main sequence as defined by both photometric studies. Having tested and eliminated the possibility that the scatter is tied to the photometric reductions, the most likely explanation is a variation in reddening equivalent to $\sim$ 0.06 mag in \ebv\ across the field of the cluster, consistent with the broad trend found among the more recent reddening maps \citep{SC11, GR19} of the same Galactic coordinates.   

With the individual stellar photometry adjusted for variable reddening to a common value tied to the cluster core, the cluster core \ebv\ was determined via multicolor analysis, consistently producing \ebv\ $\sim$ 0.25, coupled with [Fe/H] near $-0.27$. In contrast with multiple recent cluster analyses, this combination proved incompatible with the VR isochrone fits defined by Gaia DR3 parallaxes where lower reddening requires higher metallicity. While the most probable source of the discrepancy is an issue with the $U$ photometry zero point, both ours and the standards, determination of the true metallicity required a different approach. 

With Gaia astrometry and the true cluster distance modulus in hand, the lower main sequence of NGC 7789 was adjusted in \ebv\ until it matched the position of the theoretical models of fixed [Fe/H]. As the assumed [Fe/H] rises, the required \ebv\ declines. At each paired value of [Fe/H] and \ebv, one can use the MSTO region to estimate an age for the cluster. Because of the anticorrelation between reddening and metallicity, the age spread for the [Fe/H] range of -0.2 to solar remained extremely small, leading to an age of 1.46 $\pm$ 0.02 Gyr, subject to the caveat that all absolute ages are tied to a specific set of isochrones and their underlying parameterization.

The adopted reddening also impacted our derived spectroscopic abundances by altering the \teff\ derived from color-temperature relations. Using a small combined set of both red giants and slowly rotating members at the turnoff, spectroscopic abundances were derived under three assumptions for \ebv, 0.29, 0.31, and 0.33. In contrast with the isochrone fits, higher reddening led to higher [Fe/H]. Combining the two relations implied a common solution near \ebv\ = 0.30 and [Fe/H] $= -0.12$. Adopting \ebv\ = 0.30 and an initial assumption of [Fe/H] $= -0.12$, spectroscopic abundances were derived using spectra from 156 single members, dominated by giants. The contribution of the dwarf stars to the derived [Fe/H] was limited to stars in the RO$\alpha$ class, eliminating stars with rapid rotation.  
The remaining sample generated [Fe/H] $= -0.130 \pm 0.068$ (MAD) when the individual stellar abundances were averaged, and $-0.118 \pm 0.086$ (MAD) when the results from all individual lines were combined. 

With fundamental cluster properties and individual stellar parameters in hand, 
Li abundances and rotational velocity estimates were interactively measured via spectrum synthesis for 316 stars, including the rapid rotators near the turnoff. 
Dividing the sample into turnoff stars and red giants, the red giants readily separated into two distinct categories. The sample with detectable Li clearly marked out the evolutionary sequence for RGB stars, with A(Li) approaching a typical value of 1.2 to 1.3 near the red giant bump above the RGC, then declining sharply to near 0 at the red giant tip. Stars that only have upper limits to A(Li) dominated the RGC and populated a parallel sequence on the ascending giant branch, leading to the conclusion that these stars represent the cluster AGB. The color range among the RGC stars is significant, covering at least 0.3 mag in \BVp. The external Li sample available for comparison among the giants is small, but the \BVp\ trend among the residuals in A(Li) relative to NA23 implies a serious disagreement with their adopted \teff\ scale and virtually all others for the giants published to date. 

Turning to the main sequence, as one samples stars from the top of the turnoff to the limit of the sample near \Vp\ = 16.5, the A(Li) distribution can be divided roughly into distinct phases. At the top of the MSTO, \Vp$\leq 15.2 \pm 0.1$, A(Li) is dominated by Li detections with an upper limit at 3.3 but with obvious scatter to about 3.0, affected in large part by the dominance of rapid rotators. The hot edge of the Li-Dip, referred to as the Li-Wall, appears between \Vp\ = 15.2 and 15.4. Fainter than the Li-Wall, Li upper limits dominate, with the center of the Li-Dip estimated to be near 15.75 $\pm$ 0.1 if the cool side of the Li-Dip is tagged near \Vp\ = 16.0, where Li detections reemerge. Finally, the A(Li) trend rises gradually along the Li-Plateau to A(Li) around 2.7 at \Vp\ $\sim$ 16.5.

The \vrot\ distribution reveals a clear correlation with the Li pattern. This pattern is enhanced by complementary data for NGC 752, a dissolving cluster with virtually the same age and a similar metallicity to NGC 7789. Using a clean sample of only single star members, and placing NGC 752 at the same distance and reddening as NGC 7789, the Li distribution is mapped from the cool side of the Li-Wall to the Li-Plateau and beyond. More important, \vrot\ data for a larger sample of NGC 752 members than just those with A(Li) enhances the \vrot\ structure of the turnoff and its relation to A(Li). Based upon the combined NGC 7789/752 sample, it is concluded that:

a) The Li-Wall defining the start of the Li-Dip is actually the transition to the Kraft Break (KB), with the range in \vrot\ extending well above 100 \kms\ on the hot side. On the cool side, the range in A(Li) drops to a value typically below 60 \kms\, where it remains constant until \Vp\ $\sim$ 15.8, near the location of the Li-Dip center. Beyond this point, \vrot\ begins its well-known precipitous drop to $\sim$ 20 \kms\ at the Li-Plateau, the Kraft curve \citep{KR67}, and continues gradually to less than 5 \kms\ by \Vp\ = 17. 

b) The sharp transition in \vrot\ and A(Li) marked by the Li-Wall strengthens the contention first detailed in DE19 and confirmed in cluster {\citep{AT21, AT24, SU25} and field star \citep{SU26b} analyses since, that the origin of the Li-Dip is inextricably tied to the correlated effects of stellar spindown, i.e. rotationally-induced mixing and angular momentum loss near the surface. The weakening of the Li-Dip on the cool side below the Li-Dip center, despite the expectation of a greater degree of convection, is therefore a byproduct of the reduced initial range in \vrot\  (below 60 \kms) for stars arriving at the main sequence in contrast with those in the hotter half (\vrot\ above 100 \kms). 
  
c) The Li-Wall is defined as a boundary in \teff\ at approximately 6725 $\pm$ 40 K. As expected, the exact value of this transition varies depending upon the \teff\ scale adopted as well as the observational parameter used to define it, but the correlated change in the mass at the boundary with [Fe/H] among clusters from the metallicity of NGC 6253 ([Fe/H] $>$ +0.4) \citep{CU12} to [Fe/H] = -0.4 \citep{AT24} and -0.5 for NGC 2243 \citep{AT21}) signals the dominant dependence on \teff. This has significant implications for the long-term evolution of stars more massive than the Li-Wall. Evolving off the ZAMS, stars initially hotter than the Li-Wall cross into the boundary and initiate the same atmospheric processes that created the Li-Wall itself. Over time, this pattern shifts to stars of higher mass, with the cumulative effect dependent upon the comparative time spent within the Li-Dip zone versus the timescale available  to reach the subgiant branch. In broad terms, the Li-Dip is recreated by stars of higher mass with varying degrees of success due to the differences in their internal structure during the hydrogen-exhaustion phase. Over a timescale of approximately 0.4 - 0.5 Gyr, the broad \vrot\ range above the Li-Wall should collapse from above 100 \kms\ to that more typical of the cool side of the Li-Dip, i.e. 20-25 \kms\ or less. Moreover, the A(Li) distribution should become increasingly dominated by stars with reduced Li or only upper limits prior to entering the subgiant branch. This is exactly the pattern first detailed in DE19 and confirmed by additional cluster analyses since then \citep{AT21, AT24}.

d) The limited sample of stars just above the Li-Wall shows a mixed set of \vrot\ and A(Li) properties. Their velocity distribution clearly places them within the normal range for higher mass stars prior to spindown but their A(Li) structure matches stars well within the Li-Dip. This contradiction may indicate that the timescale for atmospheric mixing is shorter than that for spindown and these stars have only recently initiated the mechanisms that define the Li-Dip. With time, surface spindown will compress the \vrot\ range, after rotationally-induced mixing has continued to drive A(Li) well below its current values.

e) While NGC 7789 and NGC 752 complement each other nicely, the two clusters do exhibit differences which are both real and indicative of the underlying physical processes under discussion. Unlike NGC 752, with a well-defined and narrow single star CMD topology, NGC 7789, after removing the effects of variable reddening, still shows an excessive color range at the MSTO which is not repeated at fainter magnitudes. The cluster is now discussed as a less extreme example of the eMSTO phenomenon regularly found among Magellanic Cloud clusters \citep{DJ20}. Despite the multiple origins proposed for the eMSTO phenomenon in the massive clusters outside the Galaxy, the simplest solution for NGC 7789 is presented by the comparison with NGC 752. The significantly larger range in \vrot, possibly a byproduct of the greater cluster mass of NGC 7789 at the time of formation,  produces the spread in color among the turnoff stars. This phenomenon is nicely illustrated in the synthetic CMDs incorporating neither binaries nor age spread but with an even larger range in \vrot\ among the MSTO stars, used to evaluate the CMD of NGC 1846 \citep{LI22}. This range in \vrot\ translates to a red giant branch with a RGC spread over a wide range in color and luminosity, an indication of a range in mass/structure that combines stars undergoing He-ignition under degenerate and nondegenerate conditions in the same CMD, as originally proposed by \citet{GI00a}. The strikingly different giant branch  appearance of NGC 752 then becomes a byproduct of the narrower \vrot\ distribution found among the MSTO stars. Another example of the phenomenon may be the very similar cluster, NGC 2509, predicted to have a very narrow \vrot\ range among its turnoff stars, required to explain the tight main sequence, binaries aside \citep{DJ20}. Li analysis of the red giants in this cluster might prove informative.

\acknowledgments
NSF support for this project was provided to BJAT and BAT through NSF grant AST-1211621, and to CPD through NSF grants AST-1211699 and AST-1909456. Use was made of the WEBDA data base maintained by E. Paunzen at the University of Vienna, Austria (http://www.univie.ac.at/webda). 
  
This work has made use of data from the European Space Agency (ESA) mission {\it Gaia}, processed by the Gaia Data Processing and Analysis Consortium (DPAC). Funding for the DPAC has been provided by national institutions, in particular the institutions participating in the Gaia Multilateral Agreement.

\facilities{WIYN: 3.5m, WIYN: 0.9m}
\software{IRAF \citet{TODY}, MOOG \citet{SN73}, LACOSMIC \citet{VD01}, ROBOSPECT \citet{WH13}, TOPCAT \citet{TOPC} }

\appendix
\twocolumngrid

\section{Spatial Analysis of the CCD Photometry}
As noted in Section 2, dozens of frames in each $UBVRI$ bandpass were obtained at the WIYN 0.9-m telescope in the 2000s.  The exceptional precision of the instrumental indices, below 0.01 mag in the magnitude range of primary interest \citep{GN04}, was overshadowed by detection of spatially dependent differences in magnitude with amplitudes of 0.02 to 0.05 mag for stars in the outer regions of the CCD frames. 
In an effort to understand and correct these spatially-dependent offsets, the CCD frames were reprocessed and the magnitudes compared to the extensive set of standards generated and updated by ST00, as well as the $VI$ photometry of \citet{GI98b} in NGC 7789. 
Thanks to the excellent photometric precision of all three $VI$ photometric sources for NGC 7789, it became apparent that spatially-dependent trends in photometric residuals were not tied exclusively to the frames from the WIYN 0.9-m telescope.

Comparisons between the $VI$ photometry of ST00 and that of \citet{GI98b} exhibited spatially-dependent residuals in both $V$ and $(V-I)$ \citep{BR13} at the level of a few hundredths of a magnitude in both, with significant breaks in the patterns with (X, Y) CCD coordinate position, indicative of potential mismatches between the CCD frames mosaicked to cover the entire cluster. While \citet{BR13} made use of a smaller catalog of $\sim 500$ $VI$ standards in NGC 7789 from an earlier ST00 compilation, the residual structure of \citet{GI98b} remained when over 600 stars were included from the updated catalog in April 2013. An immediate implication of the analysis was that, assuming the standards of ST00 were truly on the \citet{LA92} system, the original data of \citet{GI98b} were systematically off in $V$ by between $-0.01$ and 0.04 mag, and in $(V-I)$ between 0.00 and $-0.07$ mag, with the offset dependent upon the position in the cluster field. Since the standard photometry in $(V-I)$ is bluer than that of \citet{GI98b}, analysis based upon the \citet{GI98b} colors would imply too high a reddening value for a given cluster age or an age too high and a distance modulus too low for an assumed reddening value. 

As concerning as possible spatial variations in the photometry were, equally troubling was the lack of a consistent trend with position when the calibrated photometry was matched with multiple photometric sources of comparable precision, implying that spatial variations were neither unique nor identical to that defined by the WOCS data. 

For a truly independent, high-precision photometric system that cannot suffer from the same spatial dependences, we turned to the most recent Gaia data release, DR3, and the broad-band data on the $B_{p}$, $G$, and $R_{p}$ system, similar but not identical to $B, V$, and $R$. In the past, we have reliably 
transformed our photometry, ($y, b$) or ($V, B$), to ($G, B_{p}$) 
if a color-dependent offset, usually defined by a polynomial, is included in the transformation. 
Such comparisons allow identification of possible coordinate misidentifications between stars on the two systems,  variables, or bad photometry. More important for the current discussion, the scatter about the mean relation of residuals with color should not exhibit spatial dependences unless caused by variable reddening.
 
After completion of the coordinate match between stars with Gaia data and CCD photometry, the CCD sample was restricted to stars with the highest precision. For each color, stars were only retained if the number of observations in the filter of interest was greater than or equal to 10 and the calculated sem for the magnitude was $\pm$0.010 mag or smaller. For the Gaia data, all stars tagged as potential variables were excluded, as were any stars with positional uncertainty in RA or DEC above 0.03\arcsec. The latter restriction was adopted after tests demonstrated that the majority of stars with excessive scatter among the photometric residuals fell within this category, a clear indication that the lower astrometric quality of the positions arose from some form of image/point-spread-function distortion that also likely impacted the measured magnitudes. 

For each $UBVRI$ filter the difference in magnitude relative to the Gaia filter with the nearest bandpass was derived, i.e. $(U - B_{p})$, $(B - B_{p})$, $(V - G)$, $(R - R_{p})$, and $(I - R_{p})$. The residuals were then plotted as a function of the Gaia index, $(B_{p} - G)$ for $U, B$, and $V$ and $(G - R_{p})$ for $R$ and $I$, and a polynomial fit to the relation. For $B$ and $V$, the Gaia color indices ranged from $(B_{p} - G)$ = 0.32 to 0.55, eliminating the sparse cluster blue stragglers as well as the red giants from the base of the giant branch and beyond. For $(R - G)$ versus $(G - R_{p})$, the comparable range is $(G - R_{p})$ = 0.49 to 0.71. For $U$, the magnitude most discrepant in wavelength relative to the Gaia comparison, the color index range was restricted to $(B_{p} - G)$ = 0.32 to 0.45. Because of this restriction, there were fewer stars defining the residual plots and the scatter was larger but the results proved consistent within the errors across all five filters. For the radial estimate, the final sample of residuals amounted to 897, 1104, 1029, 1005, and 1047 for $U, B, V, R,$ and $I$, respectively.

Once the residuals in magnitude were calculated from the polynomial, they were tested for dependences in both X (RA) and Y (DEC) CCD coordinates. It was immediately apparent, most reliably defined by the $V$ magnitudes, that positionally-dependent and correlated offsets were present in all the filters. The best fit to the patterns was a radial quadratic function with a center at (X, Y) = (900, 1100). 
For all filters, the general pattern with position is very similar, with the stars near the origin typically 0.05 mag fainter than expected relative to those in the outer regions of the frame. The discrepancy declines with radial distance, bottoming out typically between 600 and 800 pixels from the origin. What these trends reveal is that while $V$ magnitudes need to be corrected for the effect prior to calibration on the standard system, the impact on color indices should be minimal, i.e., the transformations between the instrumental and the standard system color indices do not require an {\it a priori} radial adjustment of the instrumental color indices.

Before deriving the calibration between the new photometry and the standard system of ST00 (Appendix B), it was logical to similarly investigate the standard system of ST00 for spatial dependences, especially given the residual structure discernible in the earliest installments of $VI$ data tied to the mosaicked frames of \citet{GI98b}. To check, the $U, B, V,$ and $I$ magnitudes of ST00 were run through the same procedure outlined above relative to the comparable Gaia system. There are only a few dozen $R$ standards, insufficient to map the system residuals. Since our $(V-R)$ data should lack spatial trends, we will only use the ST00 data below to tie down the zero-point of the transformation derived in conjunction with a much larger data set from NGC 188.

One of the advantages of the ST00 data is the expanded area coverage compared to our CCD frames in NGC 7789. On the 0-2000 (X, Y) pixel scale of our frames, the ST00 field extends  from $-680$ to 2800 in X and $-680$ to 2800 in Y (0 to 2800 in X for $B$). From 509 $V$ magnitudes, no spatially-dependent residual is found in X (RA). For Y (DEC), from the edge of the field to the center ($\sim 1750$ pixels), a difference of 0.022 mag exists, with the central stars being too faint relative to the edge, a pattern well fit by a quadratic. Fortunately, over the Y range of our CCD frames, this gradient reduces to only 0.010 mag from center to edge. 
For $B$, the pattern is virtually identical to that of $V$, with no trend in X (RA), but a gradient between the center and edge in Y (DEC) amounting to 0.022 mag, or 0.011 mag over the range of our CCD frames.

For $I$, using 509 stars, for the first time a trend was found in the X (RA) direction. From X $= -650$ to 550 the offset residual grows linearly from $-0.032$ mag to +0.005. Beyond this point to the edge of the field in X, the offset remains constant at +0.005 mag. With the residuals from the X pattern removed from the data, the residual trend in Y reduces to a range and form similar to that for the $V$ and $B$ frames.

For $U$, tied to a smaller set of observations only recently added to the data base, no discernible pattern is seen in the residuals with X or Y for 383 stars, though the scatter in the residuals is, as expected, significantly larger than for $B, V,$ and $I$.

Given the above, for final calibration purposes, the small Y positional corrections were only applied to the ST00 magnitudes prior to instrumental curve fitting for the $V$ calibration, the $B$ magnitude for the $(U-B)$ calibration, and the X term for $I$ prior to the $(V-I)$ calibration.

\section{Index-Specific Photometric Calibration Details}
For $V_{instr}$, all stars with 10 or more observations and sem below 0.010 were retained after the match with the standards of ST00. Corrections were applied for the spatial trends detailed in Appendix A and all stars tagged as possible variables by Gaia were eliminated. One additional star was removed after initial fits produced an unusually large residual in the comparisons. The final 503 standards, ranging from $V$ = 11.53 to 18.99 and $(B-V)$ = 0.22 to 1.81, produced a marginal color term in $(B-V)_{instr}$, optimally fit by a quadratic such that, with $a$ = 1.0, $b = -0.0203$, $c = 0.0256$, and 
$d = -2.973$, as defined by the terms of Equation (1). More important, the residuals about the final mean relation exhibited a dispersion of $\pm$ 0.0073 mag. 

For $(B-V)$, the $V$ calibration pattern was repeated, with the limit on the errors in $(B-V)_{instr}$ set to include only stars with sem below 0.010 mag and only stars with 10 or more observations in both filters. With the removal of possible Gaia variables and one star with larger than expected residuals, the final sample contained 479 stars.
The decline in numbers is defined by the sample of ST00 stars which have $V$ but no $B$ observations, not by the identical sem limit on $(B-V)_{instr}$.
Again, because of the large number of precision standards over a wide range in color, the best fit for the $(B-V)$ calibration proved to a cubic polynomial of the form
\begin{eqnarray}
(B-V) =  0.06021\cdot (B-V)_{instr} ^3 - 0.19019\cdot (B-V)_{instr}^2 \nonumber\\
+ 1.22274\cdot(B-V)_{instr} - 0.1701 \nonumber \\
\end{eqnarray}
\noindent
The dispersion of the residuals about the final mean relation is $\pm$0.0065 mag.

For $(V-I)$, overlap with ST00 amounted to 513 stars, but preliminary analysis led to the elimination of one star with anomalous residuals. For reasons discussed in Appendix A, no positional adjustment was applied to our instrumental color indices or to those of ST00 in the Y coordinate; the unique X dependence was removed from the ST00 data prior to the calibration. From 512 stars, the final calibration was:

\begin{eqnarray}
(V-I) =   0.9514*(V-I)_{instr} + 0.5389 \nonumber \\ 
\end{eqnarray}
\noindent
The residual scatter around this linear relation is $\pm$0.0072 mag.

The calibration of the $(U-B)$ index in CCD photometry is invariably the most challenging of the traditional broad-band indices given the range in ultraviolet sensitivity among differing CCD chips, both in an absolute sense and across the bandpass of the traditional $U$ filter. Until recently there were no $U$ standards on the ST00 system within NGC 7789; that has now changed, though the precision is lower than for $BVI$ due to the reduced number of observations. The initial match with ST00 produced 562 stars with $U$ data. Constraining the comparison to our CCD stars with at least 10 observations in both $U$ and $B$ and sem$_{(U-B)} < 0.010$ mag lowered the count to 501. Of these, 21 had no $B$ observations on the ST00 system and 19 were tagged as possible variables in Gaia. Finally, a $V$ = 17 cut was made to the standards in light of the increasing growth in the errors for stars below this magnitude level as seen in Figure 1, leaving a final sample of 449 standards.

An initial plot of the differences in $(U-B)$ between the standard and instrumental values as a function of the instrumental indices, both $(U-B)_{instr}$ and $(B-V)_{instr}$,  demonstrated a distinct change in the correlation between the $(U-B)$ indices for redder versus bluer stars, with the transition occurring near $(B-V)$ = 0.75. The calibration was split into two samples with 240 stars having $(B-V) \leq 0.75$ and 209 above 0.75. For the cooler sample, a quadratic supplied the optimal fit.

\begin{eqnarray}
(U-B)_{stand} = 0.01936*(U-B)_{instr}^2 + \nonumber \\
 0.97677*(U-B)_{instr} - 2.3966 \nonumber \\
\end{eqnarray}
\noindent
Eliminating one star with anomalous residuals, the dispersion about the mean relation amounted to $\pm$0.0169 mag.

For the blue side of the calibration, the optimal fit included a sizable component defined by $(B-V)$:
\begin{eqnarray}
(U-B)_{stand} = 1.1678*(U-B)_{instr} \nonumber \\
+ 2.2175*(B-V)_{instr}^3  - 4.2561*(B-V)_{instr}^2 \nonumber \\
+ 2.3574*(B-V)_{instr} - 3.0614  \nonumber \\
\end{eqnarray}
\noindent
Eliminating two stars with anomalous residuals, the remaining 238 scatter about the mean relation with a dispersion of $\pm$0.0149 mag. For deriving the final photometry with a smooth transition between the blue and red zones, $(U-B)$ was calculated using both relations and averaged for stars between $(B-V)$ = 0.72 and 0.78.

The $(V-R)$ calibration presents its own challenges due to the paucity of $R$ standards in ST00. Only 28 ST00 $(V-R)$ standards within the cluster overlap with our data set, covering the $(V-R)$ range from 0.32 to 0.93. Over this color range, the residuals, $(V-R)$ - $(V-R)_{instr}$, show no trend with color and a constant offset of -0.240 $\pm$ 0.019 (sd) mag. Fortunately, when the NGC 7789 frames were taken, a number of clusters within the ST00 sample were observed as a means of boosting the total number of standards and better defining the transformation relations. In $R$, other than NGC 7789, only NGC 188 was observed, with typically three to four frames in each filter. The initial overlap between the two NGC 188 samples identified 202 common stars. Eliminating stars with fewer than three observations in either filter and/or sem in $(V-R)_{instr}$ greater than 0.015 mag reduced the sample to 177 stars. A cubic fit to the data generates a curve with residuals that scatter about the mean relation at the $\pm 0.010$ (sd) mag level. However, a closer examination of the data reveals that curvature of the relation is dominated by 5 stars beyond $(V-R)$ = 0.92. Exclusion of these stars produces a statistically flat curve with a mean offset of $-0.245 \pm 0.010$ mag. The agreement with the NGC 7789 result is encouraging and the simple offset of $-0.240$ mag was applied to all of our NGC 7789 $(V-R)_{instr}$ data. While it is possible that the gradual change in the offset with increasing color found among the reddest stars in NGC 188 also applies to NGC 7789, without a means of explicitly testing this proposition for NGC 7789, where the reddening in $E(B-V)$ is approximately 0.2 mag larger than in NGC 188, we defer to the fixed offset. Thus, caution is urged in use of the photometry beyond $(V-R)$ = 0.93; it is possible that the indices are too large by an amount that increases with increasing color, reaching 0.10 mag near $(V-R)$ = 1.07.

\section{Variable Reddening}
Given the cluster angular size, distance, and position within 10\arcdeg\ of the Galactic plane, a reddening range across the cluster face would appear to be a plausible assumption, though the observed tightness of the main sequence distribution has consistently been used to infer, at worst, a modest range in \ebv\ (see, e.g., the discussion in \citet{GI98b}). The reddening issue was placed on more solid ground with the release of the first all-sky reddening maps \citep{SC98}. Over the coordinate range covered by our CCD frames, the cluster field exhibited a clear north-south gradient, declining by between 0.04 and 0.07 mag in \ebv, with a weaker dependence on RA. While the mean cluster reddening is lower by 0.06 mag in \ebv, a similar differential pattern emerged from the revised map by \citet{SC11}. The interpretation of these variations presents a challenge since they measure the full reddening along the line-of-sight, rather than \ebv\ at the distance of the cluster. The more recent work by \citet{GR19}, incorporating a mixture of  survey information leading to three-dimensional reddening maps, qualitatively corroborates the gradient in \ebv\ with increasing reddening to the north. The range varies depending upon whether one chooses the full line-of-sight ($\Delta$\ebv\ = 0.03 to 0.07) \ebv\ or just the value to the distance of the cluster ($\Delta$\ebv\ = 0.06 to 0.08). However, the absolute \ebv\ at the cluster distance is significantly smaller than the derived line-of-sight values in \citet{SC11} by $\Delta$\ebv = 0.15 mag. Even the full reddening along the line-of-sight is reduced by $\Delta$\ebv\ = 0.07 mag.

To place the reddening variability on a positional scale appropriate for the CCD data of Table 1, use was made of the precision photometry of Table 1 and the correlated Gaia photometric data discussed in Appendix A. Using only cluster members below $V$ = 15.2 that do not lie far enough from the main sequence to be highly probable binaries, as assumed for the red points in Figure 5, a quadratic is fit to $(B-V)$ as a function of $V$ and $(B_{P}-R_{P})$ as a function of $G$. In both cases, the range at a given $V$ or $G$ in the appropriate color index is typically 0.05 to 0.07 mag. The residuals for each star fainter than $V$ = 15.2, whether classed as a probable binary or not, relative to the mean relation for $V$ and $G$ were derived and plotted relative to each other in the sense $\delta$$(B-V)$ as a function of $\delta(B_{P}-R_{P})$. Eliminating 6 stars with $\delta(B_{P}-R_{P}) < -0.030$ mag, i.e. they appear too blue relative to the mean relation, and one extreme star with $\delta(B_{P}-R_{P}) > 0.10$, the remaining 412 stars are well fit by a quadratic function between $\delta(B_{P}-R_{P}) = -0.03$ and $+0.08$. The asymmetry arises because of the inclusion of the probable binaries in the correlation, stars which populate the $\delta(B_{P}-R_{P}) > 0.030$ portion of the distribution. It should be emphasized that even if the stars with $\delta(B_{P}-R_{P}) > 0.035$ are eliminated, the correlation between the photometric offsets still remains, confirming the claims of high precision for both data sets and the lack of a positional dependence in the $(B-V)$ zero-point as discussed earlier.

Given the relation between $\delta(B_{P}-R_{P})$ and $\delta(B-V)$, the former values for each star were transformed to the latter system and averaged, generating an offset $\delta(B-V)$ for each star on the unevolved main sequence. The offsets for the single stars of Figure 5 were plotted as a function of position on the sky and the trends across the field were smoothed to generate a two-dimensional differential map of the reddening relative to the cluster mean value as defined by stars with $\delta(B-V)$ and $\delta(B_{P}-R_{P})$ = 0.0. A color plot of the reddening variation is supplied in Figure 13. Encouragingly, the trends seen in Figure 13 are qualitatively an excellent match to the broad trends detailed in the reddening maps published to date. The dominant pattern remains the reddening gradient in the North/South direction, with stars near the northern edge of the cluster typically reddened by 0.04 to 0.06 mag more than stars at the southern boundary.

\begin{figure}
\figurenum{13}
\includegraphics[angle=0,width=\linewidth]{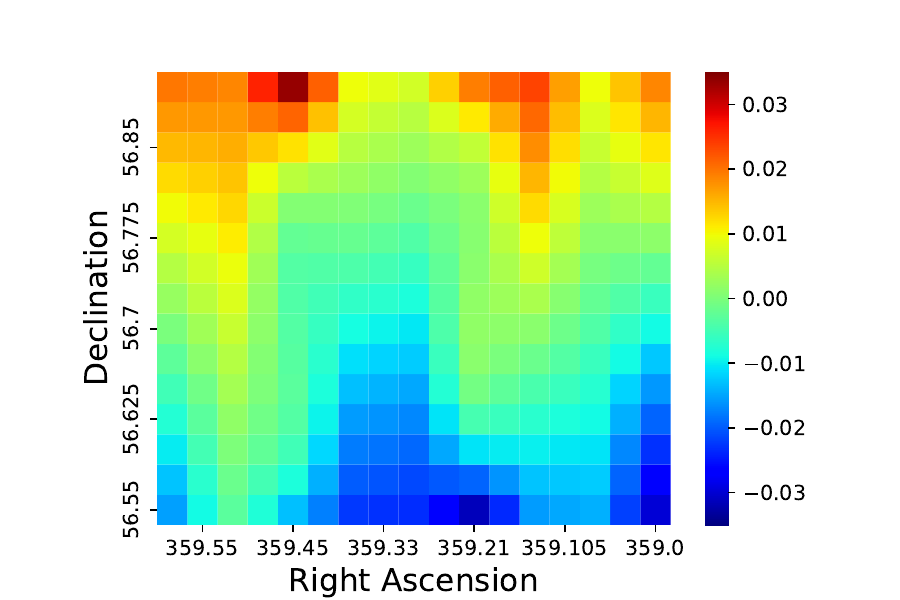}
\caption{Reddening variation across the CCD field of NGC 7789. Differential \ebv\ is defined relative to the cluster mean and a positive residual implies higher than average reddening.}
\end{figure}

\begin{thebibliography}{}
\bibitem[Abdurro'uf et al. (2022)]{DR17} Abdurro'uf, Accetta, K., Aerts C., et al. 2022, \apjs, 259, 35
\bibitem[Anthony-Twarog et al. (2018a)]{AT18a} Anthony-Twarog, B.~J., Deliyannis, C.~P., Harmer, D., et al. 2018a, \aj, 156, 37
\bibitem[Anthony-Twarog et al. (2024)]{AT24} Anthony-Twarog, B.~J., Deliyannis, C.~P., Sun, Q., \& Twarog, B.~A. 2024, \aj, 168, 103 
\bibitem[Anthony-Twarog et al. (2014)]{AT14} Anthony-Twarog, B.~J., Deliyannis, C.~P., \& Twarog, B.~A. 2014, \aj, 148, 51
\bibitem[Anthony-Twarog et al. (2016)]{AT16} Anthony-Twarog, B.~J., Deliyannis, C.~P., \& Twarog, B.~A. 2016, \aj, 152, 192
\bibitem[Anthony-Twarog et al. (2021)]{AT21} Anthony-Twarog, B.~J., Deliyannis, C.~P., \& Twarog, B.~A. 2021, \aj, 161, 159
\bibitem[Anthony-Twarog et al. (2009)]{AT09} Anthony-Twarog, B.~J., Deliyannis, C.~P., Twarog, B.~A., Croxall, K.~V., \& Cummings, J.~D. 2009, \aj, 138, 1171
\bibitem[Anthony-Twarog et al. (2018b)]{AT18b} Anthony-Twarog, B.~J., Lee-Brown, D., Deliyannis, C.~P., \& Twarog, B.~A. 2018b, \aj, 155, 138
\bibitem[Anthony-Twarog \& Twarog (2004)]{AT04} Anthony-Twarog, B.~J., \& Twarog, B.~A. 2004, \aj, 127, 1000
\bibitem[Arp \& Cuffey (1962)]{AC62} Arp, H.~A., \& Cuffey, J. 1962, \apj, 136, 51
\bibitem[Balachandran (1995)]{BA95} Balachandran, S. 1995, \apj, 446, 203
\bibitem[Bastian et al.(2018)]{BA18} Bastian, N., Kamann, S., Cabrera-Ziri, I., et al. 2018, \mnras, 480, 3739
\bibitem[Bertelli et al. (1994)]{BE94} Bertelli, G., Bressan, A., Chiosi, C., Fagotto, F., \& Nasi, E. 1994, \aaps, 106, 275
\bibitem[Beyer \& White (2024)] {BE24} Beyer, A.~C., \& White, R.~J. 2024, \apj, 973, 28E.
\bibitem[Bhattacharya et al. (2021)]{BH21} Bhattacharya, S., Agarwal, M., Rao, K.~K., \& Vaidya, K. 2021, \mnras, 505, 1607
\bibitem[B\"ocek Topcu et al. (2015)]{BO15} B\"ocek Topcu, G., Af\c{s}ar, M., Schaeuble, M., \& Sneden, C. 2015, \mnras, 446, 3562
\bibitem[B\"ocek Topcu et al. (2020)]{BO20} B\"ocek Topcu, G., Af\c{s}ar, M., Sneden, C., et al. 2020, \mnras, 491, 544
\bibitem[Boesgaard (1987)]{BO87} Boesgaard, A.~M. 1987, \pasp, 99, 1067
\bibitem[Boesgaard et al. (2004)]{BO04} Boesgaard, A.~M., Armengaud, E., King, J.~R., Deliyannis, C.~P., \& Stephens, A. 2004, \apj, 613, 1202 
\bibitem[Boesgaard et al. (2005)]{BO05} Boesgaard, A.~M., Deliyannis, C.~P., \& Steinhauer, A. 2005, \apj, 621, 991
\bibitem[Boesgaard et al. (2022)]{BO22} Boesgaard, A.~M., Lum, M.~G., Chontos, A., \& Deliyannis, C.~P. 2022, \apj, 927, 118
\bibitem[Boesgaard et al. (2016)]{BO16} Boesgaard, A.~M., Lum, M.~G., Deliyannis, C.~P., et al. 2016, \apj, 830, 49
\bibitem[Boesgaard \& Tripicco (1986)]{BT86} Boesgaard, A.~M., \& Tripicco, M.~J. 1986, \apjl, 302, 49
\bibitem[Boffin et al. (2022)]{BF22} Boffin, H.~M.~J., Jerabkova, T., Beccari, G., \& Wang, L. 2022, \mnras, 514, 3579
\bibitem[Bouvier et al. (2018)]{BO18} Bouvier, J., Barrado, D., Moraux, E., et al. 2018, \aap, 613, A63
\bibitem[Brandt \& Huang (2015a)]{BR15a} Brandt, T.~D., \& Huang, C.~X. 2015a, \apj, 807, 24
\bibitem[Brandt \& Huang (2015b)]{BR15b} Brandt, T.~D., \& Huang, C.~X. 2015b, \apj, 807, 25
\bibitem[Brandt \& Huang (2015c)]{BR15c} Brandt, T.~D., \& Huang, C.~X. 2015c, \apj, 807, 58
\bibitem[Brunker et al. (2013)]{BR13} Brunker, S.~W., Anthony-Twarog, B.~J., Deliyannis, C., \& Twarog, B.~A. 2013, BAAS 221, 250.28
\bibitem[Bruntt et al. (2012)]{BR12} Bruntt, H., Basu, S., Smalley, B., et al. 2012, \mnras, 423, 122
\bibitem[Cantat-Gaudin et al. (2018)]{CA18} Cantat-Gaudin, T., Jordi, C., Vallenari, A., et al. 2018, \aap, 618, 93
\bibitem[Cao et al. (2022)]{CA22} Cao, L., Pinsonneault, M.~H., Hillenbrand, L.~A., \& Kuhn, M.~A. 2022, \apj, 924, 84
\bibitem[Castro et al. (2016)]{CA16} Castro, M., Duarte, T., Pace, G., \& do Nascimento Jr., J.-D. 2016, \aap, 590, A94
\bibitem[Cayrel de Strobel (1988)]{CA88} Cayrel de Strobel, G. 1988, in {\it IAU Symposium 132, The Impact of Very High S/N Spectroscopy on Stellar Physics}, ed. G. Cayrel de Strobel \& M. Spite (Dordrecht: Kluwer), 345
\bibitem[Chen et al. (2024)]{CH24} Chen, J., Li, Z., Zhang, S., Zhao, W., \& Wu, Y. 2024, \aj, 167, 44
\bibitem[Cordoni et al. (2024)]{CO24} Cordoni, G., Casagrande, L., Yu, J., et al. 2024, \mnras, 532, 1547
\bibitem[Cordoni et al. (2018)]{CO18} Cordoni, G., Milone, A.~P., Marino, A.~F., et al. 2018, \apj, 869, 139
\bibitem[Cummings et al. (2012)]{CU12} Cummings, J.~D., Deliyannis, C.~P., Anthony-Twarog, B.~J., Twarog, B.~A., \& Maderak, R.~M. 2012, \aj, 144, 137
\bibitem[Cummings et al. (2017)]{CU17} Cummings, J.~D., Deliyannis, C.~P., Maderak, R.~M., \& Steinhauer, A. 2017, \aj, 153, 128
\bibitem[Daniel et al. (1994)]{DA94} Daniel, S.~A., Latham, D.~W., Mathieu, R.~D., \& Twarog, B.~A. 1994, \pasp, 106, 281
\bibitem[de Juan Ovelar et al. (2020)]{DJ20} de Juan Ovelar, M., Gossage, S., Kamann, S., et al. 2019, \mnras, 491, 2129 
\bibitem[Deliyannis et al. (2019)]{DE19} Deliyannis, C.~P., Anthony-Twarog, B.~J., Lee-Brown, D.~B., \& Twarog, B.~A. 2019, \aj, 158, 163 (DE19)
\bibitem[Deliyannis et al. (1998)]{DE98} Deliyannis, C.~P., Boesgaard, A.~M., Stephens, A., et al. 1998, \apjl, 498, 147
\bibitem[Deliyannis \& Pinsonneault (1997)]{DE97} Deliyannis, C.~P., \& Pinsonneault, M.~H. 1997, \apj, 488, 836
\bibitem[Deliyannis et al. (1993)]{DP93} Deliyannis, C.~P., Pinsonneault, M.~H., \& Duncan, D.~K. 1993, \apj, 414, 740
\bibitem[Deng \& Li (2024)]{DE24} Deng, Y.-Y., \& Li, Z.-M. 2024, {\it Res. in Astron. \& Astrophys.}, 24, 065004
\bibitem[Friel et al. (2002)]{FR02} Friel, E.~D., Janes, K.~A., Tavarez, M., et al. 2002, \aj, 124, 2693
\bibitem[Gaia Collaboration et al. (2016)]{GA16} Gaia Collaboration, Brown, A.~G.~A., Vallenari, A., et al. 2016, \aap, 595, A2
\bibitem[Gaia Collaboration et al. (2018)]{GA18} Gaia Collaboration, Brown. A.~G.~A., Vallenari, A., et al. 2018, \aap, 616, A1 (DR2)
\bibitem[Gaia Collaboration et al. (2021)]{GA21} Gaia Collaboration, Brown. A.~G.~A., Vallenari, A., et al. 2021, \aap, 649, A1 (EDR3)
\bibitem[Gaia Collaboration et al. (2022)]{GA22} Gaia Collaboration, Brown. A.~G.~A., Vallenari, A., et al. 2022, arXiv:2208.00211 (DR3) 
\bibitem[Garc\'{i}a P\'{e}rez et al. (2016)]{GP16} Garc\'{i}a P\'{e}rez, A.~E., Allende-Prieto,C., Holtzman, J.~A., et al. 2016, \aj, 151, 144
\bibitem[Geller et al. (2015)]{GE15} Geller, A.~M., Latham, D.~W., \& Mathieu, R.~D. 2015, \aj, 150, 97
\bibitem[Geller et al. (2021)]{GE21} Geller, A.~M., Mathieu, R.~D., Latham, D.~W., et al. 2021, \aj, 161, 190
\bibitem[Gilmore et al. (2012)]{GI12} Gilmore, G., Randich, S., Asplund, M., et al. 2012, Msngr, 147, 25 
\bibitem[Gilroy (1989)]{GI89} Gilroy, K.~K. 1989, \apj, 347, 835
\bibitem[Gim et al. (1998a)]{GI98a} Gim, M., Hesser, J.~E., McClure, R.~D., \& Stetson, P.~B. 1998, \pasp, 110, 1172 
\bibitem[Gim et al. (1998b)]{GI98b} Gim, M., VandenBerg, D.~A., Stetson, P.~B., Hesser, J.~E., \& Zurek, D.~R. 1998, \pasp, 110, 1318
\bibitem[Girardi (1999)]{GI99} Girardi, L. 1999, \mnras, 308, 818
\bibitem[Girardi (2016)]{GI16} Girardi, L. 2016, \araa, 54, 95
\bibitem[Girardi et al. (2002)]{GI02} Girardi, L., Bertelli, G., Bressan, A., et al. 2002, \aap, 291, 195
\bibitem[Girardi et al. (2000b)]{GI00b} Girardi, L., Bressan, A., Bertelli, G., \& Chiosi, C. 2000b, \aap, 141, 371
\bibitem[Girardi et al. (2013)]{GI13} Girardi, L., Goudfrooij, P., Kalirai, J.~S., et al. 2013, \mnras, 431, 3501
\bibitem[Girardi et al. (1998)]{GR98} Girardi, L., Groenewegen, M.~A.~T., Weiss, A., \& Salaris, M. 1998, \mnras, 301, 149
\bibitem[Girardi et al. (2000a)]{GI00a} Girardi, L., Mermilliod, J.~-C., \& Carraro, G. 2000a, \aap, 354, 892
\bibitem[Gneiser et al. (2004)]{GN04} Gneiser, H., Windschitl, J.~L., Deliyannis, C.~P., Sarajedini, A., \& Platais, I. 2004, \baas, 36, 1380
\bibitem[Gossage et al. (2018)]{GO18} Gossage, S., Conroy, C., Dotter, A., et al. 2018, \apj, 863, 67
\bibitem[Goudfrooij, Girardi, \& Correnti (2017)]{GO17} Goudfrooij, P., Girardi, L., \& Correnti, M. 2017, \apj, 846, 22
\bibitem[Goudfrooij et al. (2014)]{GO14} Goudfrooij, P., Girardi, L., Kozhurina-Platais, V., et al. 2014, \apj, 797, 35
\bibitem[Goudfrooij et al. (2009)]{GO09} Goudfrooij, P., Puzia, T.~H., Kozhurina-Platais, V., \& Chandar, R. 2009, \aj, 137, 4988
\bibitem[Green et al. (2019)]{GR19} Green, G.~M., Schlafly, E.~F., Zucker, C., Speagle, J.~S., \& Finkbeiner, D.~P. 2019, \apj, 887, 93
\bibitem[Heinemann (1926)]{HE26} Heinemann, K. 1926, AN, 227, 193
\bibitem[Hobbs \& Pilachowski (1986)]{HP86} Hobbs, L.~M. \& Pilachowski, C. 1986, \apj, 309, L17
\bibitem[Huang et al. (2015)]{HU15} Huang, Y., Liu, X.~W., Yuan, H.~B., Xiang, M.~S., \& Chen, B.~Q. 2015, \mnras, 454, 2863
\bibitem[Jacobson et al. (2011)]{JA11} Jacobson, H.~R., Pilachowski, C.~A., \& Friel, E.~D. 2011, \aj, 142, 59
\bibitem[Jadhav, Subramaniam, \& Sagar (2024)]{JA24} Jadhav, V.,  Subramaniam, A., \& Sagar, R. 2024, \aap, 688, A152
\bibitem[Jeffries et al. (2021)]{JE21} Jeffries, R.~D., Jackson, R.~J., Sun, Q., \& Deliyannis, C.~P. 2021, \mnras, 500, 1158
\bibitem[Jeffries et al. (2002)]{JE02} Jeffries, R.~D., Totten, E.~J., Harmer, S., \& Deliyannis, C.~P. 2002, \mnras, 336, 1109
\bibitem[Kamann et al. (2020)]{KA20} Kamann, S., Bastian, N., Gossage, F., et al. 2020, \mnras, 492, 2177
\bibitem[Kinman (1965)]{KI65} Kinman, T.~D. 1965, \apj, 142, 655
\bibitem[Kraft (1967)]{KR67} Kraft, R.~P. 1967, \apj, 150, 551
\bibitem[Kurucz (1995)]{KU95} Kurucz, R.~L. 1995, in IAU Symp. 149, The Stellar Populations of Galaxies, ed. B. Barbuy \& A. Renzini (Dordrecht: Kluwer), 225
\bibitem[Landolt (1992)]{LA92} Landolt, A.~U. 1992, \aj, 104, 340
\bibitem[Lee-Brown et al. (2015)]{LB15} Lee-Brown, D.~B., Anthony-Twarog, B.~J., Deliyannis, C.~P., Rich, E., \& Twarog, B.~A. 2015, \aj, 149, 121
\bibitem[Li (2021)]{LI21} Li, C. 2021, \apj, 921, 171
\bibitem[Li et al. (2023)]{LI23} Li, C., Zhong, J., Qin, S., \& Chen, L. 2023, \aap, 672, A81
\bibitem[Li et al. (2024)]{LI24} Li, C., Zhong, J., Qin, S., et al. 2024, \aap, 686, A215 
\bibitem[Lim et al. (2019)]{LM19} Lim, B., Rauw, G., Ya\"el, N., et al. 2019, {\it Nature Astron.}, 3, 76
\bibitem[Lipatov et al. (2022)]{LI22} Lipatov, M., Brandt, T.~D., \& Gossage, S. 2022, \apj, 934, 105
\bibitem[Llorente et al. (2021)]{LL21} Llorente de Andrés, F., Chavero, C., de la Reza, R., Roca-F\'{a}brega, S., \& Cifuentes, C. 2021, \aap, 654, 137 
\bibitem[Mackey \& Broby Nielsen (2007)]{MB07} Mackey, A.~D., \& Broby Nielsen, P. 2007, \mnras, 379, 151
\bibitem[Maderak et al. (2013)]{MA13} Maderak, R.~M., Deliyannis, C.~P., King, J.~R., \& Cummings, J.~D. 2013, \aj, 146, 143
\bibitem[Margheim (2007)]{MA07} Margheim, S.~J. 2007, Ph.D. Thesis, Indiana Univ. Bloomington
\bibitem[Marino et al. (2018)]{MA18} Marino, A.~F., Milone, A.~P., Casagrande, L., et al. 2018, \apjl, 863, L35
\bibitem[Mathieu (2000)]{MA00} Mathieu, R.~D. 2000, in ASP Conf. Series 198, {\it Stellar Clusters and Associations: Convection, Rotation, and Dynamos}, ed. R. Pallavicini, G. Micela, \& S. Sciortino (San Francisco, CA: ASP), 517
\bibitem[Maurya et al. (2024)]{MA24} Maurya, J., Samal, M.~R., Amard, L., et al. 2024, \mnras, 532, 1212
\bibitem[McNamara \& Solomon (1981)]{MS81} McNamara, B.~J., \& Solomon, S.~J. 1981, \aaps, 43, 337
\bibitem[Mermilliod et al. (1998)]{ME98} Mermilliod, J.-C., Mathieu, R.~D., Latham, D.~W., \& Mayor, M. 1998, \aap, 339, 423
\bibitem[Mermilliod, Mayor, \& Udry (2008)]{ME08} Mermilliod, J.~C., Mayor, M., \& Udry, S. 2008, \aap, 485, 303 
\bibitem[Mermilliod, Mayor, \& Udry (2009)]{ME09} Mermilliod, J.~C., Mayor, M., \& Udry, S. 2009, \aap, 498, 949 
\bibitem[Milone et al. (2009)]{MI09} Milone, A.~P., Bedin, L.~R., Piotto, G., \& Anderson, J. 2009, \aap, 497, 755
\bibitem[Nagarajan et al. (2023)]{NA23} Nagarajan, N., Sneden, C., Asfar, M, \& Pilachowski, C. 2023, \aj, 165, 245 (NA23)
\bibitem[Netopil et al. (2016)]{NE16} Netopil, M., Paunzen, E., Heiter, U., \& Soubiran, C. 2016, \aap, 585, 150
\bibitem[Niederhofer et al. (2016)]{NI16} Niederhofer, F., Bastian, N., Kozhurina-Platais, V., et al. 2016, \aap, 586, A148
\bibitem[Nine et al. (2020)]{NI20} Nine, A.~C., Milliman, K.~E., Mathieu, R.~D., et al. 2020, \aj, 160, 169 (NI20)
\bibitem[Nordstr{\"o}m et al. (1997)]{NO96} Nordstr{\"o}m, B., Andersen, J., \& Andersen, M.~I. 1996, \aaps, 118, 407
\bibitem[Oh et al. (2023)]{OH23} Oh, W.~S., Nordlander, T., Da Costa, G.~S., \& Mackey, A.~D. 2023, \mnras, 519, 831
\bibitem[Overbeek et al. (2015)]{OV15} Overbeek, J.~C., Friel, E.~D., Jacobson, H.~R., et al. 2015, \aj, 149, 15
\bibitem[Pancino et al. (2010)]{PC10} Pancino, E., Carrera, R., Rossetti, E., \& Gallart, C. 2010, \aap, 511, A56 
\bibitem[Piatti \& Bonatto (2019)]{PI19} Piatti, A., \& Bonatto, C. 2019, \mnras, 490, 2414
\bibitem[Pilachowski (1986)]{PI86} Pilachowski, C. 1986, \apj, 300, 289
\bibitem[Pilachowski, Saha, \& Hobbs (1988)]{PI88} Pilachowski, C.~A., Saha, A., \& Hobbs, L.~M. 1988, \pasp, 100, 474
\bibitem[Pinsonneault et al. (1990)]{PI90} Pinsonneault, M.~H., Kawaler, S.~D., \& Demarque, P. 1990, \apjs, 74, 501
\bibitem[Pinsonneault et al. (1989)]{PI89} Pinsonneault, M.~H., Kawaler, S.~D., Sofia, S., \& Demarque, P. 1989  \apj, 338, 424
\bibitem[Platais (1991)]{PL91} Platais, I. 1991, \aaps, 87, 69
\bibitem[Randich et al. (2007)]{RA07} Randich, S., Primas, F., Pasquini, L., Sestito, P., \& Pallavicini, R. 2007, \aap, 469, 163 
\bibitem[Rubele et al. (2013)]{RU13} Rubele, S., Girardi, L., Kozhurina-Platais, V., et al. 2013, \mnras, 430, 2774
\bibitem[Sandquist et al. (2023)]{SA23} Sandquist, E.~L., Buckner, A.~J. Shetrone, M.~D., et al. 2023, \aj, 165, 6
\bibitem[Schlafly \& Finkbeiner (2011)]{SC11}  Schlafly, E.~F., \&  Finkbeiner, D.~P. 2011, \apj, 737, 103
\bibitem[Schlegel, Finkbeiner, \& Davis (1998)]{SC98} Schlegel, D.~J., Finkbeiner, D.~P., \& Davis, M. 1998, \apj, 500, 525
\bibitem[Sestito, Randich, \& Pallavicini (2004)]{SE04} Sestito, P., Randich, S., \& Palavicini, R. 2004, \aap, 426, 809
\bibitem[Sills \& Deliyannis (2000)]{SD00} Sills, A., \& Deliyannis, C.~P. 2000, \apj, 544, 944
\bibitem[Sneden (1973)]{SN73} Sneden, C. 1973, \apj, 184, 839
\bibitem[Sneden et al. (2022)]{SN22} Sneden, C., Af\c{s}ar, M., Bozkurt, Z., et al. 2022, \apj, 940, 12 SU
\bibitem[Steinhauer (2003)]{ST03} Steinhauer, A. 2003, PhD thesis, Indiana University
\bibitem[Steinhauer \& Deliyannis (2004)]{ST04} Steinhauer, A., \& Deliyannis, C.~P. 2004, \apjl, 614, L65
\bibitem[Stetson (2000)]{ST00} Stetson, P.~B. 2000, \pasp, 112, 925 (ST00)
\bibitem[Sun et al. (2019)]{SU19} Sun, W., de Gris, R., Deng, L., \& Albrow, M.~D. 2019, \apj, 876, 113
\bibitem[Sun et al. (2025a)]{SU25} Sun, Q., Deliyannis, C.~P., Anthony-Twarog, B.~J., et al. 2025a, {\it Nature Comm.}, in press
\bibitem[Sun et al. (2023)]{SU23} Sun, Q., Deliyannis, C.~P., Steinhauer, A., Anthony-Twarog, B.~J., \& Twarog, B.~A. 2023, \apj, 952, 71
\bibitem[Sun et al. (2020)]{SU20} Sun, Q., Deliyannis, C.~P., Twarog, B.~A., Anthony-Twarog, B.~J., \& Steinhauer, A. 2020, \aj, 129, 246
\bibitem[Sun et al. (2022)]{SU22} Sun, Q., Deliyannis, C.~P., Twarog, B.~A., et al. 2022, \mnras, 513, 5387 
\bibitem[Sun et al. (2025b)]{SU26a} Sun, Q., Deliyannis, C.~P., Twarog, B.~A., Anthony-Twarog, B.~J., \& Steinhauer, A. 2025b, \apj, 992, 75 
\bibitem[Sun et al. (2025c)]{SU26b} Sun, Q., Ting, Y.-S., Anthony-Twarog, B.~J., et al. 2025c, \apj, 991, 185
\bibitem[Tautvai\u{s}ien\.{e} et al. (2005)]{TA05} Tautvai\u{s}ien\.{e}, G., Edvardsson, B., Puzeras, E., \& Ilyin, I. 2005, \aap, 431, 933
\bibitem[Taylor (2005)]{TOPC} Taylor, M.~B. (2005), {\it Publications of the Astronomical Society of the Pacific}, 347, 297, ed. P. Shopbell, M. Britton, R. Ebert 
\bibitem[Terndrup et al. (2002)]{TE02} Terndrup, D.~M., Pinsonneault, M., Jeffries, R.~D., et al. 2002, \apj, 576, 950
\bibitem[Tody (1986)]{TODY} Tody, D. 1986, \procspie, 627, 733
\bibitem[Twarog \& Anthony-Twarog (1996)]{TW96}  Twarog, B.~A. \& Anthony-Twarog, B.~J. 1996, \aj, 112, 1500
\bibitem[Twarog et al. (2023)]{TW23} Twarog, B.~A., Anthony-Twarog, B.~J., \& Deliyannis, C.~P. 2023, \aj, 165, 105
\bibitem[Twarog et al. (2020)]{TW20} Twarog, B.~A., Anthony-Twarog, B.~J., Deliyannis, C.~P., \& Steinhauer, A. 2020, in {\it Li in the Universe: to Be or not to Be}, \memsai, 91, 74
\bibitem[Twarog et al. (2015)]{TW15} Twarog, B.~A., Anthony-Twarog, B.~J., Deliyannis, C.~P., \& Thomas, D.~T. 2015, \aj, 150, 134
\bibitem[Twarog et al. (1997)]{TW97} Twarog, B.~A., Ashman, K., \& Anthony-Twarog, B.~A. 1997, \aj, 114, 2556
\bibitem[Twarog \& Tyson (1985)] {TT85} Twarog, B.~A. \& Tyson, N. 1985, \aj, 90, 1247
\bibitem[Vallenari et al. (2000)]{VA00} Vallenari, A., Carraro, G., \& Richichi, A. 2000, \aap, 353, 147
\bibitem[van Dokkum (2001)]{VD01} van Dokkum, P.~G. 2001, \pasp, 113, 1420
\bibitem[VandenBerg (1985)]{VA85} VandenBerg, D.~A. 1985, \apjs, 58, 711
\bibitem[VandenBerg, Bergbusch, \& Dowler (2006)]{VA06} VandenBerg, D.~A., Bergbusch, P.~A., \& Dowler, P.~D. 2006, \apjs, 162, 375 
\bibitem[VandenBerg et al. (2014)]{VA14} VandenBerg, D.~A., Bergbusch, P.~A., Ferguson, J.~W., \& Edvardsson, B. 2014, \apj, 794, 72
\bibitem[Waters \& Hollek (2013)]{WH13} Waters, C.~Z., \& Hollek, J.~K. 2013, \pasp, 125, 1164
\bibitem[Wu et al. (2016)]{WU16} Wu, X., Li, C., de Grijs, R., \& Deng, L. 2016, \apjl, 826, L14
\bibitem[Wu et al. (2007)]{WU07} Wu, Z-Y., Zhao, X., Ma, J., et al. 2007, \aj, 133, 2061
\bibitem[Yang et al. (2018)]{YA18} Yang, Y., Li, C., Deng, L., de Grijs, R., \& Milone, A.~P. 2018, \apj, 859, 98

\end{thebibliography}
\end{document}